\documentclass[twoside,11pt]{article}

\usepackage{jmlr2e}
\usepackage{amsmath}
\usepackage{natbib}
\usepackage{enumerate}
\usepackage{url} 
\usepackage{xr}
\usepackage{colortbl,color}
\usepackage{multirow,tabularx,amsfonts,algorithm,booktabs,algpseudocode}
\usepackage{subcaption, caption}
\usepackage{booktabs}

\newcommand{\rank}{\mathrm{rank}}

\renewcommand{\textit}{}

\newcommand{\diag}{\mathrm{diag}}
\newcommand{\argmin}{\mathrm{argmin}}
\newcommand{\argmax}{\mathrm{argmax}}

\newcommand{\var}{\mathrm{var}}

\newcommand{\bbB}{{\bf B}}
\newcommand{\bbC}{{\bf C}}
\newcommand{\bbc}{{\bf c}}
\newcommand{\bbD}{{\bf D}}

\newcommand{\bbe}{{\bf e}}
\newcommand{\bbE}{{\bf E}}

\newcommand{\bbP}{{\bf P}}

\newcommand{\bbH}{{\bf H}}

\newcommand{\bbw}{{\bf w}}
\newcommand{\bbI}{{\bf I}}

\newcommand{\bbM}{{\bf M}}

\newcommand{\bbN}{{\bf N}}

\newcommand{\bbQ}{{\bf Q}}
\newcommand{\bbq}{{\bf q}}
\newcommand{\bbO}{{\bf O}}

\newcommand{\bbS}{{\bf S}}

\newcommand{\bbU}{{\bf U}}

\newcommand{\bbV}{{\bf V}}
\newcommand{\bbv}{{\bf v}}

\newcommand{\bbW}{{\bf W}}

\newcommand{\bbX}{{\bf X}}

\newcommand{\bbx}{{\bf x}}

\newcommand{\bbY}{{\bf Y}}
\newcommand{\bby}{{\bf y}}
\newcommand{\bbZ}{{\bf Z}}

\newcommand{\non}{\nonumber \\}
\newcommand{\bbA}{{\bf A}}

\newcommand{\E}{{\rm I}\kern-0.18em{\rm E}}

\renewcommand{\hat}{\widehat}

\newcommand{\p}{{\rm I}\kern-0.18em{\rm P}}
\newcommand{\1}{{\rm 1}\kern-0.24em{\rm I}}

\newcommand{\bPi}{\mbox{\boldmath $\Pi$}}

\newcommand{\bpi}{\mbox{\boldmath $\pi$}}

\newcommand{\R}{{\rm I}\kern-0.18em{\rm R}}
\newcommand{\be}{\mbox{\bf e}}

\newtheorem{deff}{Definition}
\newtheorem{lem}{Lemma}
\newtheorem{prop}{Proposition}
\newtheorem{thm}{Theorem}
\newtheorem{coro}{Corollary}
\newtheorem{rmk}{Remark}

\newtheorem{cond}{Condition}

\usepackage{lastpage}
\jmlrheading{00}{2023}{1-\pageref{LastPage}}{1/23; Revised 0/00}{0/00}{21-0000}{Han et al}

\ShortHeadings{Partial information in social networks}{Han et al.}
\firstpageno{1}

\begin{document}

\title{Individual-centered partial information in social networks}

\author{\name Xiao Han \email xhan011@ustc.edu.cn \thanks{Han and Wang contribute equally to the work. Yang and Tong are corresponding authors. } \\
       \addr International Institute of Finance, School of Management \\
       University of Science and Technology of China\\
       Hefei, 230052, China
       \AND
       \name Y. X. Rachel Wang \email rachel.wang@sydney.edu.au \\
       \addr School of Mathematics and Statistics\\
       University of Sydney\\
       NSW, 2006, Australia
       \AND
       \name Qing Yang \email yangq@ustc.edu.cn \\
       \addr International Institute of Finance, School of Management \\
       University of Science and Technology of China
       \\
       Hefei, 230052, China
       \AND
       Xin Tong \email xint@marshall.usc.edu\\
       \addr Department of Data Sciences and Operations, Marshall School of Business\\
       University of Southern California\\
       CA, 90089, USA
       }

\editor{}

\maketitle

\begin{abstract}
	
In statistical network analysis, we often assume either the full network is available or multiple subgraphs can be sampled to estimate various global properties of the network. However, in a real social network, people frequently make decisions based on their local view of the network alone. Here, we consider a partial information framework that characterizes the local network centered at a given individual by path length $L$ and gives rise to a partial adjacency matrix. Under $L=2$, we focus on the problem of (global) community detection using the popular stochastic block model (SBM) and its degree-corrected variant (DCSBM). We derive theoretical properties of the eigenvalues and eigenvectors from the signal term of the partial adjacency matrix and propose new spectral-based community detection algorithms that achieve consistency under appropriate conditions. Our analysis also allows us to propose a new centrality measure that assesses the importance of an individual's partial information in determining global community structure. Using simulated and real networks, we demonstrate the performance of our algorithms and compare our centrality measure with other popular alternatives to show it captures unique nodal information. Our results illustrate that the partial information framework enables us to compare the viewpoints of different individuals regarding the global structure.

\end{abstract}

\begin{keywords}
  community detection, centrality measure, partial information
\end{keywords}

\section{Introduction}

Much of the statistical network literature is focused on estimating global properties of graphs by either using the whole graphs or combining information across appropriately sampled subgraphs. However, in social networks, despite the prevalence of social media tools, most individuals still have limited understanding of the information that exists beyond their local network, e.g., friends' friends. In this way, one often needs to make important decisions, such as whether or not to share sensitive information with friends belonging to different social circles, based on their limited local view of the global network. In this paper, we adopt such an individual-centered perspective to study global structures of social networks.

\begin{figure}[h]
	\begin{center}
		\includegraphics[width=0.9\textwidth]{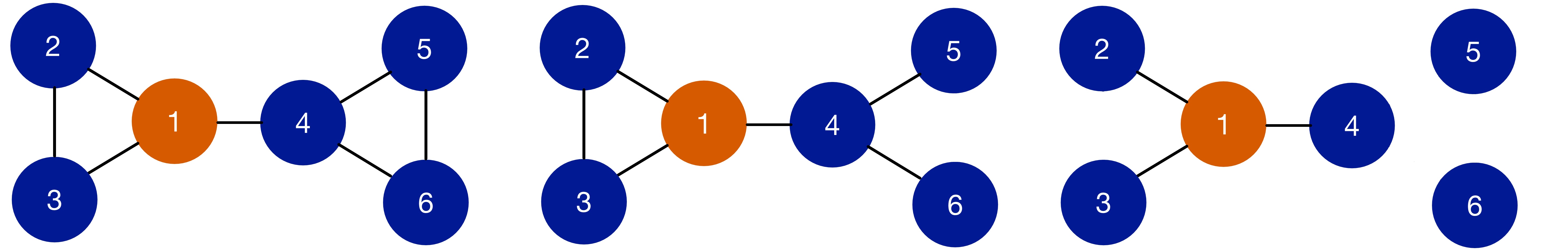}
	\end{center}
	\caption{\small A toy network of $6$ individuals. $L=3$ or \emph{full} (left); $L=2$ (middle); $L=1$ (right). \label{fig:network_partial}}
\end{figure}

To formalize an individual's local network structure, we consider an \emph{individual-centered partial information framework} that uses path length to characterize connections visible (proximal) and hidden (distant) to the individual. Concretely, let $G = (\mathcal{V}, \mathcal{E})$ denote the global network of interest, where $\mathcal{V}=\{1, \ldots, n\}$ is the index set of all individuals (nodes) and $\mathcal{E}$ is the set edges, assumed to be unweighted and undirected for simplicity, between individuals. We characterize an individual's partial knowledge of the network by their \emph{knowledge depth}: an individual $i$ has knowledge depth $L$ if all paths (starting from $i$) of length up to $L$ in the network are known to the individual. Figure  \ref{fig:network_partial} illustrates the knowledge depth concept with a toy example, where the left panel is the full network. Taking individual $1$ as the individual of interest, the left, middle and right panels correspond to their perceived networks given knowledge depths $L=3, 2, 1$,  respectively.
In choosing the value of $L$ for our study, we first note that the case $L=1$ leads to a too simplistic graph structure around the individual. On the other hand, the ``six degrees of separation" phenomenon \citep{watts1998collective} in social networks suggests moderate values of $L$ are likely to return the full network already.
Both $L=2$ and $L=3$ are more interesting cases to study and coincide with most people's experience in real social networks. We focus on $L=2$ in this paper and leave $L=3$ for future studies.

The structure of the partial network considered in our work is related to popular sampling schemes in social sciences, including egocentric sampling \citep{freeman1982centered, wasserman1994social}, snowball sampling and respondent-driven sampling (RDS) \citep{goodman1961snowball, heckathorn1997respondent, sh2004}, which all work by first selecting individuals as ``seeds'', then expanding into their neighborhoods according to certain criteria. Information across the sampled subgraphs can be combined to estimate parameters in network models and node covariates \citep{handcock2010modeling, rohe2019critical}. Egocentric sampling and other random sampling schemes have also been used to estimate community structure \citep{mukherjee2021two} and subgraph counts \citep{bhattacharyya2015subsampling}. Compared to these  works, a major conceptual difference in our framework is that we are interested in understanding the network structure visible to \emph{each} node and how these structures differ for different nodes. In addition, sampling multiple seeds may be infeasible in certain networks with restricted access (e.g., a terrorist network). These observations motivate us to consider how to infer global information using \emph{one} local network only.

In this paper, we choose community memberships to be the global network feature we aim to infer from an individual-centered partial network. Community detection is one of the most studied statistical network problems, with the stochastic block model (SBM, \cite{holland1983stochastic,abbe2017}) and its variants including the degree-corrected SBM (DCSBM, \cite{DCSBM2011}) and the mixed membership SBM (MMSBM, \cite{Airoldi:2008}) being popular generative models. There is a rich line of literature on community detection under these models; many methods are based on likelihood or spectral approaches \citep{bickel2009nonparametric, zhao2012, rohe2011, lei2015consistency, anandkumar2014tensor, jin2015, jin2017estimating}.
Much attention has also been paid to other statistical inference problems in this high-dimensional network setting,  including inference for the number of communities (e.g., \citet{Bickel2016, L16,wang2017likelihood, saldana2017many, han2019universal}) and the membership profiles (e.g., \citet{FFH18}). We adopt the spectral approach in this paper. Since the adjacency matrix of the partial network is significantly different from the global adjacency matrix, we first perform detailed theoretical analysis of its spectral properties before proposing novel spectral algorithms for inferring the global community memberships under the SBM and DCSBM models.

\begin{figure}[hbtp!]
	\captionsetup[subfigure]{aboveskip=-1pt,belowskip=-2pt}
	\centering
	\begin{subfigure}[t]{0.4\textwidth}
		\centering
		\includegraphics[width=0.8\textwidth]{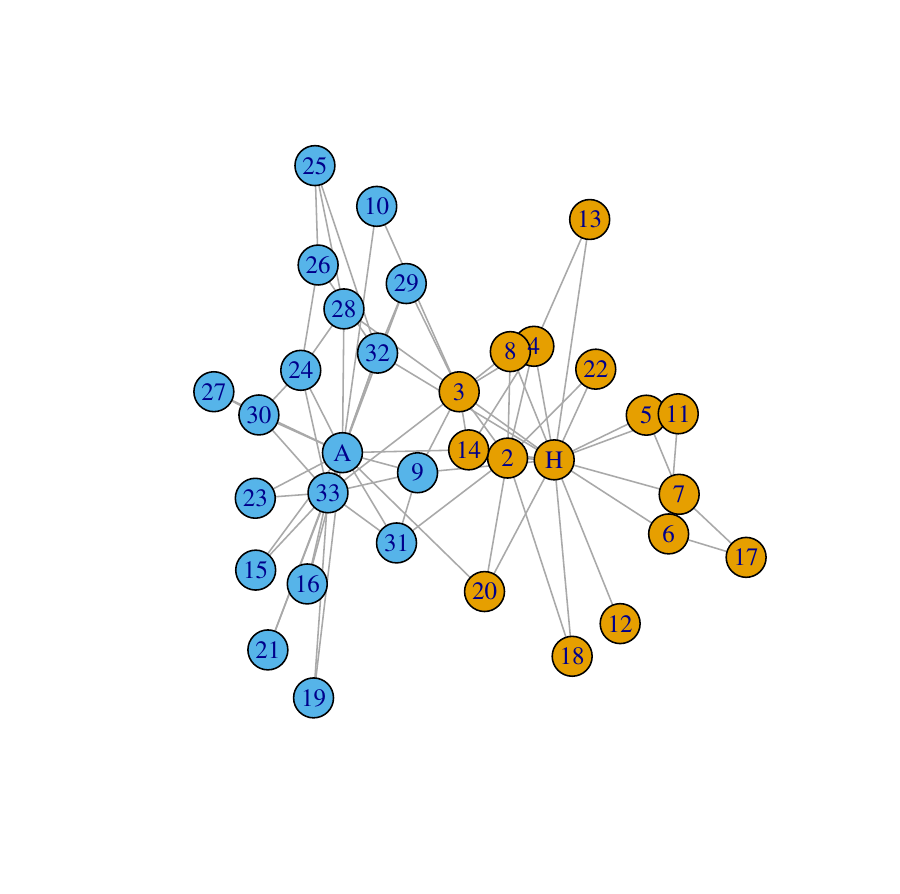}
		\caption{}
		
	\end{subfigure}%
	\begin{subfigure}[t]{0.6\textwidth}
		\centering
		\includegraphics[width=1\textwidth]{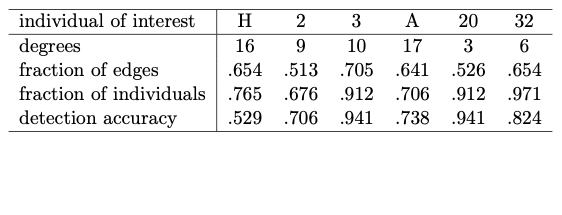}
		\caption{}
		\label{karate_table}
	\end{subfigure}
	\caption{The karate club network with nodes colored by ground truth community labels in (a) and community detection results (using Algorithm \ref{alg3}) for chosen individuals in (b).}
	\label{karate_network}
\end{figure}

Our partial information framework and new clustering algorithms enable us gain interesting insight into the relationship between global community detection and the local features perceived by a given individual. As a motivating example, take the well-known Zakary's karate club data. Figure~\ref{karate_network} shows the network colored by ground truth community labels, and the results of community detection using our algorithm for a few chosen individuals. The network has two obvious hubs as demonstrated by their degrees: the instructor ``Mr. Hi" (node ``H'' ) and the administrator ``John A." (node ``A'' ). It turns out  surprisingly that these nodes perform less well than some nodes with significantly smaller degrees (nodes 3, 20, 32), suggesting their partial networks are not the most informative when it comes to understanding the affiliation structure of the whole network. In comparison, node 20, despite their  small degree and observing a smaller fraction of edges or individuals through their $L=2$ local network compared to nodes 3 and 32, achieves one of the highest detection accuracy rates. Noting that two of the three connections of node 20 come from hubs ``H'' and ``A'', the result implies that not just how many friends one knows, but also \emph{whom} one knows, matters in inferring global structure. We describe our algorithm (Algorithm \ref{alg3}), theoretical analysis and interpretations in Section \ref{sec:DCSBM}. More analysis of this network can be found in Section \ref{sec:karate}.

Our paper makes the following main contributions:
\begin{itemize}
	\item
	We propose an individual-centered partial information framework to study social networks from \emph{an individual}'s perspective, where the person's understanding of the whole network is characterized by their knowledge depth $L$.
	\item
	We address the technical challenges brought about by the partial adjacency matrix, deriving novel results on its spectral properties for $L=2$. Different from the conventional case, the signal term of this matrix is random itself. Under some generic assumptions, we derive its rank, the explicit forms of its eigenvectors and the approximate forms of its eigenvalues, which can be of independent theoretical interest.
	\item
	Adopting the SBM as the generative model, we propose a new spectral-based algorithm for community detection and show that it achieves consistent recovery. The algorithm and theoretical guarantee can be further extended to the DCSBM, which
	allows us to interpret the convergence rate when individual (node) heterogeneity is reflected by their degrees and whom they are connected to.
	\item
	As a by-product of our analysis, we also propose a new centrality measure for assessing the importance of a node in determining global community structure. We demonstrate this measure captures unique information by comparing it with other popular centrality measures on simulated and real data.
\end{itemize}
   Finally, we note that while this paper focuses on inferring community memberships for all nodes in the network, a natural alternative is to consider only individuals reached by the partial network of knowledge depth $L$ (e.g., individuals 5 and 6 in Figure~\ref{fig:network_partial} would be excluded under $L=1$). However, in this case, the individuals included would be random, adding another layer of complexity to the problem. We leave the theoretical analysis of this case to future work, but still demonstrate empirically our current algorithms can be applied in this situation to a real dataset in Section \ref{sec:polblogs}.

The rest of the paper is organized as follows.  In Section \ref{sec:notation and main term}, we introduce the setup of our problem and $\bbB_E$, the signal term of the partial adjacency matrix with knowledge depth $L=2$.  We analyze the spectral properties of $\bbB_E$ to motivate our detection algorithms and propose a centrality measure for each node based on the magnitude of eigenvalues. Section \ref{sec:main_results} describes our algorithms and consistency properties under the SBM and DCSBM. We demonstrate the performance of the algorithms and centrality measure using a variety of simulated and real data in Sections \ref{sec:simulation} and \ref{sec:real data} respectively. Finally, we conclude with a discussion section. All the proofs and additional simulation results can be found in the Appendix.

\section{Setup and preliminary results}\label{sec:notation and main term}

\subsection{Notations}

We first introduce some notations that will be used throughout the paper. For a matrix $\bbM=(m_{ij})$, in which $m_{ij}$ is the $(i,j)$-th entry, denote the $i$th row of $\bbM$ by $\bbM(i)$. Let $\|\bbM\|_{\max}=\max_{i,j}\{|m_{ij}|\}$ and $\|\bbM\|$ be the spectral norm of $\bbM$, which is the square root of the  largest eigenvalue of $\bbM\bbM^\top$.  Moreover, we denote the Frobenious norm of $\bbM$ by $\|\bbM\|_F=\left[\text{tr}(\bbM\bbM^\top)\right]^{1/2}$. For any random matrix (or vector) $\bbM$, we use $\E\bbM$ to denote its expectation. We use $\|\cdot\|_2$ to denote the  $L_2$ norm of a vector.  If two  positive sequences $a_n$ and $b_n$ satisfy $\limsup_{n\rightarrow \infty} (a_n/b_n) <\infty$, we denote $a_n\lesssim b_n$ or $b_n\gtrsim a_n$. If $a_n\lesssim b_n$ and $b_n\lesssim a_n$, we write $a_n\sim b_n$. We write $a_n\ll b_n$ or $b_n\gg a_n$ if $\lim_{n\rightarrow \infty}(a_n/b_n)=0$. For two symmetric matrices $\bbM_n$ and $\bbN_n$, if there exists a positive constant $c$ independent of $n$, $\bbN_n$ and $\bbM_n$ such that $\bbN_n-c\bbM_n$ is a semi-positive definite matrix, then we write $\bbM_n\lesssim \bbN_n$; specifically, if $c=1$, then we write $\bbM_n\le \bbN_n$. If there exists a positive diverging sequence $c_n$ (i.e., $c_n\rightarrow \infty$) such that  $\bbN_n-c_n\bbM_n\ge 0$ for all $n$ large, we write $\bbM_n\ll\bbN_n$. We denote the $i$-th largest eigenvalue and singular value of $\bbM$ by $\lambda_i(\bbM)$ and $\sigma_i(\bbM)$, respectively. 
For any positive integer $K$, let $[K]= \{1, \ldots, K\}$, $[-K] = \{-K, \ldots, -1\}$,  and $[\pm K] = \{-K, \ldots, -1, 1, \ldots, K\}$. Standard order notations such as $O(\cdot)$, $o(\cdot)$, $O_p(\cdot)$, $o_p(\cdot)$ are also used. We use $|\cdot|$ to denote the cardinality of a set.     Throughout the paper, $c$ and  $C$ denote constants that may vary from line to line. Events \emph{with high probability} (w.h.p.) are defined as:

\begin{deff}
		We say a sequence of events $\mathcal{A}_n$ holds with high probability if for any positive constant $D$, there exists an $n_0(D)\in \mathbb{N}$ such that $\forall$ $n\ge n_0(D)$, $\p(\mathcal{A}_n)\ge 1-n^{-D}$.
	\end{deff}

\subsection{The partial adjacency matrix and its signal term}

Recall that $\bbA=(a_{ij})$ is the $n\times n$ adjacency matrix of $G=(\mathcal{V}, \mathcal{E})$, the full network, in which $a_{ij}= 1 $ if $(i,j)\in\mathcal{E}$ and 0 otherwise. Let $K = \text{rank}(\E\bbA)$ and assume that $\bbA=\bbA^\top$ and $\{a_{ij}\}_{1\le i\le j\le n}$ are independent Bernoulli random variables with expectation $\E\bbA$. We assume that $K$ is a constant and denote the (reduced form) eigen decomposition of $\E\bbA$ by
$\E\bbA=\bbV\bbD\bbV^\top\,,$
where $\bbD=\text{diag}(d_1,\ldots,d_K)$ with $d_i$ being the $i$-th largest eigenvalue (by magnitude) of $\E\bbA$ and $\bbV=(\bbv_1,\ldots,\bbv_K)$ being the corresponding eigenvector matrix\footnote{Strictly speaking, this model allows self-loops while real networks often do not.  The inclusion of self-loops does not change our conclusions, as explained in the Appendix.}.

Without loss of generality, for convenience we assume that the partial network is always \emph{centered around individual 1} in the network. Let $\bbB=(b_{ij})$ be  individual $1$'s perceived adjacency matrix with knowledge depth $L=2$. An example of $\bbA$ and $\bbB$ is illustrated in the left and middle panels of Figure~\ref{fig:network_partial}, respectively.
More generally, $b_{ij}$ takes the form
$\label{gts5}
b_{ij}=a_{ij}(1-\1(a_{1i}=0)\1(a_{1j}=0))\,, \ i,j\in  [n]\,,
$
where $\1(\cdot)$ is an indicator function.
Then it follows that
\begin{equation}\label{gba1}
	\bbB=-\bbS\bbA\bbS+\bbA\bbS+\bbS\bbA, \quad \text{where} \quad\bbS=\text{diag}(a_{11},\ldots,a_{1n})\,.
\end{equation}
We further define
\begin{equation}\label{gba1s}
	\bbB_E=-\bbS(\E\bbA)\bbS+(\E\bbA)\bbS+\bbS(\E\bbA)\,.
\end{equation}
We will next show that $\bbB_E$ is a signal term of $\bbB$ and analyze its spectral properties.

Unlike the case of full information network, where $\E \bbA$ is the signal term of $\bbA$, the partial information network has a more subtle situation.  As shown in the next lemma, the obvious candidates $\E\bbB$ and $-(\E\bbS)(\E\bbA)(\E\bbS)+(\E\bbA)(\E\bbS)+(\E\bbS)(\E\bbA)$ are much smaller than $\bbB_E$.

\begin{lem}\label{majorterm}
	In the simplest scenario that $\p(a_{ij}=1)=p_n = o(1), \text{ for } i,j \in [n]$, we have
	\begin{equation}\label{gtq18}
		\|-\E\bbS(\E\bbA)\E\bbS+(\E\bbA)\E\bbS+\E\bbS(\E\bbA)\|+\|\E\bbB\|=o_p(\|\bbB_E\|)\,.
	\end{equation}
\end{lem}
It is well known that spectral clustering relies on the leading eigenvectors.  In view of Lemma \ref{majorterm}, the leading eigenvalue of $\bbB_E$ is much larger than those of $\E\bbB$ and $-(\E\bbS)(\E\bbA)(\E\bbS)+(\E\bbA)(\E\bbS)+(\E\bbS)(\E\bbA)$ when $p_n=o(1)$, a typical asymptotic condition for large networks.  In other words, the latter two matrices do not contribute to the leading eigenvectors, and neither can be a signal term. To see why this is the case, note that $\E\bbB$ is almost equal to $-(\E\bbS)(\E\bbA)(\E\bbS)+(\E\bbA)(\E\bbS)+(\E\bbS)(\E\bbA)$ (except for differences in the diagonal entries, the first row and first column). Consider $\bbS\bbA\bbS$ and $(\E\bbS)(\E\bbA)(\E\bbS)$, the first terms in $\bbB$ and $\E\bbB$, respectively.
Intuitively, the singular values of $\bbS\bbA\bbS$ are equivalent to those of $\bbA\bbS$, given the fact that $\bbS^2=\bbS$. When $p_n=o(1)$, we can see that $(\E\bbS)^2\ll \E\bbS$, indicating multiplying $\E\bbA$ by $\E\bbS$ twice would significantly deviate it from the target matrix $\bbS\bbA\bbS$. We refer to our proof of Lemma \ref{majorterm} for more details. 

On the other hand, Lemma  \ref{errorbound}, to be introduced in the next section, shows that under appropriate conditions, the spectral norm   $\|\bbB - \bbB_E\|\ll$ the smallest (in magnitude) non-zero eigenvalue of $\bbB_E$. Thus $\bbB_E$, which is random itself, is a signal term of $\bbB$ from the spectral point of view. Intuitively, $\bbB_E$ only differs from the conditional expectation $\E[\bbB|\bbS]$ in the first row and column, thus the latter can be considered as one way to interpret $\bbB_E$. As will be shown in the next section, $\bbB_E$ has an explicit low-rank structure, lending itself to explicit spectral analysis.

\subsection{Spectral properties of  the signal term $\bbB_E$}\label{sec:B}

In this section, we present a few key theoretical properties of $\bbB_E$ and establish that $\bbB_E$ is a signal term of $\bbB$.  In addition to showing the order of its eigenvalues, we derive the exact forms of the eigenvectors and approximate forms of the eigenvalues. The results here depend only on generic assumptions about invertibility and the eigen decomposition of $\E\bbA$.

Recall $\bbS = \text{diag}(a_{11}, \ldots, a_{1n})$ and $\E\bbA = \bbV \bbD \bbV^{\top}$, where $\bbV$ is of dimensions $n\times K$. We have the following theorem regarding the spectral properties of $\bbB_E$.
\begin{thm}\label{thm2}
	Suppose that $\bbV^\top\bbS\bbV$ and $\bbI-\bbV^\top\bbS\bbV$ are invertible. Denote  $\bbH(x)=\bbI-x\bbD\bbV^\top\bbS\bbV-x^2\bbD(\bbI-\bbV^\top\bbS\bbV)\bbD\bbV^\top\bbS\bbV$.
	Then the determinant equation
	\begin{equation}\label{g0}
		\emph{det}\left(\bbH(x)\right)=0
	\end{equation}
	has $2K$   non-zero real  solutions; we denote them by $x_{-K},\ldots,x_{-1}, \, x_1,\ldots,x_K$, with $x_{i}\le x_j$ for all  $i<j$. Moreover, for $i\in[\pm K]$, let  $\bbq_{1i}$ be an eigenvector of $\bbH(x_i)$  corresponding to the zero eigenvalue, and
	$\bbq_{2i}=x_i\bbD\bbV^\top\bbS\bbV\bbq_{1i}\,.
	$
	Then $\bbq_i$ defined as
	\begin{equation}\label{gtq17}
		\bbq_i=\bbS\bbV\bbq_{1i}+(\bbI-\bbS)\bbV\bbq_{2i}
	\end{equation}
	is an eigenvector of
	$\bbB_E\,$ corresponding to the  eigenvalue $x_i^{-1}$.
	
	Conversely,
	if $\bbq_0 \neq \mathbf{0}$ is an eigenvector of
	$\bbB_E\,$ corresponding to a non-zero  eigenvalue
	$x_0^{-1}$, then $x_0$ satisfies~\eqref{g0}.
	Moreover, $\bbq_0$ can be decomposed (in the form of~\eqref{gtq17}) as $
	\bbq_0=\bbS\bbV\bbq_{10}+(\bbI-\bbS)\bbV\bbq_{20}\,,$
	where $\bbq_{10}$ is an eigenvector of $\bbH(x_0)$ corresponding to the zero eigenvalue and $\bbq_{20}=x_0\bbD\bbV^\top\bbS\bbV\bbq_{10}$.
	
	Finally, we have $\emph{rank}\left(\bbB_E\right)=2K\,.$
	
\end{thm}

\begin{rmk}\label{rmk2}
	(i) Under our setting, the invertibility assumption on $\bbV^\top\bbS\bbV$ and $\bbI-\bbV^\top\bbS\bbV$ is  not stringent. To see this, note that
	$\bbV^\top\bbS\bbV=\sum_{i=1}^na_{1i}\bbV^\top(i)\bbV(i)\,.$
	Since $n$ is large and $\bbV^\top\bbS\bbV$ is a  matrix of small dimensionality $K\times K$, the invertibility of $\bbV^\top\bbS\bbV$ should be satisfied if individual $1$ has enough neighbors. Similarly, the invertibility of $\bbI-\bbV^\top\bbS\bbV$ can be ensured if individual $1$ does not directly connect to almost everyone else in the network, which is sensible since most real networks are sparse.
	
	\noindent (ii) The rank of $\bbB_E$ and the form of eigenvectors play an important role in motivating our community detection algorithms in Section~\ref{sec:main_results}. The two terms in Eq \eqref{gtq17} naturally split the individuals into two subsets: those who are neighbors of individual 1 (represented by $\bbS$) and those who are not (represented by $\bbI-\bbS$). As we will show later, our algorithm first performs clustering on these two subsets separately, before merging them into $K$ communities.
\end{rmk}
In what follows, we introduce a few conditions to perform further analysis of our partial network.
Denote
$p_n=\max_{i,j}\p(a_{ij}=1)$.
\begin{cond}\label{cond2}
	$\min_{j\ge 2}\p(a_{1j}=1)\sim p_n$ and $1-c>p_n\gg\ \log n /n$ for some constant $c>0$.
\end{cond}
\begin{cond}\label{cond3}
	$\|\bbV\|_{\max}\le C / \sqrt n$ for some constant $C>0$.
\end{cond}
\begin{cond}\label{cond1}
	$|d_1|\sim|d_2|\sim\ldots\sim |d_K|\sim np_n$.
\end{cond}

Conditions \ref{cond2}--\ref{cond3} are sufficient to ensure the invertibility of $\bbV^\top\bbS\bbV$ and $\bbI-\bbV^\top\bbS\bbV$ with high probability, as we will prove in Lemma \ref{le2}. We note that Conditions \ref{cond2}--\ref{cond3} can be relaxed but we adopt them in the paper for convenience and simplicity. Condition \ref{cond1} is a strong condition to assume that the magnitude of the smallest non-zero eigenvalue of $\E\bbA$ has the same order as $\|\E\bbA\|_F$, which is to ensure a big enough gap between $|d_K|$ and $\|\bbA-\E\bbA\|$ for more straightforward analysis. This condition could also be relaxed by deeper analysis, but we leave it for future studies. We have the following invertibility lemma.

\begin{lem}\label{le2}
	Under Conditions \ref{cond2} and \ref{cond3}, there exists a positive constant $c$ such that w.h.p., we have
	\begin{equation}\label{eq:invertible}
		cp_n\left(1-\sqrt[4]{\frac{\log n}{np_n}}\right)\bbI\le\bbV^\top\bbS\bbV\le p_n\left(1+\sqrt[4]{\frac{\log n}{np_n}}\right)\bbI<\left(1-\frac{c}{2}\right)\bbI\,.
	\end{equation}
	Therefore w.h.p., $\bbV^\top\bbS\bbV$ and $\bbI-\bbV^\top\bbS\bbV$ are invertible.
\end{lem}

Theorem \ref{thm2}  and Lemma \ref{le2} imply the following corollary.

\begin{coro}\label{coro1}
	Under Conditions \ref{cond2} and \ref{cond3}, for suitably chosen $\bbq_{1l}$,  $l\in [\pm K]$, $\bbq_{l}$'s, as defined in Theorem \ref{thm2},  satisfy w.h.p. that
	\begin{equation}\label{gts2}
		\bbq_i^\top\bbq_j=0\,,\quad i\neq j, \ i,j\in [\pm K]\,,
	\end{equation}
	and that 
	$\emph{dim}(\emph{span}\{\bbq_l, l\in[\pm K]\})= \emph{rank}(\bbB_E)  =2K$.

\end{coro}

Corollary \ref{coro1} specifies the form of orthogonal eigenvectors in the eigen-decomposition of $\bbB_E$. Even though the eigenvalues in Theorem \ref{thm2} may have cardinality greater than 1, this does not change the eigenspace of $\bbB_E$, which is equivalent to $\bbQ\bbQ^T$. As far as spectral clustering is concerned, $\bbq_i$ do not need to be uniquely defined, and separability of eigenvectors for different nodes is typically identifiable up to an orthogonal transformation. As we develop our community detection algorithm under the SBM and DCSBM in Section~\ref{sec:main_results}, we will use the eigenvectors of the observed $\bbB$ and bound the deviation from their counterparts in $\bbB_E$.

\begin{thm}\label{thm1}
	Under Conditions \ref{cond2}-\ref{cond1}, w.h.p., we have $|x_{i}|^{-1}\sim np_n^{3/2}$ for $i\in[\pm K]$. Furthermore, if	$p_n\rightarrow 0$ and 	$\mu_i/\mu_{i+1}\ge 1+c$ for some positive constant $c$, $1\le i\le K-1$ where  $\mu_1\ge\ldots \ge \mu_K$  are the eigenvalues of $\bbV^\top(\E\bbS)\bbV\bbD(\bbI-\bbV^\top(\E\bbS)\bbV)\bbD$, it holds w.h.p. that,
	\begin{align*}
		x_{i}^{-1} & =\left(\lambda_i(\bbD(\bbI-\bbV^\top\bbS\bbV)\bbD\bbV^\top\bbS\bbV)\right)^{\frac{1}{2}}\left(1+o(\sqrt{p_n})\right)\,,\ i\in [ K]\,,\\
		x_{i}^{-1} & =-\left(\lambda_{K+i+1}(\bbD(\bbI-\bbV^\top\bbS\bbV)\bbD\bbV^\top\bbS\bbV)\right)^{\frac{1}{2}}\left(1+o(\sqrt{p_n})\right)\,,\ i\in [-K]\,.
	\end{align*}
\end{thm}

The first part of Theorem \ref{thm1} shows the order of the non-zero eigenvalues of $\bbB_E$. The second part, with more stringent conditions, finds the approximate expressions for these eigenvalues. We note in advance that  our community detection results do not depend on the additional conditions in Theorem \ref{thm1}, but the approximate expressions are of standalone interest and can be useful to study other problems under the partial information framework (e.g., in construction of centrality measure to be described in Section~\ref{subsec:centrality}).

Finally, the next lemma bounds the spectral norm of $\bbB  - \bbB_E$ and shows $\bbB_E$ is indeed a signal term of $\bbB$.

\begin{lem}\label{errorbound}
	It holds w.h.p. that
	$\label{2203.1}\|\bbB-\bbB_E\|\lesssim \sqrt{np_n}\,.
	$
	Under Conditions \ref{cond2}, \ref{cond3} and \ref{cond1}, assuming $p_n\gg\sqrt{1/n}$,  then we have w.h.p.,
	$\|\bbB-\bbB_E\|\ll \min_{i \in [2K]}\sigma_i(\bbB_E)\,.$
\end{lem}

\section{Community detection under partial information}
\label{sec:main_results}

We propose new spectral-based algorithms for performing community detection in the partial information framework, under the commonly used SBM and DCSBM settings. Using the spectral properties derived in the previous section, we provide upper bounds on the error rates and show that the recovery is almost exact under appropriate conditions. As an application utilizing the theoretical properties of $\bbB_E$, community detection requires more specialized model assumptions. Therefore the conditions we introduce in this section are sufficient conditions implying the more general Conditions \ref{cond2}--\ref{cond1}.

\subsection{Stochastic block model}\label{sec:sbm}
In the SBM \citep{holland1983stochastic}, each individual belongs to exactly one of $K$ different communities. The connection probability between two individuals depends on their community memberships.  Concretely, in the SBM with $K$ communities, $\E\bbA$ is given by
\begin{equation} \label{SBM}
	\E\bbA = \bPi\bbP\bPi^\top\,,
\end{equation}
where $\bbP=(p_{kl})$ is a symmetric $K\times K$ matrix in which $p_{kl}$ is the connection probability between communities $k$ and $l$, $\bPi = (\bpi_1,\ldots, \bpi_n)^\top \in \R^{n\times K}$ is the matrix of community membership vectors, with individual $i$'s membership vector $\bpi_i \in \{\be_1,\ldots, \be_K\}$, $\be_k\in \R^K$ is the standard basis vector with the $k$-th element being one and the other elements being zero. If individuals $i$, $j$ belong to community $k$ and $l$ respectively, we have $\E a_{ij}=\bpi_i^\top\bbP\bpi_j = p_{kl}$.

Before we present our community detection algorithm, we first need some technical results connecting the SBM with the low-rank model discussed in Section~\ref{sec:notation and main term}. Recall that $p_n=\max_{i,j}\p(a_{ij}=1)$. Without loss of generality, assume that individual 1 belongs to community 1. The following condition, which is easier to interpret in the SBM setting, implies Conditions \ref{cond2}, \ref{cond3} and \ref{cond1} in Section~\ref{sec:notation and main term}. Hence we will use this condition instead in our theoretical analysis under the SBM.

\begin{cond}\label{ncond1}
	$\min_{k\in[K]}p_{1k}\sim p_n$. $\min_{k\in[K]}\sum_{j\in[n]}\1(\bpi_j=\bbe_k)\ge c_0n$ and $\sigma_K(\bbP)\ge c_1p_n$ for some positive constants $c_0$ and $c_1$. Moreover,  for some $c>0$, $1-c\ge p_n\gg (1 / n)^{1/2}$.
\end{cond}

\begin{rmk}
\label{rmk:balanced_size}
    (i) Community detection on the entire graph typically requires $np_n \gtrsim \log n$, while our lower bound on the density $p_n$ is $1/\sqrt{n}$, which is more stringent due to significant information loss incurred by observing only a partial network. Note that the order of the largest singular value of $\E \bbA$ is essentially $np_n$, while the signal strength in our case is  $x_i^{-1}\sim np_n^{3/2}$, $i\in[\pm K]$, as shown in Theorem \ref{thm1}.

    \noindent(ii) The above condition assumes balanced community sizes to ensure the signal size in $\bbB_E$ is sufficiently large. As an example of imbalanced community sizes, consider a simple case with $\bbP=p_n\bbI_K$ and $\bPi = (\bpi_1,\ldots, \bpi_n)^\top$. Here for the first $K-1$ communities ($1\leq k \leq K-1$), $\bpi_i=\be_k$, $i \in \{n_1(k-1)+1, \dots, n_1 k\}$; for the last community, $\bpi_i=\be_K$, $i \in \{n_1(K-1)+1, \dots, n\}$. Let $n_2=n-n_1(K-1)$ and  assume $n_2\ll n$ so that the last community is much smaller than the others. As shown in Theorem \ref{thm1}, $x_K^{-1}\sim n_2p_n^{3/2}$. Since Lemma \ref{errorbound} shows the order of $\|\bbB-\bbB_E\|$ is $\sqrt{np_n}$, for signal dominance over noise, one would require that $n_2p_n^{3/2}\gg \sqrt{np_n}$, or $p_n \gg \frac{\sqrt{n}}{n_2}$, which is even more stringent than our current lower bound $1/\sqrt{n}$.
\end{rmk}

\begin{lem}\label{le4}
	Under the SBM defined in \eqref{SBM}, Condition \ref{ncond1} implies $\emph{rank}(\E\bbA) = K$ and Conditions \ref{cond2}, \ref{cond3} and \ref{cond1}. Moreover, there exists a $K\times K$ matrix $\mathcal{D}$ such that
	\begin{equation}\label{g18}
		\bbV=\bPi\mathcal{D} \quad \text{ and } \quad
		\mathcal{D}\mathcal{D}^\top\ge\frac{1}{n}\bbI\,.
	\end{equation}
\end{lem}
Thus under Condition \ref{ncond1}, Corollary \ref{coro1} allows us to write down an orthonormal eigenvector matrix $\bbQ=(\bbq_1,\ldots,\bbq_K,\bbq_{-1},\ldots,\bbq_{-K})$  such that $\bbQ^\top\bbQ=\bbI$ ($\|\bbq_i\|_2=1$).
Let  $\mathcal{Q}_i=(\bbq_{i,1},\ldots,\bbq_{i,K},\bbq_{i,-1},\ldots,\bbq_{i,-K})$ for $i=1,2$, 
then by \eqref{gtq17}, we have
\begin{equation}\label{g19}
	\bbQ=\bbS\bPi\mathcal{D}\mathcal{Q}_1+(\bbI-\bbS)\bPi\mathcal{D}\mathcal{Q}_2\,.
\end{equation}

The next lemma bounds the distances between the rows of $\bbQ$ by grouping their row indices according to (i) their community memberships and (ii) whether they are neighbors of node 1. This indicates that both information needs to be taken into account when constructing a community detection algorithm.

\begin{lem}\label{le6}
	Under Condition \ref{ncond1}, for any $2K \times 2K$ orthogonal matrix $\bbO$, it holds w.h.p. that for all $i,j\in[n]$,
	\begin{align}
		&\bpi_i\neq \bpi_j  \Longrightarrow \left\|\bbQ(i)\bbO-\bbQ(j)\bbO\right\|_2\ge \sqrt{\frac{2}{c_2n}}\,,\label{h1}\\
		&\bpi_i=\bpi_j, a_{1i}\neq a_{1j} \Longrightarrow \left\|\bbQ(i)\bbO-\bbQ(j)\bbO\right\|_2\ge \sqrt{\frac{2}{c_2n}}\,,\label{h1s}\\
		&\bpi_i=\bpi_j, a_{1i}= a_{1j}\Longrightarrow \left\|\bbQ(i)\bbO-\bbQ(j)\bbO\right\|_2=0\,.\label{h1ss}
	\end{align}
	Here $c_2$ is a positive constant defined in Lemma~\ref{le3} in the Appendix.
\end{lem}

In the case of SBM with the full network, it is well known that $\E\bbA$, the population version of $\bbA$, has exactly $K$ distinct rows in its eigenvector matrix, each corresponding to a different community. Sufficient separations between these rows and appropriate concentration of $\bbA$ around $\E\bbA$ ensure that spectral clustering works by performing eigen decomposition on $\bbA$. However, such a simple approach does not work when it comes to $\bbB$ and $\bbB_E$. Indeed, \eqref{h1s} in Lemma \ref{le6} reveals that even if $\bpi_i=\bpi_j$, $\bbQ(i)$ and $\bbQ(j)$ can be  different; \eqref{h1s} and \eqref{h1ss} suggest that we should treat the nodes separately according to whether they are direct neighbors of node 1. Motivated by Lemma \ref{le6}, we next propose our new  community detection algorithm.

\subsection{Community detection algorithm under the SBM}\label{sec:algorithm}

Let $\bbW=(\bbw_1,\ldots,\bbw_{K},\bbw_{-1},\ldots,\bbw_{-K})$ be the collection of orthonormal eigenvectors corresponding to the $2K$ largest eigenvalues (in magnitude) of $\bbB$. Roughly speaking, our approach involves first clustering the non-zero rows of $\bbS\bbW$ (neighbors of node 1) and $(\bbI-\bbS)\bbW$ (non-neighbors) separately before merging them into $K$ communities. As $\bbW$ is the empirical counterpart of $\bbQ$, the next lemma bounds the difference between $\bbW$ and $\bbQ$.

\begin{lem}\label{le5}
Under Condition \ref{ncond1}, w.h.p. we have
\begin{equation}\label{h2}
	\|\bbW-\bbQ\bbO\|_F=O\left(\frac{1}{\sqrt n p_n}\right)\,,
\end{equation}
where $\bbO=\bbU_1\bbU_2^\top$, in which $\bbU_1$ and $\bbU_2$ are from the singular value  decomposition $(\bbQ)^\top\bbW=\bbU_1\Sigma\bbU_2^\top$  such that $\Sigma$ is the diagonal matrix with singular values.
\end{lem}
\begin{rmk}
When there are repeated eigenvalues, the eigenvectors are not uniquely determined. The matrix $\bbO$ in Lemma \ref{le5} is created to handle such cases.
\end{rmk}

Recall in~\eqref{g19}, $\bbS\bPi\mathcal{D}\mathcal{Q}_1$ and $(\bbI-\bbS)\bPi\mathcal{D}\mathcal{Q}_2$ each has at most $K$ different non-zero rows, and these rows partition the node index set $[n]$ into two parts. Hence $\bbQ$ has at most $2K$ different rows.  However, $\text{rank}(\bbQ) = 2K$ w.h.p. by Corollary \ref{coro1} ,  implying $\bbQ$ has exactly $2K$ different rows (each of the two parts in~\eqref{g19} getting $K$). Since $\bbW$ is the empirical counterpart of $\bbQ$,  we will apply the $k$-means algorithm (with $k=K$) to the non-zero rows of the matrices $\bbS\bbW$ and $(\bbI-\bbS)\bbW$, respectively. Here $\bbS$ and $\bbI-\bbS$ separate the individuals into two groups, each returning $K$ clusters. We note that other clustering algorithms could be used here, and we use $k$-means for ease of analysis. We summarize this procedure in Algorithm \ref{alg1}.

\begin{algorithm}[h]
\caption{}
\begin{algorithmic}[1]
	\State Take matrices $\bbS$ and $\bbW$ as defined respectively in equation \eqref{gba1} and Lemma \ref{le5}.
	\State  Apply the $k$-means algorithm to the rows $\{\bbW(i): i \in[n], a_{1i}=1\}$ and $\{\bbW(i): i \in[n],a_{1i}=0\}$, respectively,  to separate each group into $K$ clusters.
	\State Return (i) the  $2K$ clusters and (ii) the centroid matrix $\bbC=(\bbc_1^\top,\ldots,\bbc_n^\top)^\top$ in which the length $2K$ row vector $\bbc_i$ denotes the centroid  associated with  individual $i$.
\end{algorithmic}\label{alg1}
\end{algorithm}

In the theoretical analysis of Algorithm \ref{alg1}, for simplicity we assume $\bbC$ optimizes the $k$-means objective for both $r=0$ and $r=1$, that is,
\begin{align}\label{kmean}
\{\bbc_i:  a_{1i}=r\} =\text{argmin}_{\left\{ \substack{\bbx_i: \text{$\bbx_i$ is $2K$-dimensional row vector} \\ \text{with } a_{1i}=r \text{ and } |\{\bbx_i\}_{a_{1i}=r}|\leq K}  \right\}}\sum_{i\in[n],a_{1i}=r}\|\bbW(i)-\bbx_i\|_2^2\,,
\end{align}
and also assume $|\{\bbc_l, l\in[n]\}|= 2K$. When $\bbc_i = \bbc_j$ and $a_{1i}=a_{1j}$, Algorithm \ref{alg1} assigns individuals $i$ and $j$ to the same cluster. An error bound on the clustering can be obtained by counting individuals with row in $\bbQ$ that are far (up to the permutation matrix $\bbO$) from their corresponding centroids. Similar to \cite{rohe2011}, 
we define $\mathcal{M}$,
\begin{equation}\label{eqn:M}
\mathcal{M}=\left\{i\in[n]: \|\bbc_i-\bbQ(i)\bbO\|_2\ge\frac{1}{\sqrt{2c_2n}}\right\}\,,	
\end{equation}
where $c_2$ is the positive constant in Lemma~\ref{le6}.
In Theorem \ref{t5}, we will show that $|\mathcal{M}|/n$ controls the misclustering rate in our final algorithm (Algorithm \ref{alg2}).

Algorithm \ref{alg1} returns $2K$ clusters, and we denote them by $\{\mathfrak{c}_1,\ldots,\mathfrak{c}_K\}$ and $\{\mathfrak{d}_1,\ldots,\mathfrak{d}_K\}$, for nodes in $\{\bbW(i): i\in[n], a_{1i}=1\}$ and $\{\bbW(i):i\in[n],  a_{1i}=0\}$, respectively. Noting that the clustering results are only identifiable up to label permutation, our next step is to merge them into $K$ communities. To this end, we first construct two estimates of the connection probability matrix $\bbP$  defined in \eqref{SBM}. We then merge the clusters by finding the correct label permutation that matches these two estimates. Since for $i, j > 1$ and $i\neq j$,
$\p(b_{ij}=1|a_{1i}=a_{1j}=1)=\p(b_{ij}=1|a_{1i}=1-a_{1j}=1)=\p(a_{ij}=1)\,,$
two natural estimators of $\bbP$, denoted by $\widehat\bbP^{\bbS,\bbS}$ and $\widehat\bbP^{\bbS,\bbI-\bbS}$, are defined as
\begin{equation}\label{eqn:prob estimate}
	\widehat\bbP^{\bbS,\bbS}_{kl}=\frac{1}{|\widehat\bbS^{(1)}_{k,l}|}\sum_{(i,j)\in \widehat\bbS^{(1)}_{k,l}}b_{ij}\quad \text{ and }\quad \widehat\bbP^{\bbS,\bbI-\bbS}_{kl}=\frac{1}{|\widehat\bbS^{(2)}_{k,l}|}\sum_{(i,j)\in \widehat\bbS^{(2)}_{k,l}}b_{ij}\,,
\end{equation}
where
$\widehat\bbS^{(1)}_{k,l}=\{(i,j):  a_{1i}=a_{1j}=1, i\in\mathfrak{c}_k, j\in\mathfrak{c}_l\}$, and  $\widehat\bbS^{(2)}_{k,l}=\{(i,j): \ a_{1i}=1-a_{1j}=1, i\in\mathfrak{c}_k, j\in\mathfrak{d}_l\}$, for $k,l\in[K]$. We aim to find a label permutation of the clusters $\{\mathfrak{c}_1,\ldots,\mathfrak{c}_K\}$ such that the difference between  $\widehat\bbP^{\bbS,\bbS}$ and $\widehat\bbP^{\bbS,\bbI-\bbS}$ is minimized after permutation, which will allow us to merge the clusters. This is equivalent to finding a permutation function $f: [K] \to [K]$ such that
$\widehat\bbP^{\bbS,\bbS}_{(f,f)}\approx \widehat\bbP_{(f,*)}^{\bbS,\bbI-\bbS}\,,$
where the $(i,j)$-th entry of $\widehat\bbP^{\bbS,\bbS}_{(f,f)}$ and $\widehat\bbP^{\bbS,\bbI-\bbS}_{(f,*)}$ are the $(f(i),f(j))$-th entry of $\widehat\bbP^{\bbS,\bbS}$ and the $(f(i),j)$ entry of $\widehat\bbP^{\bbS,\bbI-\bbS}$, respectively.  Algorithm~\ref{alg1} and this merging strategy together make up our final algorithm:  Algorithm \ref{alg2}. As an example of the merging Step $5$, if $\hat f_0$ returned by Step $4$ gives $\hat f_0(1)=2$, then {\color{black}$\mathfrak{c}_1$ and $\mathfrak{d}_2$ are merged into one community in Step $5$}.

\begin{algorithm}[h]
	\caption{Community detection under the SBM}
	\begin{algorithmic}[1]
		\State Take matrices $\bbS$ and $\bbW$ as defined respectively in equation \eqref{gba1} and Lemma \ref{le5}.
		\State  Apply Algorithm~\ref{alg1}.		\State Calculate the connection probability matrix estimates $\widehat\bbP^{\bbS,\bbS}$ and $\widehat\bbP^{\bbS,\bbI-\bbS}$ defined in \eqref{eqn:prob estimate}.
		\State  For all $K!$ permutations $f$, find $\widehat\bbP^{\bbS,\bbS}_{(f,f)}$ and $\widehat\bbP_{(f,*)}^{\bbS,\bbI-\bbS}$ by permuting the rows and columns of $\widehat\bbP^{\bbS,\bbS}$ and $\widehat\bbP^{\bbS,\bbI-\bbS}$.
    \State Find $\hat f_0=\arg\min_f\|\widehat\bbP^{\bbS,\bbS}_{(f,f)}-\widehat\bbP_{(f,*)}^{\bbS,\bbI-\bbS}\|_F$.
		\State  Merge the $2K$ clusters into $K$ communities using the permutation $\hat f_0$.
		\State Return the community memberships.
	\end{algorithmic}\label{alg2}
\end{algorithm}

To establish the theoretical  property of Algorithm \ref{alg2}, we need one more assumption.

\begin{cond}\label{cond6}
	There exists a positive constant $c$ such that for any $i_1\neq i_2\in[K]$, there exists $j_1\in[K]$ with $|p_{i_1j_1}-p_{i_2j_1}|\ge cp_n$.
\end{cond}
As estimators of $\widehat\bbP$ in~\eqref{eqn:prob estimate} contain noise, Condition \ref{cond6}  ensures
there are sufficient separations between different elements of $\bbP$ to overcome the noise. 
Algorithm \ref{alg2} essentially outputs an estimated membership matrix $\widehat\bPi=(\widehat\bpi_1,\ldots,\widehat\bpi_n)^\top$; we define its misclustering rate by

\begin{equation}\label{mis_rate}
	d(\hat{\bPi}, \bPi)=\min_{\{\bbZ: \text{$\bbZ$ is $K\times K$ permutation matrix}\} }\sum_{j=1}^n\1\left(\widehat\bpi_j^\top\bbZ\neq\bpi_j^\top\right)/n\,.
\end{equation}

\begin{thm}\label{t5}
	Under Conditions \ref{ncond1}--\ref{cond6} and $p_n\gg 1/\sqrt{n}$,  with probability tending to 1, Algorithm \ref{alg2} gives,
	$$d(\hat{\bPi}, \bPi)\le \frac{|\mathcal{M}|}{n} = O\left(\frac{1}{np_n^2}\right)\,.$$
	That is, Algorithm \ref{alg2} has the almost exact recovery property (c.f. Def 4 of  \cite{abbe2017}).
\end{thm}

Under some stronger conditions on the density $p_n$ and the eigenvalues of $\bbB_E$, we can extend the above result to achieve exact recovery. We refer to Section \ref{subsec:exact_recov} in the Appendix for detailed statements and proofs.

\subsection{Extension to the degree-corrected stochastic block model}\label{sec:DCSBM}

In this section, we first consider extending Algorithm~\ref{alg2} and the upper bound on misclustering error to the more general DCSBM, followed by a more specific conditional probability setting that enables us to interpret the convergence rate for different individuals in terms their connection patterns, as motivated by the Karate club network example in the Introduction.

We consider the DCSBM defined as
\begin{equation} \label{DCSBM}
	\E(\bbA|\Theta) = \Theta\bPi\bbP\bPi^\top\Theta\,,
\end{equation}
where $\Theta=\diag(\theta_1,\ldots,\theta_n)$ is the set of degree parameters associated with the nodes. Given $\Theta$, the edges in $\bbA$ are still generated independently as Bernoulli random variables. Unlike the SBM where individuals within the same community have the same connection pattern, the inclusion of $\Theta$ allows for degree heterogeneity in the model and the presence of hubs (i.e., nodes with significantly higher degrees than average nodes). For ease of modeling, we assume  $\theta_i\in (0,1]$, $i\in[n]$ are i.i.d. random variables with a distribution function $F(\cdot)$. Also we assume the mean of the distribution satisfies:

\begin{cond}\label{cond10}
	$\E(\theta_i)=\theta\sim 1$.
\end{cond}

Similar to Lemma \ref{le4}, we first show that the DCSBM \eqref{DCSBM} can be written as the low-rank model in Section \ref{sec:notation and main term}. The proof of Lemma \ref{le4} together with Lemma 4.1 of \cite{JA15} gives us the following proposition.
\begin{prop}\label{prop1}
	Under Conditions \ref{ncond1} and \ref{cond10}, $\emph{rank}(\E\bbA)=K$ and Conditions \ref{cond2}, \ref{cond3} and \ref{cond1} hold with high probability.
	Moreover, there exists a $K\times K$ orthogonal matrix  $\mathcal{D}'$ such that
	\begin{equation}\label{sg18}
		\bbV(i)=\theta_i\left(\sum_{j\in \text{Community $k$}}\theta_j^2\right)^{-1/2}\mathcal{D}'(k)\,,\qquad i\in \text{Community $k$}
	\end{equation}
	for all $i\in [n]$.
\end{prop}
Thus similar to \eqref{g19}, Theorem \ref{thm2} allows us to write $\bbQ$  as
\begin{equation}\label{sg19}
	\bbQ=\bbS\Theta\bPi\mathcal{D}\mathcal{Q}_1+(\bbI-\bbS)\Theta\bPi\mathcal{D}\mathcal{Q}_2\,,
\end{equation}
where $\mathcal{D}=\text{diag}(\sum_{j\in \text{Community $1$}}\theta_j^2,\ldots,\sum_{j\in \text{Community $K$}}\theta_j^2)^{-1/2}\mathcal{D}'$.
{\color{black}Under Conditions \ref{ncond1} and \ref{cond10}, it is easy to check the key results  in Sections~\ref{sec:sbm}-\ref{sec:algorithm}, namely  Lemma \ref{le2}, \ref{errorbound}, \ref{le5} and the first part of Theorem \ref{thm1} still hold using Proposition \ref{prop1}. We omit the proofs since they are almost identical.}

In Algorithm \ref{alg2}, following \cite{JA15}, we replace the $k$-means algorithm by the spherical $k$-median algorithm to cluster the non-zero rows of $\bbS\bbW$ and $(\bbI-\bbS)\bbW$, returning centroids $\bbC=(\bbc_1^\top,\ldots,\bbc_n^\top)^\top$ defined as
\begin{align}\label{kmedian}
	\{\bbc_i:  a_{1i}=r\} =\text{argmin}_{\left\{ \substack{\bbx_i: \text{$\bbx_i$ is $2K$-dimensional row vector} \\ \text{with } a_{1i}=r \text{ and } |\{\bbx_i\}_{a_{1i}=r}|\leq K}  \right\}}\sum_{i\in[n],a_{1i}=r, \bbW(i)\neq 0}\Big\|\bbW(i)/\|\bbW(i)\|_2-\bbx_i\Big\|_2\,
\end{align}
for $r=0,1$. $i$ and $j$ are assigned to the same cluster when $\bbc_i = \bbc_j$ and $a_{1i}=a_{1j}$. Similar to the SBM case, other clustering algorithms can be used here to return $K$ clusters in each group; we adopt the spherical $k$-median algorithm for ease of analysis.

For the merging step in Algorithm \ref{alg2}, we use the same equations as \eqref{eqn:prob estimate}. Roughly speaking, the matrices $\widehat\bbP^{\bbS,\bbS}$ and $\widehat\bbP^{\bbS,\bbI-\bbS}$ now estimate (up to label permutation) the quantities
$$\bbX:=\diag(\bar{\theta}_1^{\bbS}, \dots, \bar{\theta}_K^{\bbS})\bbP\diag(\bar{\theta}_1^{\bbS}, \dots, \bar{\theta}_K^{\bbS}) \quad\text{and} \,\,\, \bbY:=\diag(\bar{\theta}_1^{\bbS}, \dots, \bar{\theta}_K^{\bbS})\bbP\diag(\bar{\theta}_1^{\bbI-\bbS}, \dots, \bar{\theta}_K^{\bbI-\bbS}), $$
respectively, where $\bar{\theta}_k^{\bbS}$ is the average degree parameter for nodes $i\in$ Community $k$ and $a_{1i}=1$, and similarly for $\bar{\theta}_k^{\bbI-\bbS}$. To simplify our notation, define a matrix function:
\begin{deff}
	For any matrices $\bbX=(x_{ij})_{1\le i,j\le K}$ and $\bbY=(y_{ij})_{1\le i,j\le K}$, define
	$$g(\bbX,\bbY)=\emph{diag}\left(\frac{\sum_{i=1}^Kx_{i1}}{\sum_{i=1}^Ky_{i1}},\ldots,\frac{\sum_{i=1}^Kx_{iK}}{\sum_{i=1}^Ky_{iK}}\right)\,.$$
\end{deff}
It is easy to see that if $\bbX$ and $\bbY$ have matching community labels, $\bbY=\bbX g(\bbX, \bbY)$. Based on this observation, we aim to find a permutation $f$ such that $\|\widehat\bbP^{\bbS,\bbS}_{(f,f)}-\widehat\bbP_{(f,*)}^{\bbS,\bbI-\bbS}g(\widehat\bbP^{\bbS,\bbS}_{(f,f)},\widehat\bbP_{(f,*)}^{\bbS,\bbI-\bbS})\|_F$ is minimized. This leads to the following algorithm for the DCSBM case.

\begin{algorithm}[h]
	\caption{Community detection under the DCSBM}
	\begin{algorithmic}[1]
		\State Take matrices $\bbS$ and $\bbW$ as defined respectively in equation \eqref{gba1} and Lemma \ref{le5}.
		\State  Apply the spherical $k$-median algorithm \eqref{kmedian} to the non-zero rows of $\bbS\bbW$ and $(\bbI-\bbS)\bbW$ respectively.
		\State Calculate the matrix estimates $\widehat\bbP^{\bbS,\bbS}$ and $\widehat\bbP^{\bbS,\bbI-\bbS}$ defined in \eqref{eqn:prob estimate}.
        \State For all $K!$ permutations $f$, find $\widehat\bbP^{\bbS,\bbS}_{(f,f)}$ and $\widehat\bbP_{(f,*)}^{\bbS,\bbI-\bbS}$ by permuting the rows and columns of $\widehat\bbP^{\bbS,\bbS}$ and $\widehat\bbP^{\bbS,\bbI-\bbS}$.
		\State  Find $\hat f_0=\arg\min_f\|\widehat\bbP^{\bbS,\bbS}_{(f,f)}-\widehat\bbP_{(f,*)}^{\bbS,\bbI-\bbS}g(\widehat\bbP^{\bbS,\bbS}_{(f,f)},\widehat\bbP_{(f,*)}^{\bbS,\bbI-\bbS})\|_F$.
		\State  Merge the $2K$ clusters into $K$ communities using the permutation $\hat f_0$.
		\State Return the community memberships.
	\end{algorithmic}\label{alg3}
\end{algorithm}
Let $f_s=\inf\{x,F(x)>0\}$. Using arguments similar to Theorem \ref{t5} for the SBM case and Theorem 4.2 in \cite{JA15}, we can show that Algorithm \ref{alg3} returns a misclustering rate (defined in \eqref{mis_rate}) with the following upper bound.

\begin{thm}\label{consist}
	Under Conditions \ref{ncond1}--\ref{cond10} and $p_n\gg \frac{1}{\sqrt nf_s}$, with probability tending to 1, Algorithm \ref{alg3} gives
	$$d(\hat{\bPi}, \bPi)=O\left(\frac{1}{\sqrt nf_sp_n}\right).$$
\end{thm}

The above theorem extends our result in Section \ref{sec:algorithm} to DCSBM. However, since (within the same community) the degree parameters for individuals and their edges are generated identically under the full model, the unconditional probability setting still gives the same upper bound on misclustering rate for all individuals.

In what follows, we consider a setting that sheds more light on how the structure around a given individual affects their ability to discern global community memberships through their partial network. More specifically, we consider i) the conditional probability given a neighborhood $\bbS$ around individual 1, and ii) a two-component mixture distribution on $\theta_i$, where one component represents the ``hub'' nodes with denser connections than average individuals.

\begin{cond}\label{cond9} $\theta_i$'s are i.i.d. random variables with CDF $yF_1(x)+(1-y)F_2(x)$, $y \in(0,1)$.
	$\E(\theta_i)=\theta=y\mu_1+(1-y)\mu_2\sim 1$, where $\mu_2\sim 1$ and $\mu_1\le\mu_2$. The support of each $F_j$ is bounded, that is,  $1>\sup\{x, F_j(x)\in(0,1)\}>\inf\{x, F_j(x)\in(0,1)\}\sim \mu_j$, $j=1,2$.
\end{cond}

\begin{rmk}\label{rmk:hub_interpret}
	$F_1$ and $F_2$ represent degree distributions of the non-hub and hub nodes respectively. When $F_1=F_2$ (and $\mu_1=\mu_2$), we recover the original DCSBM. Note that $\mu_1\ll \mu_2$ is allowed as long as $y$ is bounded away from 1. The component $F_2$ with potentially a much larger mean $\mu_2$ captures the effect of hub nodes. We also note that although for the sake of consistency, we have generated individual 1 and their edges following the same DCSBM, it is possible to relax this constraint as our arguments are conditioned on fixed $\bbS$.
\end{rmk}

We next analyze the misclustering rate of Algorithm~\ref{alg3} conditioned on $\bbS$ and $\theta_i$'s coming from high probability events. First, it is easy to see that by concentration (Lemma \ref{le1} in the Appendix) $|\{\theta_i\sim \mu_2\}|\sim n(1-y)$ with high probability. Hence we always assume $\theta_i$'s belong to this event in the following arguments. Next, for the non-zero entries of  $\{a_{1i}\}$ in $\bbS$, let $n_{jk}=|\{i: a_{1i}=1, \theta_i \text{ generated from } F_j, \text{ and } i\in \text{Community }k\}|$ for $k=1, \dots, K$ and $j=1,2$. We assume $n_{1k}$ and $n_{2k}$ satisfy the following condition.

\begin{cond}\label{ncond2}
	$\min_{k,k'\in [K]}\frac{n_{2k}}{n_{2k'}}\sim 1$, $\min_{k\in[K]}\sum_{j\in[n]}\1(\bpi_j=\bbe_k)\ge c_0n$ and $\sigma_K(\bbP)\ge c_1p_n$ for some positive constants $c_0$ and $c_1$. Moreover,  for some $c>0$, $1-c\ge p_n\min_{k\in[K]}(n_{2k}\mu_2^2+n_{1k}\mu_1^2)/n\gg 1 / n$.
\end{cond}

We note that the last condition, $1-c\ge p_n\min_{k\in[K]}(n_{2k}\mu_2^2+n_{1k}\mu_1^2)/n\gg \frac{1}{n}$, can be reduced to $ p_n\gg \frac{1}{\sqrt{n}}$ in Condition \ref{ncond1} for SBM since $\min_{k\in[K]}(n_{2k}\mu_2^2+n_{1k}\mu_1^2)/n\sim p_n$ with high probability.

The following lemma is analogous to Lemma~\ref{le2} but for the conditional setting given $\bbS$.
\begin{lem}\label{le10}
	Under Conditions \ref{cond9} and \ref{ncond2}, for any given $\bbS$ we have
	$$\min_{1\le k\le K}\frac{n_{1k}\mu_1^2+n_{2k}\mu_2^2}{n}\bbI\lesssim \bbV^T\bbS\bbV\lesssim \max_{1\le k\le K}\frac{n_{1k}\mu_1^2+n_{2k}\mu_2^2}{n}\bbI.$$
\end{lem}

The other key results in Sections~\ref{sec:sbm}-\ref{sec:algorithm} also can be checked in a similar way and adapted to a given $\bbS$, and we leave the details to the Appendix. Finally,  we have the following bound on misclustering rate which depends on the neighborhood features in $\bbS$.
\begin{thm}\label{consist_fixedS}
		Under Conditions \ref{cond6}, \ref{cond9} and \ref{ncond2}, for $\mu_1^{-2}\frac{1}{\min_{k\in[K]}(n_{2k}\mu_2^2+n_{1k}\mu_1^2)}\ll p_n\,$, conditioned on $\bbS$ and $\Theta$, with probability tending to 1,  the misclustering rate of Algorithm \ref{alg3} is
		$$d(\hat{\bPi}, \bPi)=O\left(\mu_1^{-1}\sqrt{\frac{1}{p_n\min_{k\in[K]}(n_{2k}\mu_2^2+n_{1k}\mu_1^2)}}\right).$$
	\end{thm}
	
\begin{rmk}
		\label{rmk:conditional}
		In this conditional setting, the bound depends on the quantity $\min_{k\in[K]}n_{2k}\mu_2^2+n_{1k}\mu_1^2$ which describes the neighborhood structure around individual 1. When the total number of neighbors the individual knows in each community is fixed (i.e., $n_{1k}+n_{2k}$), a larger  $\min_{k\in [K]}n_{2k}$ leads to a smaller bound on misclustering, suggesting knowing more powerful neighbors across all communities enables the individual to have a better understanding of the global community structure.
	\end{rmk} 

\subsection{Centrality measure}\label{subsec:centrality}

Our analysis above implies individuals may have different capacities for community detection in the full network using their local information. Hence a natural question is whether we can quantify this capacity of each individual with a centrality measure. We note that here, we need to switch the perspective from a local one to a global one that ranks all the individuals. As such, the centrality measure we propose below will depend on the full network $\bbA$, like many existing commonly used centrality measures such as eigenvector centrality and betweenness centrality.

Recall the explicit form of	eigenvalues in Theorem \ref{thm1}; together with the bounds in Lemma \ref{le2}, it suggests that $\lambda_{\min} = \lambda_K(\bbV^\top\bbS\bbV)$ determines the gap between the smallest eigenvalue (in magnitude) and 0, and consequently how easy it is to perform spectral clustering. Therefore, $\lambda_{\min}$ captures the amount of clustering information available through the centered individual's partial network, measuring their importance in the whole network in the clustering context. In other words, $\lambda_{\min}$ can be seen as a centrality measure for the centered individual. Empirically, an estimate $\hat{\lambda}_{\min}$ can be computed using the empirical version of $\bbV$ from $\bbA$. It is easy to check that both $\lambda_{\min}$ and $\hat{\lambda}_{\min}$ lie between 0 and 1.

Under the conditional setting discussed in Section~\ref{sec:DCSBM}, $\lambda_{\min}$ behaves like $\min_{k\in [K]}\frac{n_{1k}\mu_1^2+n_{2k}\mu_2^2}{n}$ by Lemma \ref{le10}. Hence the same interpretation as in Remark~\ref{rmk:conditional} applies to $\lambda_{\min}$ as a measure of how ``central'' an individual is in the context of global community detection. That is, knowing more powerful neighbors across all communities makes an individual more ``important''.

Finally, we note the connection between $\lambda_{\min}$ and the well-known eigenvector centrality. It is easy to see that when $K=1$, $\lambda_{\min}=\sum_{i:a_{1i}=1}v_i^2$,  where $v_i$ is the eigenvector centrality of node $i$. In this sense, $\lambda_{\min}$ is related to both degree and eigenvector centralities. In the following sections, as part of the numerical analysis, we compare $\lambda_{\min}$ with other commonly used centrality measures and show that it correlates well with clustering accuracy.

\section{Simulation studies}\label{sec:simulation}
In this section, we consider simulating a SBM and a DCSBM with the following parameters:
\begin{itemize}
	\item Model $1$ (SBM): $\bbP=\left(
	\begin{array}{ccc}
		3q & q \\
		q& 3q \\
	\end{array}
	\right)$ and each group is of size $n/2$.
	
	\item Model $2$ (DCSBM): the same $\bbP$ and group proportions as the above are used. We simulate an i.i.d. uniform and a mixture distribution for $\theta_i$, to be specified below.
\end{itemize}
Additional results for a SBM with $K=3$ are presented in Section \ref{sec:add plots model 2} of the Appendix.

\subsection{Results for Model $1$}

For Model 1, we vary the number of individuals $n\in \{300,600,900,1200,1500,1800,2100\}$ and the edge density $q \in \{.1, \sqrt{\log n / n}, (\log n / n)^{1/4}/2, 1/\sqrt{n}\}$. For every combination of $n$ and $q$, we simulate $100$ datasets; Algorithm \ref{alg2} is applied to the partial network centered at node 1 in each dataset.

We first check the number of individuals and the fraction of edges observed within the partial network centered at individual 1. Tables \ref{tab:addlabel1} and
\ref{tab:addlabel2} in the Appendix show that although partial network can reach almost everyone in the network , the fraction of missing edges is quite significant ($>50\%$ in most cases).

\begin{figure}[hbtp!]
	\centering
	\begin{subfigure}{0.5\textwidth}
		\centering
		\includegraphics[height=1.65in]{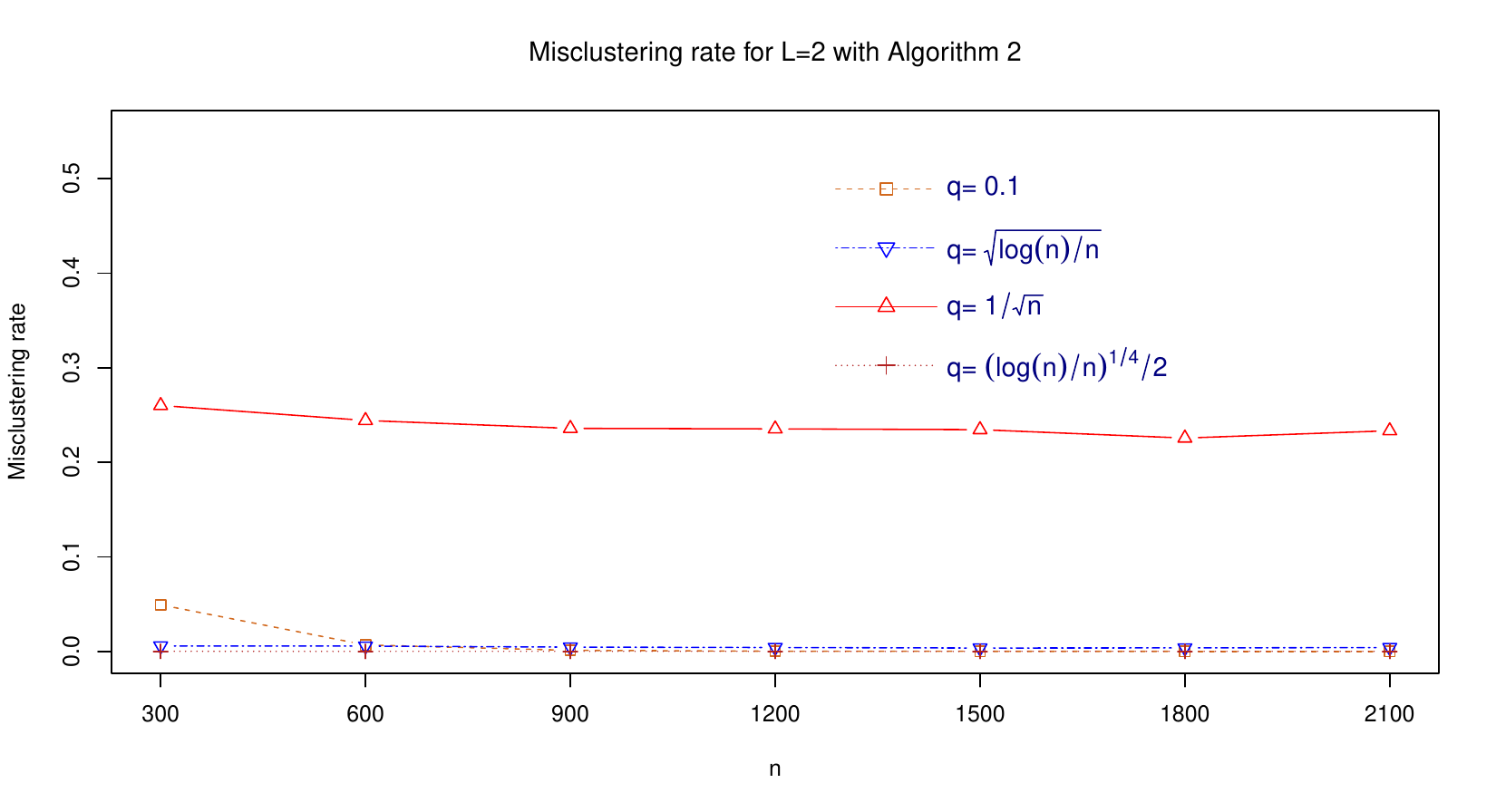}
		\caption{$L=2$}
		\label{f2}
	\end{subfigure}%
	\begin{subfigure}{0.5\textwidth}
		\centering
		\includegraphics[height=1.65in]{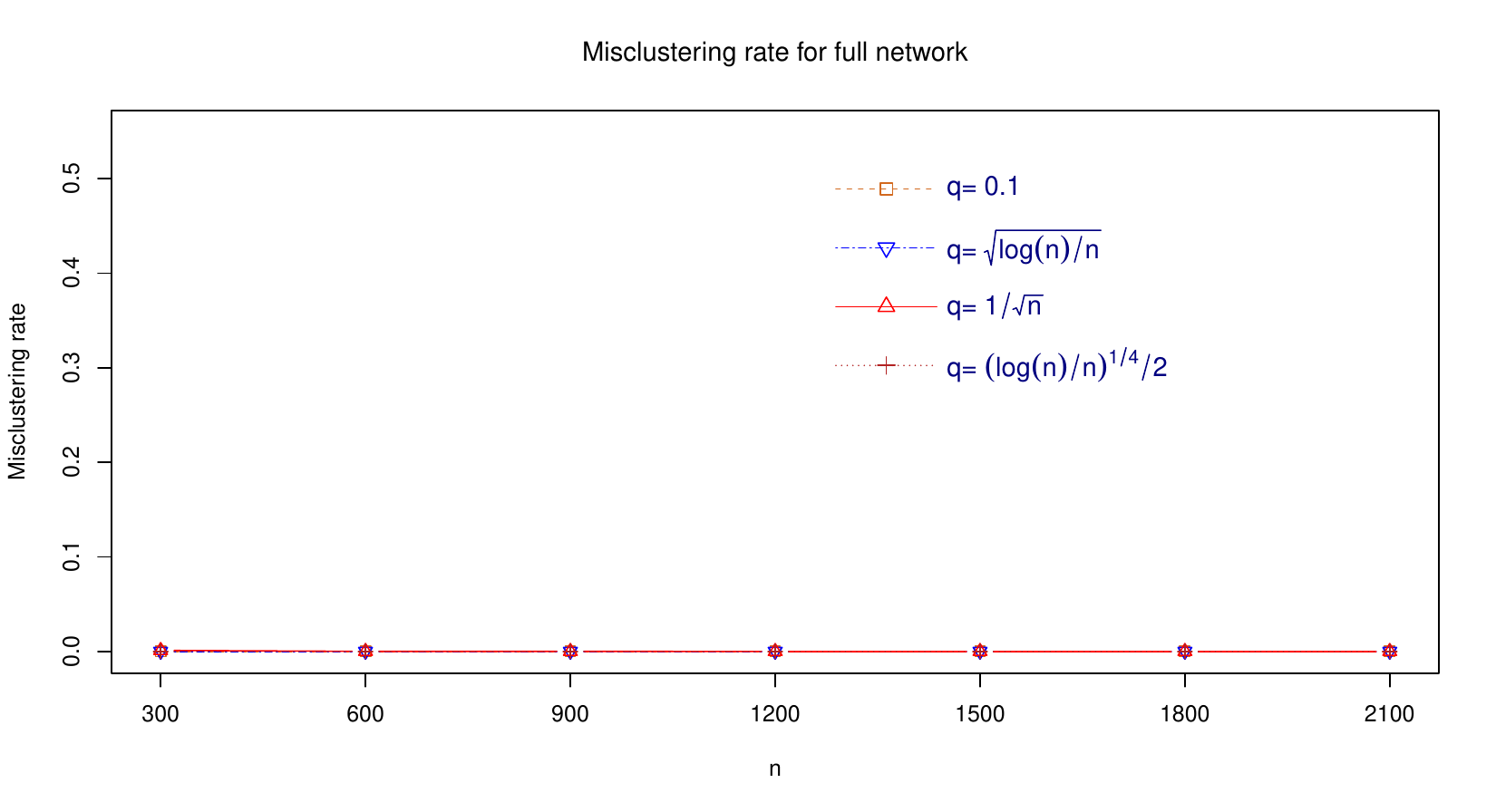}
		\caption{full network}
		\label{f1ss}
	\end{subfigure}
	\caption{Misclustering rate for Model $1$ ($L=2$), (full network), averaged over $100$ datasets for each combination.}
    \label{fig:results_model1}
\end{figure}

We calculate the mean misclustering rate for each parameter combination and report the results in Figure \ref{fig:results_model1}(a).  Except for $q = 1/\sqrt{n}$, all the other $q$'s lead to almost perfect clustering. As a comparison, we apply spectral clustering to $\bbA$, the full network, and plot the results in Figure \ref{fig:results_model1}(b). Unsurprisingly, perfect clustering is achieved for all the parameter settings since the $q$'s we consider lie in the exact recovery regime, which requires edge density to be at least $\log n /n$ \citep{abbe2017}.

In Figures \ref{f3_supp} and \ref{f4_supp} of the Appendix, we provide visualizations of two eigenvectors of an examplary $\bbB$ matrix corresponding to the two positive eigenvalues (recall that $\bbB_E$ has two positive and two negative eigenvalues for $K=2$). The nodes are separated into four clusters, based on their community memberships and whether they are neighbors of node 1, hence justifying our overall approach of clustering.

\subsection{Results for Model $2$}

\begin{figure}[hbtp!]
	\captionsetup[subfigure]{aboveskip=-1pt,belowskip=-2pt}
	\centering
	\begin{subfigure}[t]{0.53\textwidth}
		\centering
		\includegraphics[width=0.85\textwidth]{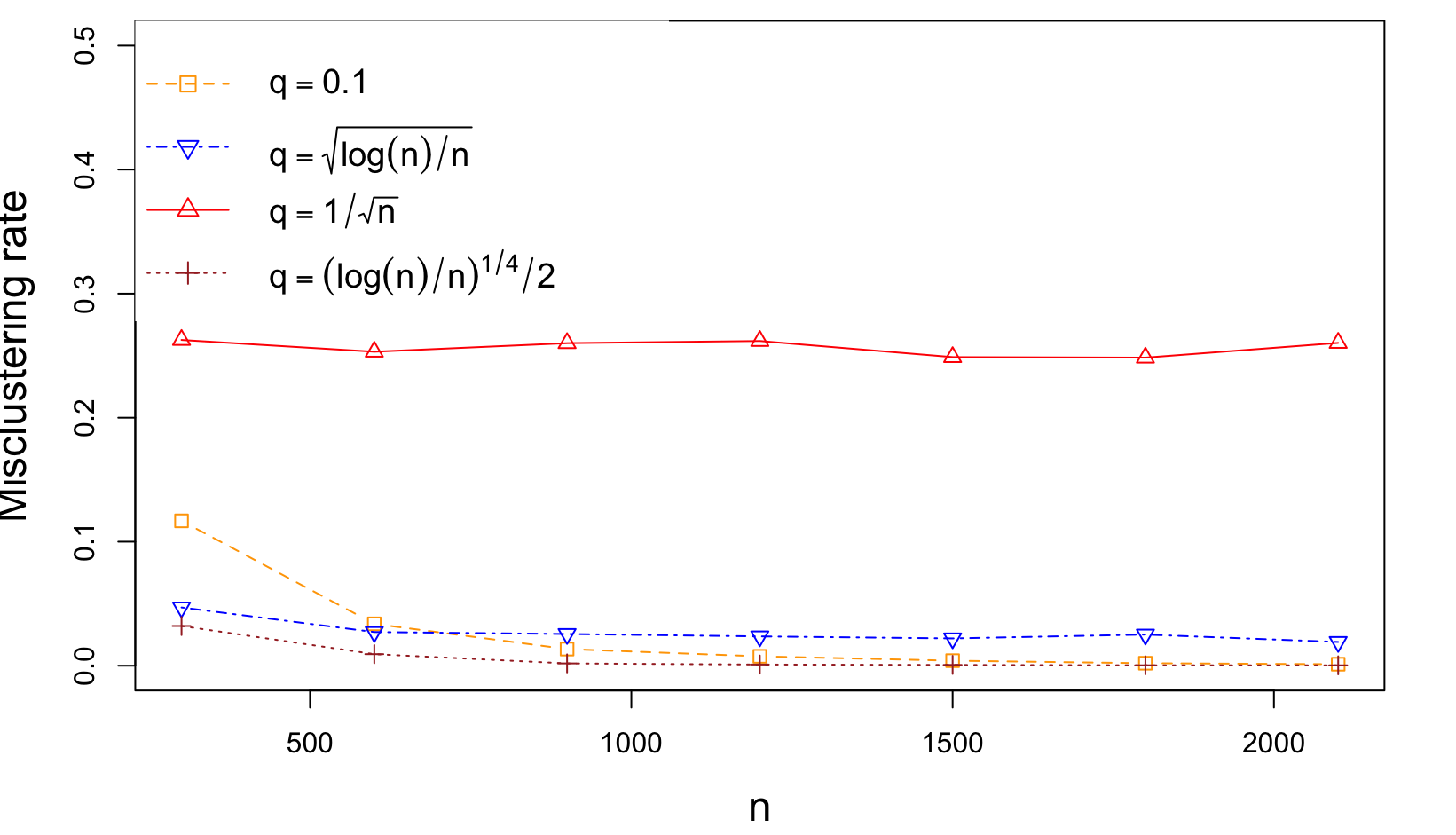}
		\caption{}
		\label{f3}
	\end{subfigure}%
	\begin{subfigure}[t]{0.45\textwidth}
		\centering
		\includegraphics[width=0.7\textwidth]{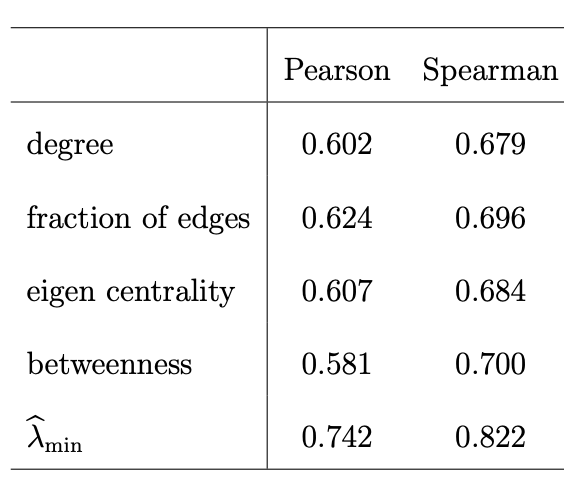}
		\caption{}
		\label{cent_cor_tab}
	\end{subfigure}
	\caption{(a) Misclustering rate for Model $2$ ($L=2$), averaged over $100$ datasets for each combination, $\theta_i$ following an i.i.d. uniform distribution. (b) Correlations between clustering accuracy and various centrality measures, $\theta_i$ following a mixture distribution.}
\end{figure}

First generating $\theta_i$ from an i.i.d. Unif(.5, 1.5) distribution, we check how the clustering accuracy varies with respect to the edge density $q$ and the number of nodes $n$, using the same ranges of values as in Model $1$. Applying Algorithm \ref{alg3} to node 1 in 100 simulated networks for each combination of $q$ and $n$, Figure~\ref{f3} shows the overall trends are consistent with Figure~\ref{f2}, although the convergence rates are slower as expected.

Next considering the mixture setting, we take $F_1$, $F_2$ as the Unif$(0.5, 0.75)$, Unif$(0.8, 1.05)$ distribution respectively, with proportions $(0.85, 0.15)$. Taking $q=0.2$, $n=600$, we generate one instance of $\bbA$ under Model $2$ and apply Algorithm \ref{alg3} to all the nodes in the network. Figure \ref{cent_cor_tab} reports the Pearson and Spearman correlations between the clustering accuracy and various centrality measures for all the nodes. For comparison with the empirical version of $\lambda_{\min}$ (defined in Section 3.4), we choose degree centrality, fraction of edges and individuals observed in the partial network, eigenvector and betweenness centrality as alternative measures. $\hat{\lambda}_{\min}$ exhibits the highest correlations with accuracy. Furthermore,  $\hat{\lambda}_{\min}$ is less correlated with the other four measures than those measures among themselves (Figure \ref{fig:cent_cor_simu} in the Appendix), which suggests it offers unique information about node importance. We note here that same as Model 1, Model 2 produces networks dense enough that all partial networks can reach almost every node in the full network. For this reason, fraction of individuals observed is not a meaningful centrality measure in this case and has almost zero correlation with all the other centrality measures.

\section{Real data analysis}\label{sec:real data}

\subsection{Zachary's karate club data}
\label{sec:karate}

The karate club network \citep{zachary1977information} mentioned in the Introduction is a well-known dataset in network analysis. The network consists of $34$ individuals, $78$ edges, and two ground truth communities which formed after conflicts between the instructor (node ``H'') and the administrator (node ``A''). As shown in Figure  \ref{karate_table}, individual 20, despite their small degree and low fraction of observed edges, outperforms the two hub nodes in detection accuracy. This trend is consistent with the empirical $\hat{\lambda}_{\min}$ values, which are $.276, .117, .120$ for node 20, ``H'' and ``A'' respectively. We have also applied Algorithm \ref{alg2} to the same set of individuals, and the results are presented in Table \ref{z2} in the Appendix. We note that even though the accuracy rates are different for some individuals if we assume the underlying model is SBM instead of DCSBM, the qualitative conclusions about node 20 and the two hub nodes still hold. We continue the rest of our discussion using DCSBM and Algorithm \ref{alg3}, since the presence of two clear hubs is indicative of degree heterogeneity.

We conjecture that the high accuracy of individual $20$ is due to their direct connections to ``H'' and ``A''. In other words, knowing powerful neighbors in both communities enables this individual to have a better understanding of the global community structure. As further evidence, we delete the edge between node $20$ and ``A''
and apply Algorithm \ref{alg3} again to the same set of individuals. There is a noticeable drop in accuracy for individual $20$, while the other individuals change by a small margin or are unaffected.

{\tiny
	\begin{table}[h!]
		\caption{\label{z3} Results for chosen individuals after deleting the edge between ``A" and $20$. }
		\centering

		\begin{tabular}{l | cccccc}
			\hline
			individual of interest   & H & 2 &3 & A & 20 & 32\\

			\hline
			detection accuracy  & .529 & .706 & .941 & .706 & .676  & .824  \\
			\hline
		\end{tabular}%
		
	\end{table}
}

\subsection{Microfinance in Indian villages}

This dataset contains information about the social interactions and the diffusion of information about a microfinance program in 43 Indian villages
\citep{banerjee2013diffusion, ChengXing}. These villages are far apart from each other and can be regarded as independent social networks. Following \citet{banerjee2013diffusion}, in each village, we take households as nodes; each edge is undirected and binary representing any of the 12 relationships collected in the survey (e.g., borrowing / lending money or material goods). We use the caste information available in the survey for households as the ground truth community labels. As shown in Figure \ref{fig:vil_networks} of the Appendix, many of the networks exhibit assortative structure with respect to these community labels. After a simple filtering step (described in Section \ref{sec:indian_supp} of the Appendix), we analyze $39$ villages, with the number of households varying between $24$-$155$ and $K$ between $2$-$4$.

\begin{figure}[hbtp!]
	\captionsetup[subfigure]{aboveskip=-1pt,belowskip=-2pt}
	\centering
	\begin{subfigure}[t]{0.4\textwidth}
		\centering
		\includegraphics[height=2.3in]{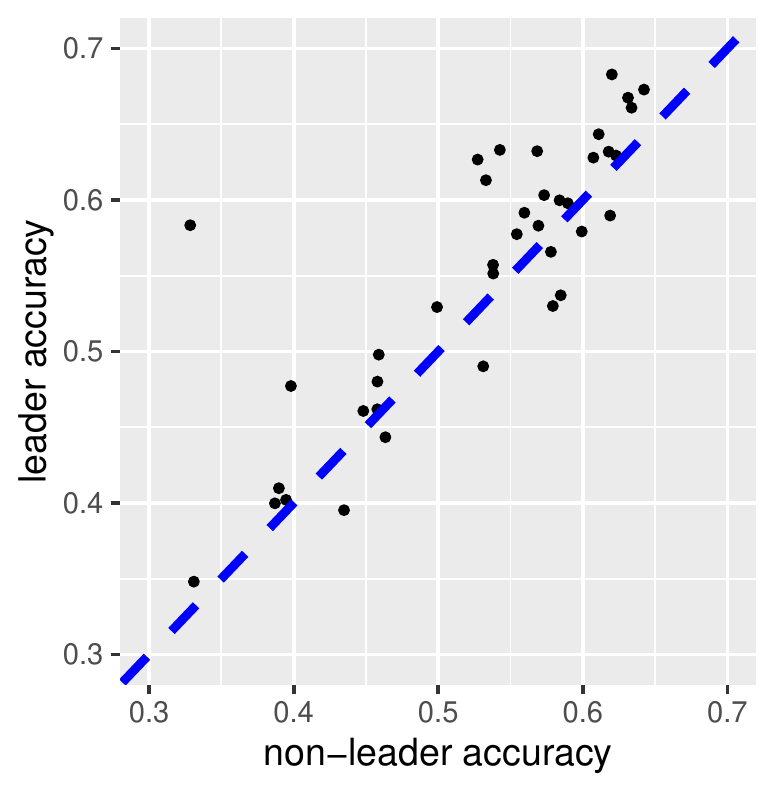}
		\caption{}
		\label{fig:accu_leader_nonleader}
	\end{subfigure}%
	\begin{subfigure}[t]{0.6\textwidth}
		\centering
		\includegraphics[height=2.8in]{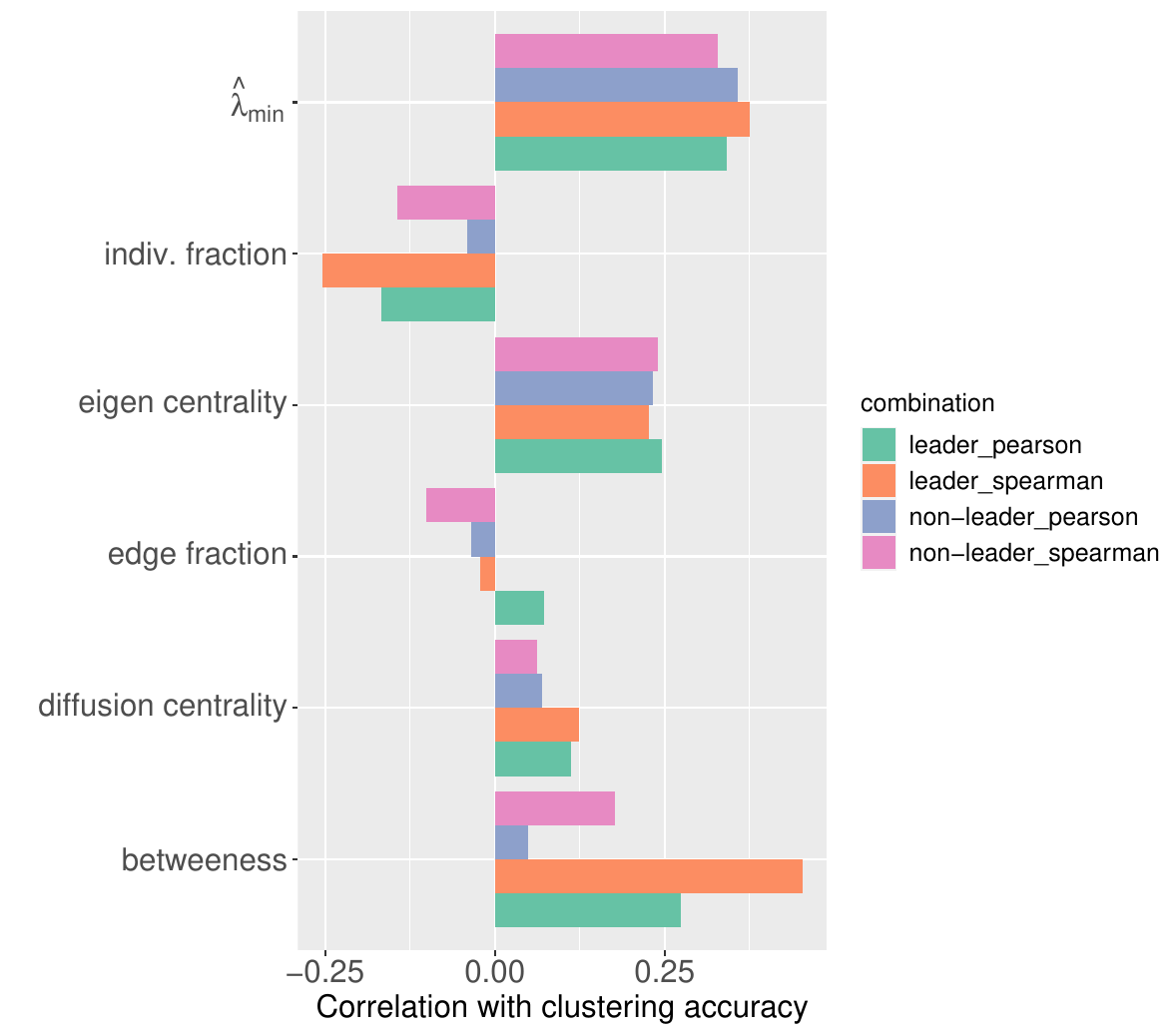}
		\caption{}
		\label{fig:centrality_accu_village}
	\end{subfigure}
	\caption{(a) Mean clustering accuracy among leaders vs. non-leaders for all villages. (b) Correlations (Pearson and Spearman) across all villages between the mean clustering accuracy and mean centrality measures for leaders and non-leaders.}
\end{figure}

We apply Algorithm \ref{alg3} to each household  in all the villages. Each village also has several predefined leaders expected to be well-connected within the village, who served as the ``injection'' points for information about the microfinance program. To compare the awareness of global community structure among leaders vs. non-leaders, we compute the mean clustering accuracy among these two groups for each village. In most villages, the mean accuracy of leaders is higher than that of the non-leaders (Figure \ref{fig:accu_leader_nonleader}), thus to an extent justifying the choice of these leaders as the injection point for spreading information.

Next we examine the usefulness of $\hat{\lambda}_{\min}$ as a centrality measure in this dataset. \citet{banerjee2013diffusion} proposed two centrality measures for assessing the efficiency of information spread from a node: the first is derived from their full structural model, while the second is a simpler diffusion centrality computed using only network topology. The two measures were shown to be highly correlated at village level; we include the diffusion centrality for comparison since it is much easier to compute. In Figure \ref{fig:centrality_accu_village}, we calculate the correlations at village level between the mean clustering accuracy and mean centrality measures among the leaders and non-leaders. $\hat{\lambda}_{\min}$ is the most correlated measure in most cases, although betweenness centrality performs better for leaders using the Spearman correlation.

\begin{figure}[hbtp!]
	\centering
	\begin{subfigure}[b]{0.5\textwidth}
		\centering
		\includegraphics[width=7cm]{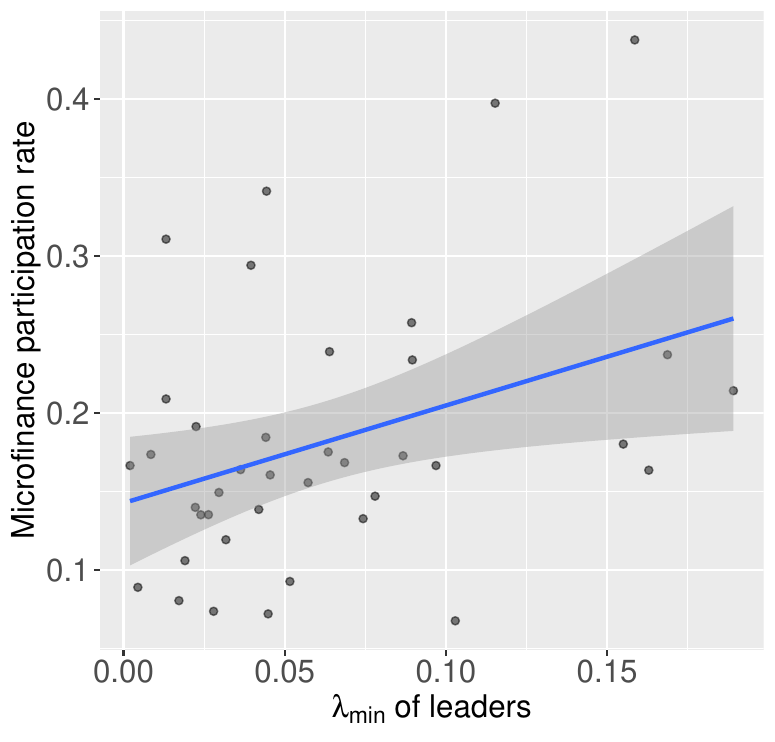}
		\caption{}
	\end{subfigure}%
	\begin{subfigure}[b]{0.5\textwidth}
		\centering
		\includegraphics[width=7cm]{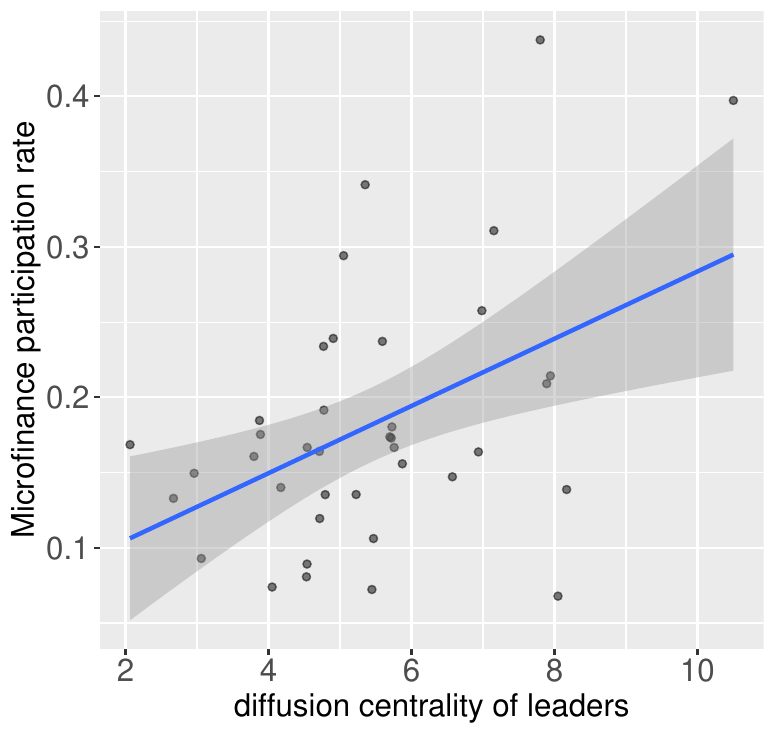}
		\caption{}
	\end{subfigure}
	\caption{Program participation rate as a function of (a) $\hat{\lambda}_{\min}$, fitted slope 0.621 with p-value 0.022; (b) diffusion centrality, fitted slope 0.022 with p-value 0.003. The centrality measures are calculated for leaders in each village and an average is taken across each village. }\label{fig:mfrate_reg}
\end{figure}

Finally, in Figure \ref{fig:mfrate_reg}, we show that $\hat{\lambda}_{\min}$ is correlated with the village-level participation rate in the program. Here, the centrality measures are computed for all leaders in each village, who were responsible for spreading the information about the program, before an average is taken for each village. The correlation suggests the information $\hat{\lambda}_{\min}$ captures extends beyond community detection and is related to information diffusion. Although unsurprisingly, the correlation is weaker than that of the diffusion centrality, which was designed to explain the participation rates in the original paper. The correlations with other centrality measures are shown in Figure \ref{fig:mfrate_reg_app} in the Appendix. While some of the other centrality measures also exhibit a positive correlation with the participation rate, $\hat{\lambda}_{\min}$ remains one of the best fitting under a linear model.

\subsection{Political blog data}
\label{sec:polblogs}
The political blog network (\cite{adamic2005political}) records hyperlinks between web blogs observed in the run-up to the $2004$ U.S. presidential election. Each blog was labeled as democratic or republican based on their political outlooks. Following most previous community detection studies using this dataset, we take the largest (weakly) connected component consisting of $1222$ nodes and treat the hyperlinks as undirected.

Similar to the karate club data, we pick six individuals (blogs) and examine their network information and clustering performance in detail. Because this is a large network with relatively sparse edges, most of the partial networks with $L=2$ can only reach a fraction of all the nodes. Therefore we apply Algorithms \ref{alg2} and \ref{alg3} to nodes reached by each partial network only. That is, for a given partial network, the nodes with no observed edges are removed. Table \ref{z4} summarizes the performance of Algorithm \ref{alg3}. The results from Algorithm \ref{alg2} are similar and presented in Table \ref{z5} of the Appendix. In this setting, the accuracy rate is calculated using only nodes included in each partial network; similarly the ratio of edges is calculated out of the subnetwork after removing isolated nodes in the partial network. These observations again demonstrate the importance of identifying which nodes contain the most powerful partial information about community structure. Moreover, it shows our algorithms can be successfully applied to recover local community structure by restricting attention to nodes reached by the partial network only, even though our theoretical analysis has focused on global community memberships.

{\tiny
	\begin{table}
		\caption{\label{z4} The network information and detection results for chosen individuals (blogs) in the political blog data using Algorithm \ref{alg3}.}
		\centering

		\begin{tabular}{l| c c c c c c}
			\hline
			node of interest   & 1073 & 1074 &1075 & 1076 & 1077 &1078\\
			
			\hline
			the ratio of the edges observed  & .1145 & .1295 & .3362  & .1553  &.2761 & .4116  \\
			$\#$ of the nodes observed & 476 & 485 & 880 & 715 & 808  & 793  \\
			clustering accuracy  & .5798 & .5505 & .9261 & .9021 & .8936 & 9177  \\
			\hline
		\end{tabular}%
		
	\end{table}%
}

\section{Discussion}

Each individual in a social network only has a local understanding of the full network, which has motivated us to study the problem of inferring global community memberships using such a local view. In contrast to the popular line of works that performs statistical inference of global network properties by patching together a large number of local subnetworks, we focus on what is attainable using one subnetwork only. This perspective has also allowed us to propose a new centrality measure that assesses the importance of an individual's partial information in determining global community structure. As shown in our analysis of simulated and real networks, this measure is capable of identifying information complementary to other commonly used centrality measures.

We have considered three different types of recovery overall. On the theoretical front, we have analyzed  weak recovery and exact recovery of community memberships for all nodes, both can be considered as a form of global recovery. In practice, when the network is sparse, we can apply our algorithms to only individuals reached by the partial network and compute the accuracy rate on this subset of nodes, as we have done in the political blog data analysis in Section \ref{sec:polblogs}. This is a form of local recovery, which is more challenging to analyze since the partial network contains random individuals. A meaningful future direction is to extend our theoretical analysis and recovery guarantees to this setting. It is also plausible to consider a hierarchical form of recovery, which becomes weaker for layers of individuals further away from the center. This would be relevant if we believe the connectivity probability decreases as $L$ increases.

Another more specific question is, can the individual of interest correctly identify their own membership? From a practical point of view, after one assigns everyone else into $K$ communities, they can decide on their own affiliation based on side information. Alternatively, we could analyze the asymptotic properties of the eigenvectors of $\bbB$ and design a correction step accordingly. This is a point that we would like to pursue in future studies.

There are many interesting topics under the individual-centered partial information framework. We end our paper with five open problems to inspire further work. (1) A data-driven way of choosing the number of communities $K$: one possible solution is to study the differences between spiked eigenvalues and non-spiked eigenvalues (e.g., \citet{fan2019estimating, cai2020limiting}), but other approaches are also possible. (2) It is possible that the individual of interest knows their edges exactly but can only identify their neighbors' edges with a high probability, then the community detection algorithm should account for the noise in these edges. (3) Extend our analysis to the more expansive $L=3$ partial information networks. (4) In addition to community detection, we can also study other problems, such as link prediction, mixed membership profile inference, subgraph counts, under the current framework. (5) How to combine multiple, but possibly finite, individuals' partial information and collectively gain a better understanding of the full network is also an interesing problem to explore.

\newpage




\newpage

\appendix

\section*{Appendix}
The appendix contains the proofs of main results and some established results we cited in the proofs for readers' convenience. Additional results from simulation and real data are also presented.

\section{Proofs of the results in Section \ref{sec:notation and main term}}

\subsection{Proof of Lemma \ref{majorterm}}

To show \eqref{gtq18}, it suffices to prove
\begin{equation}\label{gtq19}
	\|-\E\bbS(\E\bbA)\E\bbS+(\E\bbA)\E\bbS+\E\bbS(\E\bbA)\|=o_p(\|\bbB_E\|),\ \|\E\bbB\|=o_p(\|\bbB_E\|)\,.
\end{equation}
By the definition of $\bbS$, $a_{ij}$ is independent of $\bbS$ for $i
\ge 2$ and $j\ge 2$. By \eqref{gts5}, $$\E b_{ij}=\E a_{ij}(1-(1-\E a_{1i})(1-\E a_{1j})),\ 2\le i\neq j\le n\,.$$ Therefore $\E\bbB$ is equal to $-(\E\bbS)(\E\bbA)(\E\bbS)+(\E\bbA)(\E\bbS)+(\E\bbS)(\E\bbA)$ except for the diagonal entries, the first row and first column. Then we have
$$\|\E\bbB-\left(-(\E\bbS)(\E\bbA)(\E\bbS)+(\E\bbA)(\E\bbS)+(\E\bbS)(\E\bbA)\right)\|\lesssim (\sum_{j=1}^n(\E a_{1j})^2)^{1/2}+p_n\le (\sqrt n+1) p_n\,.$$
Therefore,  it suffices to prove the first inequality of \eqref{gtq19}.

First of all, we look at the  matrix
$$(\E\bbS)(\E\bbA)(\E\bbS)\,.$$ Since $(\E\bbS)(\E\bbA)(\E\bbS)=(\E\bbS)\bbV\bbD\bbV^\top(\E\bbS)$, $(\E\bbS)(\E\bbA)(\E\bbS)$ has the same non-zero eigenvalues as $\bbD\bbV^\top(\E\bbS)^2\bbV$. The  counterpart of $(\E\bbS)(\E\bbA)(\E\bbS)$ in $\bbB_E$ is $\bbS(\E\bbA)\bbS$, whose non-zero eigenvalues are the same as $\bbD\bbV^\top\bbS^2\bbV$.  In this case $\bbV^\top(\E\bbS)^2\bbV=p_n^2\bbI$ and $\E(\bbV^\top\bbS^2\bbV)= p_n\bbI$, which means that we cannot replace $\bbS$ by $\E\bbS$ for $\bbD\bbV^\top\bbS^2\bbV$. Similarly, we can show that
$\E\|(\E\bbA)\bbS\|_F^2=\text{tr}(\E\bbA\E(\bbS^2)\E\bbA)=n^2p_n^{3}\gtrsim\|(\E\bbA)\E\bbS\|_F^2=n^2p_n^4$. These insights combined with Condition $p_n = o(1)$ imply that
$$\|-\E\bbS(\E\bbA)\E\bbS+(\E\bbA)\E\bbS+\E\bbS(\E\bbA)\|=o_p(\|\bbB_E\|)\,.$$

\subsection{Proof of Theorem \ref{thm2}}

Note that $\text{det}(y^2\bbI-y\bbD\bbV^\top\bbS\bbV-\bbD(\bbI-\bbV^\top\bbS\bbV)\bbD\bbV^\top\bbS\bbV)$ is a polynomial of $y$ with degree $2K$.  Hence the equation $\text{det}(y^2\bbI-y\bbD\bbV^\top\bbS\bbV-\bbD(\bbI-\bbV^\top\bbS\bbV)\bbD\bbV^\top\bbS\bbV)= 0$ has $2K$ solutions in $y\in\mathbb{C}$. Moreover, as $\bbV^{\top} \bbS \bbV$ and $\bbI - \bbV^\top \bbS \bbV$ are invertible, $\text{det}(\bbD(\bbI-\bbV^{\top}\bbS\bbV)\bbD\bbV^{\top}\bbS\bbV)\neq  0$; hence $y=0$ is NOT a solution.  Let $x=y^{-1}$, then $x^{-2K}\times\text{det}(\bbH(x))=\text{det}(y^2\bbI-y\bbD\bbV^\top\bbS\bbV-\bbD(\bbI-\bbV^\top\bbS\bbV)\bbD\bbV^\top\bbS\bbV)$. Hence there are  $2K$ non-zero solutions to $\text{det}\left(\bbH(x)\right)=0$ (i.e.,  \eqref{g0}). Denote these solutions by $x_{-K},\ldots, x_{-1},x_1,\ldots,x_K$. 

Then it remains to prove that for each $i\in [\pm K]$,  $\bbq_i$ is an eigenvector of
$\bbB_E\,$ corresponding to the  eigenvalue $x_i^{-1}$,  $\bbq_i \neq \mathbf{0}$, and $x_i\in\R$. By the definitions of $\bbq_{1i}$ and $\bbq_{2i}$,  we have
\begin{eqnarray}\label{g17}
	&&\Big(-\bbS(\E \bbA)\bbS+(\E \bbA)\bbS+\bbS(\E \bbA)\Big)(\bbS\bbV\bbq_{1i}+(\bbI-\bbS)\bbV\bbq_{2i})\non
	&&=\bbV\bbD\bbV^\top\bbS\bbV\bbq_{1i}+x_i\bbS\bbV\bbD(\bbI-\bbV^\top\bbS\bbV)\bbD\bbV^\top\bbS\bbV\bbq_{1i}\non
	&&=x_i^{-1}(\bbV\bbq_{2i}+\bbS\bbV(\bbq_{1i}-\bbq_{2i}))=x_i^{-1}(\bbS\bbV\bbq_{1i}+(\bbI-\bbS)\bbV\bbq_{2i}),
\end{eqnarray}
where the second equation follows from
\begin{equation}\label{gts18}
	\bbq_{1i}-x_i\bbD\bbV^\top\bbS\bbV\bbq_{1i}-x_i^2\bbD(\bbI-\bbV^\top\bbS\bbV)\bbD\bbV^\top\bbS\bbV\bbq_{1i}=\mathbf{0}\,.
\end{equation}
Therefore $\bbq_i = \bbS\bbV\bbq_{1i}+(\bbI-\bbS)\bbV\bbq_{2i}$ is the eigenvector of $-\bbS(\E \bbA)\bbS+(\E \bbA)\bbS+\bbS(\E \bbA)$ corresponding to the eigenvalue $x_i^{-1}$ if $\bbq_i\neq \mathbf{0}$.  We prove $\bbq_i\neq \mathbf{0}$ by contradiction. Actually, if $\bbq_i=0$, by the definition of $\bbq_i$, we have
$$0=\|\bbq_i\|^2_2\ge \bbq_{1i}^\top\bbV^\top\bbS\bbV\bbq_{1i}=0\,.$$
Then $(\bbV^\top\bbS\bbV)^{1/2}\bbq_{1i}=\mathbf{0}$ and by \eqref{gts18}, we have
$$ \bbq_{1i}=x_i\bbD\bbV^\top\bbS\bbV\bbq_{1i}+x_i^2\bbD(\bbI-\bbV^\top\bbS\bbV)\bbD\bbV^\top\bbS\bbV\bbq_{1i}=\mathbf{0}\,,$$
which contradicts with $\|\bbq_{1i}\|_2=1$! Finally, since $-\bbS(\E \bbA)\bbS+(\E \bbA)\bbS+\bbS(\E \bbA)$ is a real symmetric  matrix, its eigenvalues $x_{-K},\ldots, x_{-1},x_1,\ldots,x_K$ are real numbers. Without loss of generality, we can take $x_{i}\le x_j$ for all  $i<j$.

Recalling that $\E\bbA=\bbV\bbD\bbV^\top$ we have
\begin{equation}\label{gtq15}
	\E \bbA-(\bbI-\bbS)\E \bbA(\bbI-\bbS)=\left(\bbV,(\bbI-\bbS)\bbV\right)\diag(\bbD,-\bbD)\left(\bbV,(\bbI-\bbS)\bbV\right)^\top\,.
\end{equation}
By simple algebra, the non-zero eigenvalues of $\left(\bbV,(\bbI-\bbS)\bbV\right)\diag(\bbD,-\bbD)\left(\bbV,(\bbI-\bbS)\bbV\right)^\top$ are equal to the non-zero eigenvalues of
$$\diag(\bbD,-\bbD)\left(\bbV,(\bbI-\bbS)\bbV\right)^\top\left(\bbV,(\bbI-\bbS)\bbV\right)=\left(
\begin{array}{ccc}
	\bbD &0 \\
	0 & -\bbD\\
\end{array}
\right)\left(
\begin{array}{ccc}
	\bbI&\bbV^\top(\bbI-\bbS)\bbV \\
	\bbV^\top(\bbI-\bbS)\bbV & \bbV^\top(\bbI-\bbS)\bbV\\
\end{array}
\right) \,.$$
Since $\bbV^\top\bbS\bbV$ and $\bbV^\top(\bbI-\bbS)\bbV$ are invertible , we have
\begin{align*}
	&\det\left(\left(
	\begin{array}{ccc}
		\bbI &\bbV^\top(\bbI-\bbS)\bbV \\
		\bbV^\top(\bbI-\bbS)\bbV & \bbV^\top(\bbI-\bbS)\bbV\\
	\end{array}
	\right)\right)=\det(\bbV^\top(\bbI-\bbS)\bbV-(\bbV^\top(\bbI-\bbS)\bbV)^2)\non
	&=\det(\bbV^\top\bbS\bbV\bbV^\top(\bbI-\bbS)\bbV)\neq 0\,.
\end{align*}
Combining this with  \eqref{gtq15}, we have
$$\text{rank}\Big(-\bbS(\E \bbA)\bbS+(\E \bbA)\bbS+\bbS(\E \bbA)\Big)=2K\,.$$

We will show that there exists $\bbq_{10}$ and $\bbq_{20}$ such that
\begin{equation}\label{gtq14}
	\bbq_0=\bbS\bbV\bbq_{10}+(\bbI-\bbS)\bbV\bbq_{20}\,.
\end{equation}
Indeed, we have
\begin{equation}\label{gtq11}
	x_0^{-1}\bbq_0=\bbB_E\bbq_0=\left(-\bbS\bbV\bbD\bbV^\top\bbS+\bbV\bbD\bbV^\top\bbS+\bbS\bbV\bbD\bbV^\top\right)\bbq_0\,.
\end{equation}
Multiplying both sides of \eqref{gtq11} by $\bbS$, we have
\begin{equation}\label{gtq12}
	x_0^{-1}\bbS\bbq_0=\bbS\bbV(\bbD\bbV^\top\bbq_0)\,.
\end{equation}
Similarly, multiplying both sides of \eqref{gtq11} by $(\bbI-\bbS)$, we have
\begin{equation}\label{gtq13}
	x_0^{-1}(\bbI-\bbS)\bbq_0=(\bbI-\bbS)\bbV(\bbD\bbV^\top\bbS\bbq_0)\,.
\end{equation}
Notice that $\bbq_0=\bbS\bbq_0+(\bbI-\bbS)\bbq_0$, by \eqref{gtq12} and \eqref{gtq13}, \eqref{gtq14} holds by defining $\bbq_{10}=x_0\bbD\bbV^\top\bbq_0$ and $\bbq_{20}=x_0\bbD\bbV^\top\bbS\bbq_0$.
Now it is ready for us to show  the statement below \eqref{gtq17} hold.  Substituting $\bbq_0=\bbS\bbV\bbq_{10}+(\bbI-\bbS)\bbV\bbq_{20}$ and $\E\bbA=\bbV\bbD\bbV^\top$ into the eigenvalue definition
$$\bbB_E\bbq_0=x_{0}^{-1}\bbq_0\,,$$
we have the following equality
\begin{equation}\label{gts17}
	\bbV\bbD\bbV^\top\bbS\bbV\bbq_{10}+\bbS\bbV\bbD(\bbI-\bbV^\top\bbS\bbV)\bbq_{20}=x_{0}^{-1}\left[\bbS\bbV\bbq_{10}+(\bbI-\bbS)\bbV\bbq_{20}\right]\,.
\end{equation}
Multiplying $\bbV^\top(\bbI-\bbS)$ to both sides of \eqref{gts17}, we have
$$\bbV^\top(\bbI-\bbS)\bbV\bbD\bbV^\top\bbS\bbV\bbq_{10}=x_0^{-1}\bbV^\top(\bbI-\bbS)\bbV\bbq_{20}\,.$$
This means that $\bbq_{20}=x_0\bbD\bbV^\top\bbS\bbV\bbq_{10}$  if $\bbV^\top(\bbI-\bbS)\bbV =\bbI - \bbV^{\top}\bbS \bbV$ is invertible. Substituting $\bbq_{20}=x_0\bbD\bbV^\top\bbS\bbV\bbq_{10}$ into \eqref{gts17} and multiply both sides of \eqref{gts17} by $\bbV^\top\bbS$, we see that $\bbq_{10}$ is should be an eigenvector of $\bbH(x_0)$ corresponding to the zero eigenvalue if $\bbV^\top\bbS\bbV$ is invertible, therefore $\det(\bbH(x_0))=0$.  

\subsection{Proof of Lemma \ref{le2}}

By Condition \ref{cond2}, for sufficiently large $n$, there exists a positive constant $c$ such that
\begin{equation*}
	\bbV^\top(\E\bbS)\bbV\le  p_n\bbI\, \text{ }\text{ }\text{ and }\text{ }\text{ } 2cp_n\bbI \le \bbV^\top(\E\bbS+(p_n-\E a_{11})\bbe_1\bbe_1^\top)\bbV\,,
\end{equation*}
in which $\bbe_1 = (1, 0, \ldots, 0)^\top \in \R^{n}$.  Condition \ref{cond3} implies that
$$\|\bbV^\top(p_n-\E a_{11}))\bbe_1\bbe_1^\top\bbV\|_F\le \frac{p_nKC^2}{n}\,.$$
Combining the three above inequalities together, we have
\begin{equation}\label{g11}
	cp_n\bbI \le \bbV^\top(\E\bbS)\bbV\le  p_n\bbI\,.
\end{equation}
Then we study the relation between $\bbV^\top(\E\bbS)\bbV$ and $\bbV^\top\bbS\bbV$.  By  Lemma \ref{le1}, we have for any $t>0$,   
\begin{equation}\label{g15}
	\p\left(\left|\bbv_i^\top(\bbS-\E\bbS)\bbv_j\right|\ge t\right)\le \exp\left(-\frac{t^2/2}{\sum_{l=1}^n\bbv_{il}^2\bbv_{jl}^2\var(S_{ll})+\frac{Lt}{3}}\right)\,,
\end{equation}
in which $L=\frac{C^2}{n}$ by Condition \ref{cond3}.   It follows from Conditions \ref{cond2} and \ref{cond3} that, $$\sum_{l=1}^n\bbv_{il}^2\bbv_{jl}^2\var(S_{ll}) \leq  p_n\sum_{l=1}^n \bbv_{il}^2 \bbv_{jl}^2\le \frac{C^2p_n}{n}\,.$$
We choose  $t=2c_1\sqrt{\log n}\cdot\sqrt{p_n\frac{C^2}{n}}$ for some constant $c_1>1$. With this choice of $t$, under Condition \ref{cond2}, it holds for sufficiently large $n$ that
$$\frac{t^2/2}{\sum_{l=1}^n\bbv_{il}^2\bbv_{jl}^2\var(S_{ll})+\frac{Lt}{3}}\ge c_1\log n\,.$$

It follows from Condition \ref{cond2} that $p_n\sqrt[4]{\frac{\log n}{np_n}}\gg t$.
Moreover, since $c_1$ can any positive constant, by \eqref{g15}, with high probability we have
\begin{equation}\label{g3b}
	-cp_n\left(\sqrt[4]{\frac{\log n}{np_n}}\right)\bbI\le\bbV^\top(\bbS-\E\bbS)\bbV\le p_n\left(\sqrt[4]{\frac{\log n}{np_n}}\right)\bbI\,.
\end{equation}
This combined with \eqref{g11} implies that  with high probability,
\begin{equation*}
	cp_n\left(1-\sqrt[4]{\frac{\log n}{np_n}}\right)\bbI\le\bbV^\top\bbS\bbV\le p_n\left(1+\sqrt[4]{\frac{\log n}{np_n}}\right)\bbI<\left(1-\frac{c}{2}\right)\bbI\,.
\end{equation*}

\subsection{Proof of Corollary \ref{coro1}}

By Lemma \ref{le2}, with high probability, both $\bbV^\top \bbS \bbV$ and $\bbI - \bbV^\top \bbS \bbV$ are invertible and the inequalities \eqref{eq:invertible} hold. In the rest, we restrict ourselves to this high probability event $\mathcal{A}_1$.

We will show that for suitably chosen $\bbq_l$,  $\text{dim}(\text{span}\{\bbq_l, l\in[\pm K]\})=2K$.  Concretely, for any given pair $i_0\neq j_0$,   we consider two scenarios \textbf{(I)} and \textbf{(II)}.

\textbf{(I)}  $x_{i_0}\neq x_{j_0}$. Note that $-\bbS(\E \bbA)\bbS+(\E \bbA)\bbS+\bbS\E \bbA$ is a real symmetric matrix and $\bbq_{i_0}$ and $\bbq_{j_0}$ are eigenvectors corresponding to distinct eigenvalues,  then $\bbq_{i_0}^\top\bbq_{j_0}=0\,.$

\textbf{(II)}  $x_{i_0}=x_{j_0}$. In this scenario, the argument above does not directly apply.  However, we can perturb the entries of $\bbA$ and $\bbS$ and replicate the argument, and then make the perturbation vanish in the limit.  Concretely,  we replace $a_{ij}$ by $\widehat a_{ij}=a_{ij}+e^{-m}g_{ij}$, where $m\ge n$, $g_{ij}=g_{ji}$ and $g_{ij}$ follows i.i.d. standard guassian distribution for $i\le j$. Then the entries of $\widehat\bbA=(\widehat a_{ij})$ are absolute continuous random variables.  Then the entries of the matrix $-\widehat\bbS(\E \widehat\bbA)\widehat\bbS+(\E \widehat\bbA)\widehat\bbS+\widehat\bbS\E \widehat\bbA$, where $\widehat \bbS = \text{diag}(\widehat a_{11}, \ldots, \widehat a_{1n})$,  are absolute continuous random variables, and its nonzero eigenvalues are not equal almost surely (c.f. \cite{AJ13}).   Similar to $\E\bbA$, we write $\E\widehat\bbA=\E\bbA=\bbV\bbD\bbV^\top$. By the tail probability of standard guassian distribution,  $\max_{j}|\widehat a_{1j}-a_{1j}|\le \frac{p_n}{\log n}$ with high probability. This combined with with  Lemma \ref{le2} implies that $\bbV^\top\widehat\bbS\bbV$ and $\bbI-\bbV^\top\widehat\bbS\bbV$ are invertible with high probability. Denote this high probability event by $\mathcal{A}_2$.  In the following, we restrict ourselves to $\mathcal{A}_1\cap \mathcal{A}_2$. Note that with $a_{ij}$ replaced by $(\widehat a_{ij})$, counterparts of Theorem \ref{thm2} holds by following  exact the same proof, and we use notations $\widehat \bbq_{1i}$, $\widehat \bbq_{2i}$, and $\widehat \bbq_{i}$ accordingly. By Theorem \ref{thm2}, $\|\widehat\bbq_{1i}\|_2\neq 0$, $i\in [\pm K]$. Without loss of generality,  we assume $\|\widehat\bbq_{1i}\|_2=1$, $i\in [\pm K]$. We denote the non-zero eigenvalues of  $-\widehat\bbS(\E \widehat\bbA)\widehat\bbS+(\E \widehat\bbA)\widehat\bbS+\widehat\bbS\E \widehat\bbA$ by $\widehat x_{-j}$, $j\in\{\pm 1, \ldots, \pm K\}$ and $\widehat x_i^{-1} \leq  \widehat x_{j}^{-1}$ for $i < j$.  Moreover, as almost surely, the non-zero real eigenvalues of  $-\widehat\bbS(\E \widehat\bbA)\widehat\bbS+(\E \widehat\bbA)\widehat\bbS+\widehat\bbS\E \widehat\bbA$ are not equal, we can just ignore the measure zero set and take $\widehat x_i^{-1}  <  \widehat x_{j}^{-1}$ for $i < j$.  For the particular indexes $i_0$ and $j_0$, we have $\widehat \bbq_{i_0} ^{\top} \widehat \bbq_{j_0} = 0$.

By Weyl's inequality,  $\lim_{m\rightarrow \infty}\widehat x_{i_0}=x_{i_0}$ and $\lim_{m\rightarrow \infty}\widehat x_{j_0}=x_{j_0}$.    Without loss of generality, assume that the limits $\lim_{m\rightarrow \infty}\widehat\bbq_{1i_0}$ and $\lim_{m\rightarrow \infty}\widehat\bbq_{1j_0}$ exist.  Otherwise, because $\|\widehat\bbq_{1i_0}\|_2=\|\widehat\bbq_{1j_0}\|_2=1<\infty$, we can always find a subsequence of $\{m,m+1,\ldots\}$ and take the limits on this subsequence. 
Denote by  $\bbq_{1i_0}=\lim_{m\rightarrow \infty}\widehat\bbq_{1i_0}$ and $\bbq_{1j_0}=\lim_{m\rightarrow \infty}\widehat\bbq_{1j_0}$. It can be shown easily that $\bbq_{i_0}$ and  $\bbq_{j_0}$ are unit eigenvectors of $\bbH(x_{i_0}) $ and $\bbH(x_{j_0}) $, respectively.    Let    $\bbq_{i_0} = \bbS\bbV\bbq_{1i_0}+(\bbI-\bbS)\bbV\bbq_{2i_0}$ and $\bbq_{j_0} = \bbS\bbV\bbq_{1j_0}+(\bbI-\bbS)\bbV\bbq_{2j_0}$. By the definition of $\widehat\bbq_{i_0}$, we have
$$[-\widehat\bbS(\E \widehat\bbA)\widehat\bbS+(\E \widehat\bbA)\widehat\bbS+\widehat\bbS\E \widehat\bbA]\widehat\bbq_{i_0}=\widehat x_{i_0}^{-1}\widehat\bbq_{i_0}\,.$$
Then
\begin{align*}
	&[-\bbS(\E \bbA)\bbS+(\E \bbA)\bbS+\bbS\E \bbA]\bbq_{i_0}=\lim_{m\rightarrow \infty}[-\widehat\bbS(\E \widehat\bbA)\widehat\bbS+(\E \widehat\bbA)\widehat\bbS+\widehat\bbS\E \widehat\bbA]\widehat\bbq_{i_0}\non
	&=\lim_{m\rightarrow \infty}\left[\widehat x_{i_0}^{-1}\widehat\bbq_{i_0}\right]=x_{i_0}^{-1}\bbq_{i_0}\,.
\end{align*}

Combining this with Lemma \ref{le2}, $\bbq_{i_0}$ ($\bbq_{j_0}$) is not equal to $\mathbf{0}$ and it is the eigenvector of $-\bbS(\E \bbA)\bbS+(\E \bbA)\bbS+\bbS(\E \bbA)$ corresponding to $x_{i_0}^{-1}$ ($x_{j_0}^{-1}$). Moreover,  
\begin{align*}
	&\bbq_{i_0}^\top\bbq_{j_0}= (\lim_{m\rightarrow  \infty} \widehat \bbq_{i_0})^\top(\lim_{m\rightarrow  \infty} \widehat \bbq_{j_0})=\lim_{m\rightarrow  \infty} \widehat \bbq_{i_0}^\top\widehat \bbq_{j_0} = 0, \ a.s\,.
\end{align*}
In the above,  one should note that the limit is taken on $m$ while $n$ is fixed.

Therefore we can finish our proof of the first statement \eqref{gts2}. The second statement follows from \eqref{gts2}  directly.

\subsection{Proof of Theorem \ref{thm1}}

Because  $\text{det}(\bbI-\bbA\bbB)=\text{det}(\bbI-\bbB\bbA)$, $\text{det}(\bbH(x)) = 0$ (i.e., \eqref{g0}) is equivalent to
\begin{eqnarray}\label{g7}
	\text{det}(\bbI-x(\bbV^\top\bbS\bbV)^{1/2}\bbD(\bbV^\top\bbS\bbV)^{1/2}-x^2(\bbV^\top\bbS\bbV)^{1/2}\bbD(\bbI-\bbV^\top\bbS\bbV)\bbD(\bbV^\top\bbS\bbV)^{1/2})=0\,.
\end{eqnarray}
Let $\bbA_1=(\bbV^\top\bbS\bbV)^{1/2}\bbD(\bbV^\top\bbS\bbV)^{1/2}$,  $\bbA_2=(\bbV^\top\bbS\bbV)^{1/2}\bbD(\bbI-\bbV^\top\bbS\bbV)\bbD(\bbV^\top\bbS\bbV)^{1/2}$ and $y=1/x$.   Then \eqref{g7} becomes
\begin{eqnarray}\label{g8}
	\text{det}(y^2\bbI-y\bbA_1-\bbA_2)=0\,.
\end{eqnarray}

By Lemma \ref{le2} and Condition \ref{cond1}, with high probability we have
\begin{equation}\label{g1b}
	\|\bbA_1\|\le |d_1|\cdot\|\bbV^\top\bbS\bbV\|\le |d_1|\cdot p_n\left(1+\sqrt[4]{\frac{\log n}{np_n}}\right)\lesssim np_n^2\,.
\end{equation}
Also by  Lemma \ref{le2} and Condition \ref{cond1}, it holds with high probability that

\begin{equation}\label{g13}
	n^2p_n^3\bbI\lesssim \lambda_{K}(\bbV^\top\bbS\bbV)\lambda_{K}(\bbD^2)\bbI\lesssim \bbA_2 \le \lambda_1(\bbV^\top\bbS\bbV)\lambda_1(\bbD^2)\bbI\lesssim n^2p_n^3\bbI\,.
\end{equation}
Combining \eqref{g1b} and \eqref{g13}, we have with high probability,
\begin{equation}\label{g14}
	\bbA_1\lesssim np_n^2\bbI\lesssim \frac{1}{np_n}\bbA_2\,.
\end{equation}
Similarly, we also have
\begin{equation}\label{g14a}
	\bbA_1\gtrsim np_n^2\bbI\,.
\end{equation}
For $y$ to be a solution for \eqref{g8}, it must hold that
$$\lambda_K(y\bbA_1 + \bbA_2) \bbI\le y^2\bbI\le \lambda_1(y\bbA_1 + \bbA_2) \bbI\,.$$
Combining this with  \eqref{g13} and \eqref{g14}, we have
$$ -|y|np_n^2\bbI+n^2p_n^3\bbI\lesssim y^2\bbI\lesssim |y|np_n^2\bbI+n^2p_n^3\bbI\,.$$
Therefore we have
\begin{equation}\label{g2b}
	-|y_i|np_n^2+n^2p_n^3\lesssim y_i^2\lesssim |y_i|np_n^2+n^2p_n^3, \ i=\pm 1,\ldots,\pm K\,.
\end{equation}
For the right inequality that $y_i^2\lesssim |y_i|np_n^2+n^2p_n^3$, by the solutions to quadratic equation in one unknown, we imply that
$$|y_i|\lesssim np_n^2+\sqrt{n^2p_n^4+4n^2p_n^3}\lesssim np_n^{3/2}\,.$$
Similarly considering the left inequality of \eqref{g2b}, we have
$$|y_i|\gtrsim -np_n^2+\sqrt{n^2p_n^4+4n^2p_n^3}\gtrsim np_n^{3/2}\,.$$
It follows from the two above inequalities that  $|x_i|^{-1}=|y_i|\sim np_n^{3/2}$.  This conclude the proof of the first claim.

When $p_n\rightarrow 0$, by \eqref{g14}, we have $|y_i|\bbA_1\lesssim p_n^{1/2}\bbA_2\ll \bbA_2$. Then to approximately solve for $y$ in \eqref{g8},  we solve $z$ in   the following determinant equation
\begin{eqnarray}\label{g8.1}
	\text{det}(z^2\bbI-\bbA_2)=0\,.
\end{eqnarray}
The left hand side of \eqref{g8.1} is a $2K$ polynomial of $z$; therefore \eqref{g8.1} has  $2K$ solutions in $z$. A straightforward calculation shows that \eqref{g8.1}
has $2K$ solutions $\pm\sqrt{\lambda_i(\bbA_2)}$, $i\in[K]$.   By Lemma \ref{le2} and an intermediate step \eqref{g3b} in its proof, it holds with high probability that
\begin{align*}
	&|\lambda_i(\bbV^\top(\E\bbS)\bbV\bbD(\bbI-\bbV^\top(\E\bbS)\bbV)\bbD)-\lambda_i(\bbA_2)|\non
	&=|\lambda_i(\bbV^\top(\E\bbS)\bbV\bbD(\bbI-\bbV^\top(\E\bbS)\bbV)\bbD)-\lambda_i(\bbV^\top\bbS\bbV\bbD(\bbI-\bbV^\top\bbS\bbV)\bbD)|
	&\lesssim n^2p_n^3 \left(\sqrt[4]{\frac{\log n}{np_n}}\right)\,.
\end{align*}
Combining this with eigenvalue separation condition $\mu_i$, we have
\begin{equation}\label{g3b.1}
	1+c_1\le \frac{\lambda_{i-1}(\bbA_2)}{\lambda_i(\bbA_2)},\quad i=2,\ldots,K\,,
\end{equation}
for some positive constant $c_1$.
For $i=1,\ldots, K$, let $y_{1i}=\sqrt{\lambda_i(\bbA_2)}+p_n^{5/6}n^{1/2}\sqrt{\|\bbA_1\|}$ and $y_{2i}=\sqrt{\lambda_i(\bbA_2)}-p_n^{5/6}n^{1/2}\sqrt{\|\bbA_1\|}$, by \eqref{g13}, \eqref{g14}, \eqref{g14a}, \eqref{g3b.1} and Weyl's inequality, it can be shown that with high probability

\begin{align}\label{g5a}
	\lambda_{i-1}(y_{li}\bbA_1+\bbA_2)\ge \lambda_{i-1}(\bbA_2)-|y_{li}|\|\bbA_1\|&\ge(1+c_1)\lambda_i(\bbA_2)-n^2p_n^{7/2-1/100}>y_{li}^2\non
&\sim \lambda_i(\bbA_2)-(-1)^ln^2p_n^{10/3}\,,
\end{align}
\begin{align}\label{g5b}
	\lambda_{i+1}(y_{li}\bbA_1+\bbA_2)\le \lambda_{i+1}(\bbA_2)+|y_{li}|\|\bbA_1\|&\le (1+c_1)^{-1}\lambda_i(\bbA_2)+n^2p_n^{7/2-1/100}<y_{li}^2\non
	&\sim \lambda_i(\bbA_2)-(-1)^ln^2p_n^{10/3},  l=1,2\,.
\end{align}
By Weyl's inequality and \eqref{g5a}-\eqref{g5b}, we have
\begin{align}\label{g5c}\lambda_i(y_{1i}\bbA_1+\bbA_2)\le \lambda_i(\bbA_2)+|y_{1i}|\|\bbA_1\|<y_{1i}^2, \quad \lambda_i(y_{2i}\bbA_1+\bbA_2)>\lambda_i(\bbA_2)-|y_{2i}|\|\bbA_1\|>y_{2i}^2\,.
\end{align}
It follows from \eqref{g5a}-\eqref{g5c} that
\begin{align}
	&\text{det}(y_{1i}^2\bbI-y_{1i}\bbA_1-\bbA_2)\cdot \text{det}(y_{2i}^2\bbI-y_{2i}\bbA_1-\bbA_2)\non
	&=\prod_{j=1}^K\left(y_{1i}^2-\lambda_j(y_{1i}\bbA_1+\bbA_2)\right)\prod_{j=1}^K\left(y_{2i}^2-\lambda_j(y_{2i}\bbA_1+\bbA_2)\right)< 0\,.
\end{align}
Since $\text{det}(y^2\bbI-y\bbA_1-\bbA_2)$ is a continuous function of $y$,
there exists one $y_i\in[y_{2i},y_{1i}]$ satisfying \eqref{g7}, or equivalently \eqref{g8}.  Moreover, by  \eqref{g13},\eqref{g14}, \eqref{g14a}  and \eqref{g3b.1}, the intervals $[y_{2i}, y_{1i}]$ are non overlapping for different $i$. Hence the second claim of Theorem \ref{thm1} holds for $i\in[K]$. Similarly,  the second claim of Theorem \ref{thm1} holds for $i<0$ by almost the same proof if we define  $y_{1i}=-\sqrt{\lambda_{K+i+1}(\bbA_2)}+p_n^{5/6}n^{1/2}\sqrt{\|\bbA_1\|}$ and $y_{2i}=-\sqrt{\lambda_{K+i+1}(\bbA_2)}-p_n^{5/6}n^{1/2}\sqrt{\|\bbA_1\|}$.

\subsection{Proof of Lemma \ref{errorbound}}
By Theorem 5.2 of \cite{JA15}, with high probability we have
\begin{equation}\label{t1}
	\|\bbA-\E\bbA\|\lesssim \sqrt{np_n}\,.
\end{equation}
Combining this with the fact that $\|\bbS\|\le 1$, it holds with high probability that
$$\|\bbB-\bbB_E\|\lesssim \sqrt{ np_n}\,.$$
\eqref{2203.1} and Theorem \ref{thm1} (first part) imply the second part of this Lemma, whose proof we omit. Therefore we complete the proof of this Lemma.

\subsection{The matrix $\bbA$ and no-self loop}\label{sec:self-loop}

In network analysis, if there is no self-loop, then the diagonal entries of the adjacency matrix are $0$'s. In this case we should analyze $\widetilde\bbA=\bbA-\text{diag}(a_{11},\ldots,a_{nn})$ instead of $\bbA$ and therefore individual $1$'s perceived adjacency  matrix is $\widetilde \bbB=-\widetilde\bbS\widetilde\bbA\widetilde\bbS+\widetilde\bbA\widetilde\bbS+\widetilde\bbS\widetilde\bbA$, where $\widetilde\bbS=\bbS-\text{diag}(a_{11},0,\ldots,0)$. In this case, $\E\widetilde\bbA$ may not be a low-rank matrix, while $\E\bbA$ is. For instance, we look at a simple  SBM with $n=4$, where the corresponding $4\times 4$ expected adjacency matrix with self-loop can be expressed as
$$\E\bbA=\left(
\begin{array}{cccc}
	p_{11} & p_{11} & p_{12} &p_{12}\\
	p_{11} & p_{11} & p_{12}&p_{12}\\
	p_{21} & p_{21} & p_{22}&p_{22}\\
	p_{21} & p_{21} & p_{22}&p_{22}\\
\end{array}
\right)\,,$$
in which $p_{12}=p_{21}< p_{11}=p_{22}$, and $p_{ij}\in (0,1)$. Clearly $\text{rank}(\E\bbA)=2$.  In contrast, the expectation of the matrix $\widetilde \bbA$ is
$$\E\widetilde\bbA=\left(
\begin{array}{cccc}
	0& p_{11} & p_{12} &p_{12}\\
	p_{11} & 0& p_{12}&p_{12}\\
	p_{21} & p_{21} & 0&p_{22}\\
	p_{21} & p_{21} & p_{22}&0\\
\end{array}
\right)\,.$$
It is easy to show that $\text{rank}(\E\widetilde\bbA)=4$, so $\E\widetilde\bbA$ has the full rank.

There is a rich line of network literature that assumes a low-rank structure of $\bbA$, including \cite{zhao2012}, \cite{abbe2017}, \cite{AFWZ17}, and \cite{YEJ15}.  Moreover, in this paper we consider the case that $\|\widetilde\bbA-\E\widetilde\bbA\|\rightarrow \infty$, which is a common assumption in network models. Loosely, this assumption is to avoid extremely sparse networks (e.g., we do not deal with the case that the largest degree of the nodes are bounded). By the equation  $\widetilde\bbA=\E \bbA+(\widetilde\bbA-\E\bbA)$ and $\|\widetilde\bbA-\E\bbA\|\sim\|\widetilde\bbA-\E\widetilde\bbA\|\rightarrow \infty$, $\widetilde\bbA-\E\bbA$ can be regarded as the ``noise matrix'' of the model.  Hence the noise level (measured by spectral norm) of  $\widetilde \bbA$ is not changed compared to $\bbA$. In other words, the signal matrix of $\widetilde\bbA$  is essentially $\E\bbA$.  Therefore a major term of $\widetilde \bbB$ is also a major term of $\bbB$, and Theorems \ref{thm2}--\ref{thm1} can be applied too.  On the other hand, by the definition of bernoulli random variables, $\|\text{diag}(a_{11},a_{22},\ldots,a_{nn})\|\le 1$. By checking the proofs carefully, the community detection results from  Theorem \ref{t5} to Theorem \ref{consist_fixedS} mainly rely on the order of the gap between  $\|\widetilde\bbB-\bbB_E\|$ and the smallest non-zero  eigenvalue (in magnitude) of $\bbB_E$. It is  essentially  the same  as the gap between $\|\bbB-\bbB_E\|$ and the smallest non-zero  eigenvalue (in magnitude) of $\bbB_E$, where the difference between $\bbS$ and $\widetilde \bbS$ can be shown to have a negligible effect on this gap. Therefore Theorems \ref{t5}--\ref{consist_fixedS} hold for the stochastic block model without self-loop.  Given the above arguments, throughout this paper, we only consider $\bbA$ (instead of $\widetilde\bbA$) for convenience.

\section{Proofs of the results in Section \ref{sec:main_results}}

\subsection{Proof of Lemma \ref{le4}}


As we have assumed that individual $1$ belongs to community $1$,   $\p(a_{1l}=1)=\bpi_1^\top\bbP\bpi_l = \bbe_1^\top \bbP\bpi_l$, for $l\in[n]$. Then in view of the definition of $p_n$ and $\min_{k\in[K]}p_{1k}\sim p_n$, it follows that   $\min_{l\ge 2}\p(a_{1l}=1)\sim p_n$. Moreover, $1-c\ge p_n\gg (1/ n)^{1/2} \gg \log n / n$.  Therefore Condition \ref{cond2} is validated.

By $\min_{k\in[K]}\sum_{j=1}^n\mathbf{1}(\bpi_j=\bbe_k)\ge c_0n$, $\sigma_K(\bbP)\ge c_1p_n$,  we have 
$$\rank(\E\bbA) = \rank (\bPi\bbP\bPi^\top) = K\,.$$

Recall the eigen decomposition $\E\bbA=\bbV\bbD\bbV^\top$, in which $\bbV=(\bbv_1,\ldots,\bbv_K)$. By the structure of the stochastic block model and $\text{rank}(\E\bbA)=K$, there are  $K$ different rows in $\bbV$ corresponding to the communities and hence there are $K^2$ different values in $\bbV$ at most. Indeed, let $\bbv_k(l)$ be the $l$-th entry of $\bbv_k$, by the definition of eigenvector, we have
$$(\E\bbA)\bbv_k=d_k\bbv_k\,, \quad \text{ for } k \in [K]\,,$$
and therefore
\begin{equation*}
	\sum_{j=1}^n\left(\bpi_l^\top\bbP\bpi_j\bbv_k(j)\right)=d_k\bbv_k(l),\quad \text{ for } k \in [K] \text{ and } l\in[n]\,.
\end{equation*}
For any  $l_1\neq l_2$ with $\bpi_{l_1}=\bpi_{l_2}$, we have $\bpi_{l_1}^\top\bbP\bpi_j=\bpi_{l_2}^\top\bbP\bpi_j$, $j=1,\ldots,n$ and therefore
$$\sum_{j=1}^n\left(\bpi_{l_1}^\top\bbP\bpi_j\bbv_k(j)\right)=\sum_{j=1}^n\left(\bpi_{l_2}^\top\bbP\bpi_j\bbv_k(j)\right)=d_k\bbv_k(l_1)=d_k\bbv_k(l_2)\,.$$
Notice that $\bpi_i\in \{\bbe_1,\ldots,\bbe_K\}$, $i\in [n]$.  Then we conclude that $\bbV$ has at most $K$ different rows  and $\{\bbv_k(l), k\in[K],l\in[n]\}$ only has at most $K^2$ distinct values.  Moreover, as $\bpi_{l_1}\neq \bpi_{l_2}$ means that $l_1$ and $l_2$ belong to different communities, the rows of $\bbV$ with distinct values are corresponding to different communities. Since $\text{rank}(\E\bbA)=\text{rank}(\bbV\bbD\bbV^\top)=K$, $\text{rank}(\bbV)=K$ and therefore $\bbV$ contains exactly $K$ different rows.

Without loss of generality, assume that the first $K$ rows of $\bbV$ are different and we denote this $K\times K$ matrix by $\bbV^{(K)}$. Since distinct row values are corresponding to different communities, the first $K$ rows of $\bPi$ are different. Noticing that $\bpi_i\in \{\bbe_1,\ldots,\bbe_K\}$, $i\in [n]$; without loss of generality, we assume the first $K$ rows of $\bPi$ equal to $\bbI$. Let $\mathcal{D}=\bbV^{(K)}$.
Then it follows that 
\begin{equation}\label{g18a}
	\bbV=\bPi\mathcal{D}\,.
\end{equation}

Because $\mathcal{D}^\top\bPi^\top\bPi\mathcal{D}=\bbV^\top\bbV=\bbI$, we have $\mathcal{D}\mathcal{D}^\top\ge \frac{1}{P}\bbI$,
where $P=\max_{k\in[K]}\sum_{j=1}^n\mathbf{1}(\bpi_j=\bbe_i)$. Therefore, \eqref{g18} is proved. Moreover, by the condition that $\min_{k\in[K]}\sum_{j=1}^n\mathbf{1}(\bpi_j=\bbe_k)\ge c_0n$, we have

\begin{equation*}
	\mathcal{D}\mathcal{D}^\top\le\frac{1}{c_0n}\bbI\,.
\end{equation*}
Therefore we have

$$\|\mathcal{D}\|_{\max}\le  \frac{1}{\sqrt{c_0n}}\,.$$
Combining this with \eqref{g18a},  Condition \ref{cond3} holds. Now we prove Condition \ref{cond1}. Notice that $\bPi^\top\bPi = \bpi_1 \bpi_1^\top + \ldots, + \bpi_n \bpi_n^\top$ is a $K\times K$ diagonal matrix whose diagonal elements are $\sum_{j=1}^n\mathbf{1}(\bpi_j=\bbe_k)$, $k\in[K]$. By Condition \ref{ncond1}, we have
\begin{align*}
	&d_K^2=\sigma^2_{K}(\E\bbA)=\sigma^2_{K}(\bPi\bbP\bPi^\top)=\lambda_{K}(\bPi\bbP\bPi^\top\bPi\bbP\bPi^\top)\non
	&=\lambda_{K}(\bbP\bPi^\top\bPi\bbP\bPi^\top\bPi)\ge \lambda_K(\bPi^\top\bPi)\lambda_{K}(\bbP\bPi^\top\bPi\bbP)\non
	&\ge \lambda^2_K(\bPi^\top\bPi)\lambda_{K}(\bbP^2)=\sigma^2_K(\bPi^\top\bPi)\sigma^2_K(\bbP)\gtrsim n^2w^2_n\,.
\end{align*}

This, combined with
$$\|\E\bbA\|_F = \left[\sum_{j=1}^n \sum_{i=1}^n \left(\p(a_{ij}=1)\right)^2\right]^{1/2}\le np_n\,,$$

implies that $|d_1|\sim\ldots\sim |d_K|\sim np_n$, which is Condition \ref{cond1}.

\subsection{Proof of Lemma \ref{le6}}

First note that \eqref{g19} can be equivalently expressed as
\begin{equation}\label{g21}
	\bbQ=\left(\frac{\bbS\bPi}{\sqrt{p_n}},(\bbI-\bbS)\bPi\right)\mathcal{Q},\quad \text{where}\quad \mathcal{Q}=\left(
	\begin{array}{ccc}
		\sqrt{p_n}\mathcal{D}\mathcal{Q}_1 \\
		\mathcal{D}\mathcal{Q}_2 \\
	\end{array}
	\right)\,.
\end{equation}
We have the following lower bounds.
\begin{lem}\label{le3}
	Under Condition \ref{ncond1}, with high probability, there exists some positive constant $c_2$ such that
	\begin{equation}\label{g24}
		\mathcal{Q}\mathcal{Q}^\top \ge (c_2 n)^{-1}\bbI\,,\quad \mathcal{D}\mathcal{Q}_2\mathcal{Q}_2^\top \mathcal{D}^\top\ge (c_2 n)^{-1}\bbI\,,\quad \text{and} \quad p_n \mathcal{D}\mathcal{Q}_1\mathcal{Q}_1^\top\mathcal{D}^\top \ge (c_2 n)^{-1}\bbI\,.
	\end{equation}

\end{lem}

\begin{proof}
	Substituting \eqref{g21} into $\bbQ^\top\bbQ=\bbI$ we have,
	\begin{equation}\label{g22}
		\bbI=\bbQ^\top\bbQ=\mathcal{Q}^\top \text{diag}\left(\bPi^\top\frac{\bbS}{p_n}\bPi,\bPi^\top(\bbI-\bbS)\bPi\right)\mathcal{Q}\,.
	\end{equation}
	By  Lemma \ref{le2} and \eqref{g18} in Lemma \ref{le4}, there exists a positive constant $c_2$ such that with high probability,
	\begin{align}\label{g23}
		\text{diag}\left(\bPi^\top\frac{\bbS}{p_n}\bPi,\bPi^\top(\bbI-\bbS)\bPi\right)&=\text{diag}(\mathcal{D}^\top,\mathcal{D}^\top)^{-1}\text{diag}\left(\bbV^\top\frac{\bbS}{p_n}\bbV,\bbV^\top(\bbI-\bbS)\bbV\right)\text{diag}(\mathcal{D},\mathcal{D})^{-1}\non
&\le c_2 n\bbI\,.
	\end{align}
	By \eqref{g22} and \eqref{g23}, with high prbability we have
	\begin{equation*}
		\mathcal{Q}\mathcal{Q}^\top \ge (c_2 n)^{-1}\bbI\,.
	\end{equation*}
	Then by the definition of $\mathcal{Q}$ in \eqref{g21}, we have
	\begin{equation*}
		\mathcal{D}\mathcal{Q}_2\mathcal{Q}_2^\top \mathcal{D}^\top\ge (c_2 n)^{-1}\bbI\quad \text{ and } \quad p_n \mathcal{D}\mathcal{Q}_1\mathcal{Q}_1^\top\mathcal{D}^\top \ge (c_2 n)^{-1}\bbI\,.
	\end{equation*}
	
\end{proof}

\begin{proof}[Proof of Lemma \ref{le6}]
	By \eqref{g19} and Corollary \ref{coro1}, $\bbS\bPi$ and $(\bbI-\bbS)\bPi$ contain $K$ different non-zero rows in each matrix; without loss of generality, these rows can be rearranged  as  $2K\times 2K$ identity matrix $\bbI_{2K}=\text{diag}(\bbI_K,\bbI_K)$, where the two $\bbI_K$'s correspond to the different non-zero rows of $\bbS\bPi$ and $(\bbI-\bbS)\bPi$ respectively.   For \eqref{h1}, without loss of generality, assume that $\bpi_i=\bbe_1$ and $\bpi_j=\bbe_2$, $a_{1i}=a_{1j}=1$. The other cases $a_{1i}\neq a_{1j}$ and $a_{1i}=a_{1j}=0$ can be proved similarly.  
	
	By \eqref{g24} in Lemma \ref{le3}, with high probability, uniformly for $j$ we have
	\begin{align*}
		&\left\|\bbQ(i)\bbO-\bbQ(j)\bbO\right\|_2=\left\|(\bbe_i^{(n)}-\bbe_j^{(n)})^\top\left(\frac{\bbS\bPi}{\sqrt{p_n}},(\bbI-\bbS)\bPi\right)\mathcal{Q}\bbO\right\|_2\non
		&\ge \|(\bbe^{(2K)}_1-\bbe_2^{(2K)})\|_2\times\sigma_{2K}\left(\mathcal{Q}\bbO\right)\ge \frac{\|\bbe_1^{(2K)}-\bbe_2^{(2K)}\|_2}{\sqrt{c_2n}}=\sqrt{\frac{2}{c_2n}}\,.
	\end{align*}
	For \eqref{h1s}, without loss of generality, assume that $\bpi_i=\bpi_j=\bbe_1$, $a_{1i}=1-a_{1j}=1$. Similar to the inequality above, we have
	\begin{align*}
		&\left\|\bbQ(i)\bbO-\bbQ(j)\bbO\right\|_2=\left\|(\bbe_i^{(n)}-\bbe_j^{(n)})^\top\left(\frac{\bbS\bPi}{\sqrt{p_n}},(\bbI-\bbS)\bPi\right)\mathcal{Q}\bbO\right\|_2\non
		&\ge \|(\bbe^{(2K)}_1-\bbe_{K+1}^{(2K)})\|_2\times\sigma_{2K}\left(\mathcal{Q}\bbO\right)\ge \frac{\|\bbe_1^{(2K)}-\bbe_{K+1}^{(2K)}\|_2}{\sqrt{c_2n}}=\sqrt{\frac{2}{c_2n}}\,.
	\end{align*}
	
	The implication \eqref{h1ss} follows from the expression of $\bbQ$ in \eqref{g19} and the fact that $\|\bbQ(i)\bbO- \bbQ(j)
	\bbO\|_2$ = $\|\bbQ(i) - \bbQ(j)\|_2$ for an orthogonal matrix $\bbO$.
\end{proof}

\subsection{Proof of Lemma \ref{le5}}
By Lemma \ref{errorbound}, with high probability we have
$$\|\bbB-\bbB_E\|\lesssim \sqrt{np_n}\,.$$
Note that $\bbQ$ consists of unit eigenvectors of $\bbB_E$ and Lemma \ref{le4} validates the first statement in Theorem \ref{thm1}. Then by Davis-Kahan theorem in \cite{davis1970rotation} (c.f. Theorem 10 in \cite{cai2013sparse}), with high probability we have

$$\|\bbW\bbW^\top-\bbQ\bbQ^\top\|_F=O\left(\frac{\sqrt{np_n}}{ n p^{3/2}_n}\right)=O\left(\frac{1}{\sqrt n p_n}\right)\,.$$
Moreover,  it follows from the definition of $\bbO$ that
\begin{align}
	&\|\bbW-\bbQ\bbO\|_F^2=\text{tr}\left[(\bbW-\bbQ\bbU_1\bbU_2^\top)^\top(\bbW-\bbQ\bbU_1\bbU_2^\top)\right]\non
	&=\text{tr}\left(\bbW^\top\bbW+\bbU_2\bbU_1^\top\bbQ^\top\bbQ\bbU_1\bbU_2^\top-2\bbU_2\bbU_1^\top\bbQ^\top\bbW\right)\non
	&=4K-2\text{tr}(\bbU_2\bbU_1^\top\bbQ^\top\bbW)\le 4K-2\text{tr}(\bbQ^\top\bbW\bbW^\top\bbQ)\non
	&=\|\bbW\bbW^\top-\bbQ\bbQ^\top\|_F^2\,,
\end{align}
where the inequality follows from $\|\Sigma\|=\|\bbQ^\top\bbW\|\le 1$ and $\text{tr}(\bbU_2\bbU_1^\top\bbQ^\top\bbW)=\text{tr}(\Sigma)\ge \text{tr}(\Sigma^2)=\text{tr}(\bbQ^\top\bbW\bbW^\top\bbQ)$. Therefore \eqref{h2} is proved.

\subsection{Proof of Theorem \ref{t5}}

To prove Theorem \ref{t5}, we need a few more auxilliary results.
The next lemma gives some theoretical properties of Algorithm \ref{alg1}.

\begin{lem}\label{lem2.1}
	Under Condition \ref{ncond1}, for any $1\times 2K$ row vector $\bbc$ and $i\in[n]$, if $\|\bbc-\bbQ(i)\bbO\|_2<1/\sqrt{2c_2n}$, where $c_2$ is the same as in Lemma \ref{le3} and $\bbO$ is defined in Lemma \ref{le5},  then  with high probability, $\|\bbc-\bbQ(j)\bbO\|_2>1/\sqrt{2c_2n}$ for all $j\in[n]$ such that $\bpi_j\neq \bpi_i$.  
\end{lem}
\begin{proof}
	In view of Lemma \ref{le6}, we have
	\begin{align*}
		\|\bbc-\bbQ(j)\bbO\|_2\ge \|\bbQ(i)\bbO-\bbQ(j)\bbO\|_2-\|\bbc-\bbQ(i)\bbO\|_2> \sqrt{\frac{2}{c_2n}}-\frac{1}{\sqrt{2c_2n}}
		=\frac{1}{\sqrt{2c_2n}}\,.
	\end{align*}
	
\end{proof}

Next we bound the set $\mathcal{M}$, which in turn will be used to give an upper bound on the final misclustering rate.

\begin{lem}\label{t3}
	Under Condition \ref{ncond1}, let $\{\bbc_i, i\in[n]\}$ be the centroids returned by \eqref{kmean},    then for $\mathcal{M}$ defined in \eqref{eqn:M}, it holds with high probability that
	$|\mathcal{M}|=O\left(1/ p^2_n\right)\,.$
\end{lem}

First, we show that $\bbS \bPi$ has $K$ different nonzero rows with high probability. Concretely, we will show that
\begin{equation}\label{g28}
	\p(\exists k\in[K]: a_{1i}=0 \text{ for all } i\in[n] \text{ such that }  \bpi_i=\bbe_k)=O(n^{-D})\,,
\end{equation}
for sufficiently large $n$, where $D$ is some positive constant. In fact, by Condition \ref{ncond1}, for sufficiently large $n$ depending on $D$, \eqref{g28} follows from the inequality that
\begin{align*}
	& \text{L.H.S of } \eqref{g28}\le \sum_{k\in[K]}\p\left(a_{1i}=0 \text{ for all } i\in[n] \text{ such that }  \bpi_i=\bbe_k\right)\non
	&\le  \sum_{k\in[K]}\left[1-\min_i\p(a_{1i}=1)\right]^{\sum_{i=1}^n\1(\bpi_i=\bbe_k)}=O\left(e^{-\frac{\sum_{i=1}^n\1(\bpi_i=\bbe_k)}{\sqrt n}}\right)=O(n^{-D})\,.
\end{align*}
By the decomposition equation \eqref{g19} of $\bbQ$, we have $\bbS \bbQ = \bbS \bPi \mathcal{D}\mathcal{Q}_1$. In view of Lemma \ref{le3}, $\mathcal{D}\mathcal{Q}_1(\mathcal{D}\mathcal{Q}_1)^{\top}$ is a $K\times K$ positive definite matrix, and so $\bbS \bbQ$ has $K$ different non-zero rows with high probability.  Then, as $\bbO$ is an orthogonal matrix, $\bbS \bbQ \bbO$ has the same property.  Similarly, we can show that  $(\bbI-\bbS)\bbQ\bbO$ has $K$ different non-zero rows with high probability. 

Then, as  $\{\bbc_i, i\in[n]\}$ are the centroids returned by \eqref{kmean}, we have with high probability,
$$\|\bbS\bbW-\bbS\bbC\|_F\le \|\bbS\bbW-\bbS\bbQ\bbO\|_F\,,$$
and
$$\|(\bbI-\bbS)\bbW-(\bbI-\bbS)\bbC\|_F\le \|(\bbI-\bbS)\bbW-(\bbI-\bbS)\bbQ\bbO\|_F\,.$$
Then it follows
$$\|\bbS\bbC-\bbS\bbQ\bbO\|_F\le \|\bbS\bbW-\bbS\bbC\|_F+\|\bbS\bbW-\bbS\bbQ\bbO\|_F\le 2\|\bbS\bbW-\bbS\bbQ\bbO\|_F\,,$$
and
\begin{align*}
	\|(\bbI-\bbS)\bbC-(\bbI-\bbS)\bbQ\bbO\|_F &\le \|(\bbI-\bbS)\bbW-(\bbI-\bbS)\bbC\|_F+\|(\bbI-\bbS)\bbW-(\bbI-\bbS)\bbQ\bbO\|_F\\
	&\le 2\|(\bbI-\bbS)\bbW-(\bbI-\bbS)\bbQ\bbO\|_F\,.
\end{align*}
Combining this with Lemma \ref{le5}, we conclude that with high probability
\begin{align}\label{230711.1}
	|\mathcal{M}|=\sum_{i\in \mathcal{M}}1 &\le 2c_0n\sum_{i\in \mathcal{M}}\|\bbc_i-\bbQ(i)\bbO\|_2^2 \leq 2c_0n\sum_{i\in [n]}\|\bbc_i-\bbQ(i)\bbO\|_2^2 \non
	&= 2c_0n\left(\|\bbS\bbC-\bbS\bbQ\bbO\|_F^2+\|(\bbI-\bbS)\bbC-(\bbI-\bbS)\bbQ\bbO\|_F^2\right)\non
	&\le 8c_0n\left(\|\bbS\bbW-\bbS\bbQ\bbO\|_F^2+\|(\bbI-\bbS)\bbW-(\bbI-\bbS)\bbQ\bbO\|_F^2\right)\non
	&=O\left(n(\|\bbS\bbW-\bbS\bbQ\bbO\|_F^2+\|(\bbI-\bbS)\bbW-(\bbI-\bbS)\bbQ\bbO\|_F^2)\right)\non
	&=O\left(n\|\bbW-\bbQ\bbO\|_F^2\right)\non
	&=O\left(\frac{1}{ p^2_n}\right).
\end{align}

Finally, we need the following guarantee for our merging strategy in Algorithm \ref{alg2}.

\begin{prop}\label{t4}
	Let $\mathfrak{C}=\{\mathfrak{c}_1,\ldots,\mathfrak{c}_K\}$ and $\mathfrak{D}=\{\mathfrak{d}_1,\ldots,\mathfrak{d}_K\}$ be two ordered sets of size $K$ and let $f_0:[K]\rightarrow [K]$ be the unique permutation such that $\mathfrak{c}_{f_0(i)}=\mathfrak{d}_i$ for all $i\in[K]$.  Suppose that the $K$ elements in these ordered sets  have been endowed with pairwise connection probabilities. Let $\bbC=(c_{ij})$ be a $K\times K$ matrix such that the $(i,j)$-th entry $c_{ij}$ is the connection probability between $\mathfrak{c}_i$ and  $\mathfrak{c}_j$, and $\bbD=(d_{ij})$ be a $K\times K$ matrix such that the $(i,j)$-th entry $d_{ij}$ is the connection probability between  $\mathfrak{c}_i$ and  $\mathfrak{d}_j$.
	If $\emph{rank}(\bbC)=K$, then for a permutation function $f:[K]\rightarrow [K]$,
	\begin{equation*}\label{g29}
		f=f_0\Longleftrightarrow \bbC_{(f,f)}=\bbD_{(f,*)}\,,
	\end{equation*}
	where the $(i,j)$-th entries of  $\bbC_{(f,f)}$ and $\bbD_{(f,*)}$ are  $c_{f(i)f(j)}$ and $d_{f(i)j}$, respectively.
\end{prop}
\begin{proof}
	As $\mathfrak{c}_{f_0(i)}=\mathfrak{d}_i$ for $i\in[K]$, the matrix $\bbD$ is formed by a column permutation of $\bbC$. Therefore, we have
	\begin{equation*}\label{g30}
		\text{rank}(\bbD)=\text{rank}(\bbC)=K\,,
	\end{equation*}
	which implies that columns of $\bbD$ are distinct.
	By the definition of $f_0$, clearly we have
	
	\begin{equation*}\label{g31}
		f=f_0\Longrightarrow \bbC_{(f_0,f_0)}=\bbD_{(f_0,*)}\,.
	\end{equation*}
	On the other hand, if $f\neq f_0$, there exists a $j_0\in [K]$ such that
	\begin{equation*}\label{gtq4}
		f(j_0)\neq f_0(j_0)\,.
	\end{equation*}
	This, combined with the fact the columns of $\bbD$ are all distinct,  implies
	$$(c_{f(1)f(j_0)},\ldots,c_{f(K)f(j_0)})^{\top}\neq (d_{f(1)j_0},\ldots,d_{f(K)j_0})^{\top}\,.$$
	Hence there exists $i_0\in [K]$ such that
	$$c_{f(i_0)f(j_0)}\neq d_{f(i_0)j_0}\,.$$
	In other words,  $\bbC_{(f,f)}\neq\bbD_{(f,*)}$. 
\end{proof}

Combining the results above, we now prove Theorem  \ref{t5}.

\begin{proof}[Proof of Theorem \ref{t5}]

For $i,j\in \mathcal{M}^c$, by Lemma \ref{lem2.1}, with high probability we have
	\begin{equation}\label{gtq3}
		\bbc_i\neq \bbc_j, a_{1i}=a_{1j}\Leftrightarrow \bpi_i\neq \bpi_j, a_{1i}=a_{1j} \,,
	\end{equation}
	which means that in Step 2, for $i,j\in \mathcal{M}^c$, with high probability $i$ and $j$ are assigned to the same cluster if and only if  $a_{1i}=a_{1j}$ and $\bpi_i=\bpi_j$.   By \eqref{g19}, $\bbQ$ has $2K$ different rows $$\mathrm{R}=\{\bbe_1^{\top}\mathcal{D}\mathcal{Q}_1,\bbe_1^{\top}\mathcal{D}\mathcal{Q}_2,\ldots,\bbe_K^{\top}\mathcal{D}\mathcal{Q}_1,\bbe_K^{\top}\mathcal{D}\mathcal{Q}_2\}\,.$$
	Notice that, for $l = 1, 2$,  respectively,  $\bbe_k$ reflects the membership of individual $i$ if $\bpi_i=\bbe_k$; hence $\{\bbe_1^{\top}\mathcal{D}\mathcal{Q}_l,\ldots,\bbe_K^{\top}\mathcal{D}\mathcal{Q}_l\}\subset \mathrm{R}$ can be regarded as  the ``membership'' vectors for the individuals $\{i\in[n]:a_{1i}=2-l\}$. By \eqref{gtq3}, with high probability, in step $2$ of Algorithm \ref{alg2}, $i\in \mathcal{M}^c\cap \{i\in{n}:a_{1i} = 2-l\}$ is assigned to the cluster associated with  $\bbe_k^{\top}\mathcal{D}\mathcal{Q}_l$ if $\bpi_i=\bbe_k.$
	
	Therefore, with high probability we have
	\begin{equation*}\label{gts21}
		|i\in[n]: \{\text{In step 2,}\ i\  \text{is not assigned to the cluster associated with} \ \bbe_k^{\top}\mathcal{D}\mathcal{Q}_1 \ \text{or}\ \bbe_k^{\top}\mathcal{D}\mathcal{Q}_2\}|\le |\mathcal{M}|\,.
	\end{equation*}

	By Lemma \ref{t3}, under Condition \ref{ncond1},  we have
	\begin{equation}\label{gtq5}
		|\mathcal{M}|=O\left(\frac{1}{ p^2_n}\right)\,.
	\end{equation}
	We say the $1$st group is $\{i\in[n]: a_{1i}=1\}$ and the $2$nd group is $\{i\in[n]: a_{1i}=0\}$.
	Let $\bbS_l(k)$ be the collection of individuals belonging to $k$-th  community in the $l$-th group, and $\widehat\bbS_l(k)$ be the $k$-th cluster of the $l$-th group returned by step $2$. Let $\widehat p^{(1)}_{k_1k_2}$ and $\widehat p^{(2)}_{k_1k_2}$ be the $(k_1,k_2)$-th entry of $\widehat\bbP^{\bbS,\bbS}$ and $\widehat\bbP^{\bbS,\bbI-\bbS}$ (defined by equation \eqref{eqn:prob estimate}) respectively. Let $\bbP^{\bbS,\bbS}=(p_{k_1k_2}^{(1)})$ and $\bbP^{\bbS,\bbI-\bbS}=(p_{k_1k_2}^{(2)})$ be the corresponding population versions. Note that there exists a unique permutation function $f_0$ such that
	\begin{equation}\label{gts22}
		\bbP^{\bbS,\bbS}_{f_0,f_0}=\bbP^{\bbS,\bbI-\bbS}_{f_0,*}\,.
	\end{equation}
	Without loss of generality, we assume $\bbP^{\bbS,\bbI-\bbS}=\bbP$. 

	\begin{align}\label{eqn:decomposition diff}
		\widehat p^{(l)}_{k_1k_2}-p^{(l)}_{k_1k_2}=&\frac{1}{|\widehat\bbS_1(k_1)||\widehat\bbS_l(k_2)|}\Big[\sum_{i\in \bbS_1(k_1)\cap \widehat\bbS_1(k_1), j\in \bbS_l(k_2)\cap \widehat\bbS_l(k_2)}(b_{ij}-p^{(l)}_{k_1k_2})\non
		&+\sum_{i\in \widehat\bbS_1(k_1)\setminus\bbS_1(k_1),\ j\in \bbS_l(k_2)\cap \widehat\bbS_l(k_2)}(b_{ij}-p^{(l)}_{k_1k_2})\non
		&+\sum_{i\in \bbS_1(k_1)\cap \widehat\bbS_1(k_1),\ j\in \widehat\bbS_l(k_2)\setminus\bbS_l(k_2)}(b_{ij}-p^{(l)}_{k_1k_2})\non
		&+\sum_{i\in \widehat\bbS_1(k_1)\setminus\bbS_1(k_1),\ j\in \widehat\bbS_l(k_2)\setminus\bbS_l(k_2)}(b_{ij}-p^{(l)}_{k_1k_2})\Big]\,,
	\end{align}
	and
	\begin{align}\label{eqn:decomposition diffs}
		&\frac{1}{|\widehat\bbS_1(k_1)||\widehat\bbS_l(k_2)|}\sum_{i\in \bbS_1(k_1), j\in \bbS_l(k_2)}(b_{ij}-p^{(l)}_{k_1k_2})\non
&=\frac{1}{|\widehat\bbS_1(k_1)||\widehat\bbS_l(k_2)|}\Big[\sum_{i\in \bbS_1(k_1)\cap \widehat\bbS_1(k_1), j\in \bbS_l(k_2)\cap \widehat\bbS_l(k_2)}(b_{ij}-p^{(l)}_{k_1k_2})\non
		&+\sum_{i\in \bbS_1(k_1)\setminus\widehat\bbS_1(k_1),\ j\in \bbS_l(k_2)\cap \widehat\bbS_l(k_2)}(b_{ij}-p^{(l)}_{k_1k_2})\non
		&+\sum_{i\in \bbS_1(k_1)\cap \widehat\bbS_1(k_1),\ j\in \bbS_l(k_2)\setminus\widehat\bbS_l(k_2)}(b_{ij}-p^{(l)}_{k_1k_2})\non
		&+\sum_{i\in \bbS_1(k_1)\setminus\widehat\bbS_1(k_1),\ j\in \bbS_l(k_2)\setminus\widehat\bbS_l(k_2)}(b_{ij}-p^{(l)}_{k_1k_2})\Big]\,.
	\end{align}

	By Lemma \ref{le1}, with high probability, we have
	\begin{align}\label{t6e3}
		|\bbS_1(k)|=\sum_{i\in[n]: \bpi_i=\bbe_{k}}a_{1i}\sim  \sum_{i\in[n]: \bpi_i=\bbe_{k}}\E a_{1i}\sim np_n, k \in [K]\,,
	\end{align}
	and
	\begin{align}\label{t6e4}
		|\bbS_2(k)|=\sum_{i\in[n]: \bpi_i=\bbe_k}(1-a_{1i})\sim \sum_{i\in[n]: \bpi_i=\bbe_{k}}\E (1-a_{1i})\sim n, k \in [K]\,,
	\end{align}
	Moreover, by  \eqref{gtq5}, with high probability we have
	\begin{equation}\label{t6e5}
		\left||\widehat\bbS_l(k)|-|\bbS_l(k)|\right|\le |\widehat\bbS_l(k)\setminus\bbS_l(k)|+|\bbS_l(k)\setminus\widehat\bbS_l(k)|\le |\mathcal{M}|=O\left(\frac{1}{p_n^2}\right),l=1,2,\ k\in [K]\,.
	\end{equation}

	It follows from \eqref{t6e3}--\eqref{t6e5} that with high probability,
	\begin{equation}\label{gtq6}
		\frac{1}{|\widehat\bbS_1(k_1)||\widehat\bbS_l(k_2)|}=\frac{1}{|\bbS_1(k_1)||\bbS_l(k_2)|}+O\left(\frac{1}{n^3p_n^5}\right)\,.
	\end{equation}

	By the condition $p_n\gg (\frac{1}{n})^{1/2}$ and \eqref{gtq6},  with  probability tending to 1, the first term in \eqref{eqn:decomposition diff} is bounded from above by
	\begin{equation}\label{gtq7}
		\frac{1}{|\widehat\bbS_1(k_1)||\widehat\bbS_l(k_2)|}\sum_{i\in \bbS_1(k_1), j\in \bbS_l(k_2)}(b_{ij}-p^{(l)}_{k_1k_2})=O\left(\frac{1}{np_n}\right)\ll p_n\,.
	\end{equation}
	
	By \eqref{230711.1}, we have
$$\sum_{a_{1i}=1,i\in \bbS_1}1=O(\|\bbS\bbC-\bbS\bbQ\bbO\|_F^2).$$
Therefore, we can see that
\begin{equation}\label{0712.1}
\E\sum_{a_{1i}=1,i\in M}1=O(\frac{1}{p_n}).
\end{equation}
	
	First of all, we have
	\begin{align}
		\label{gtq81}
		&\frac{1}{|\widehat\bbS_1(k_1)||\widehat\bbS_l(k_2)|}\sum_{i\in \widehat\bbS_1(k_1)\setminus\bbS_1(k_1),\ j\in \bbS_l(k_2)}|b_{ij}-p^{(l)}_{k_1k_2}|\non
&=\frac{1}{|\widehat\bbS_1(k_1)||\widehat\bbS_l(k_2)|}\sum_{i\in \widehat\bbS_1(k_1)\setminus\bbS_1(k_1),\ j\in \bbS_l(k_2)}|1-p^{(l)}_{k_1k_2}|I(b_{ij}=1)\non
&+\frac{1}{|\widehat\bbS_1(k_1)||\widehat\bbS_l(k_2)|}\sum_{i\in \widehat\bbS_1(k_1)\setminus\bbS_1(k_1),\ j\in \bbS_l(k_2)}p^{(l)}_{k_1k_2}I(b_{ij}=0).
	\end{align}
For the first term of the right hand side of \eqref{gtq81}, we have
\begin{align}
		\label{gtqhx1}\E\sum_{i\in \widehat\bbS_1(k_1)\setminus\bbS_1(k_1),\ j\in \bbS_l(k_2)}|1-p^{(l)}_{k_1k_2}|I(b_{ij}=1)\le \sum_{a_{1i}=1\in \mathcal {M},\ j\in \bbS_l(k_2)}(\E I(b_{ij}=1))=O(|S_l(k_2)|).
\end{align}
Moreover, considering the second term of the right hand side of \eqref{gtq81}, by \eqref{0712.1} we have
\begin{align}
		\label{gtqhx2}\E\sum_{i\in \widehat\bbS_1(k_1)\setminus\bbS_1(k_1),\ j\in \bbS_l(k_2)}p^{(l)}_{k_1k_2}I(b_{ij}=0)&=O(p_n \sum_{a_{1i}=1, i \in \mathcal{M},\ j\in \bbS_l(k_2)}(\E I(b_{ij}=0)))\non
&=O(|S_l(k_2)|).
\end{align}
Therefore, we imply that
\begin{align}
		\label{gtq82}
		&\frac{1}{|\widehat\bbS_1(k_1)||\widehat\bbS_l(k_2)|}\sum_{i\in \widehat\bbS_1(k_1)\setminus\bbS_1(k_1),\ j\in \bbS_l(k_2)}|b_{ij}-p^{(l)}_{k_1k_2}|=O_p\left(\frac{1}{np_n}\right)\ll p_n\,.\end{align}
	Similar to \eqref{gtq82}, we have
	\begin{equation}\label{gtq9}
		\frac{1}{|\widehat\bbS_1(k_1)||\widehat\bbS_l(k_2)|}\sum_{i\in \bbS_1(k_1),\ j\in \widehat\bbS_l(k_2)\setminus\bbS_l(k_2)}|b_{ij}-p_{k_1k_2}|=O_p\left(\frac{1}{np_n}\right)\ll p_n\,,
	\end{equation}
	
	\begin{equation}\label{gtq10}
		\frac{1}{|\widehat\bbS_1(k_1)||\widehat\bbS_l(k_2)|}\sum_{i\in \widehat\bbS_1(k_1)\setminus\bbS_1(k_1),\ j\in \widehat\bbS_l(k_2)\setminus\bbS_l(k_2)}|b_{ij}-p_{k_1k_2}|=O_p\left(\frac{1}{np_n}\right)\ll p_n\,.
	\end{equation}
	and
	\begin{align}\label{gtq10s}
		&\frac{1}{|\widehat\bbS_1(k_1)||\widehat\bbS_l(k_2)|}\Big[\sum_{i\in \bbS_1(k_1)\setminus\widehat\bbS_1(k_1),\ j\in \bbS_l(k_2)\cap \widehat\bbS_l(k_2)}|b_{ij}-p^{(l)}_{k_1k_2}|\non
		&+\sum_{i\in \bbS_1(k_1)\cap \widehat\bbS_1(k_1),\ j\in \bbS_l(k_2)\setminus\widehat\bbS_l(k_2)}|b_{ij}-p^{(l)}_{k_1k_2}|+\sum_{i\in \bbS_1(k_1)\setminus\widehat\bbS_1(k_1),\ j\in \bbS_l(k_2)\setminus\widehat\bbS_l(k_2)}|b_{ij}-p^{(l)}_{k_1k_2}|\Big]\non
&=O_p\left(\frac{1}{np_n}\right)\ll p_n\,.
	\end{align}
	By \eqref{eqn:decomposition diffs} and \eqref{gtq10s}, we have
	\begin{align}\label{gtq10ss}
		&\frac{1}{|\widehat\bbS_1(k_1)||\widehat\bbS_l(k_2)|}\left|\sum_{i\in \bbS_1(k_1)\cap \widehat\bbS_1(k_1), j\in \bbS_l(k_2)\cap \widehat\bbS_l(k_2)}(b_{ij}-p^{(l)}_{k_1k_2})\right|\ll p_n\,.
	\end{align}

	Therefore, by \eqref{eqn:decomposition diff},  \eqref{gtq7}--\eqref{gtq10} and \eqref{gtq10ss}, we have $\max_{i,j\in [K]}|\widehat p^{(l)}_{k_1k_2}-p^{(l)}_{k_1k_2}|\ll p_n$. In view of this result,  if $\widehat f_0=f_0$, it follows from \eqref{gts22} that
	$$\|\widehat\bbP^{\bbS,\bbS}_{(f_0,f_0)}-\widehat\bbP_{(f_0,*)}^{\bbS,\bbI-\bbS}\|_F\ll p_n\,.$$
	Otherwise if $\widehat f_0\neq f_0$, by Condition \ref{cond6} and Proposition \ref{t4}, we have
	$$\|\widehat\bbP^{\bbS,\bbS}_{(\widehat f_0,\widehat f_0)}-\widehat\bbP_{(\widehat f_0,*)}^{\bbS,\bbI-\bbS}\|_F\sim p_n\,.$$

	Recall that $\widehat f_0$ is defined by $\widehat f_0=\arg\min_f\|\widehat\bbP^{\bbS,\bbS}_{(f,f)}-\widehat\bbP_{(f,*)}^{\bbS,\bbI-\bbS}\|_F$. Hence,  with probability tending to 1, we have $\widehat f_0=f_0$. Therefore, by Algorithm \ref{alg2} with  probability tending to 1, the set  $\mathrm{R}$ can be merged into
	$$
	\widetilde{\mathrm{R}}=\left\{\{\bbe_1^{\top}\mathcal{D}\mathcal{Q}_1,\bbe_1^{\top}\mathcal{D}\mathcal{Q}_2\},\ldots,\{\bbe_K^{\top}\mathcal{D}\mathcal{Q}_1,\bbe_K^{\top}\mathcal{D}\mathcal{Q}_2\}\right\}\,.
	$$
	
	Notice that the $2K$ clusters are merged into $K$ communities According to $\widetilde{\mathrm{R}}$.  Then it follows from \eqref{gtq3} that with  probability tending to 1, for $i,j\in \mathcal{M}^c$,
	\begin{equation*}
		\text{Individuals } i \ \text{and} \ j\ \text{are assigned to the same community}\Leftrightarrow\bpi_i = \bpi_j\,.
	\end{equation*}

	Furthermore, it holds with  probability tending to 1 that
	\begin{align*}\label{gts20}
		&1-\text{Misclustering rate}\ge \frac{|\{i\in\mathcal{M}^c: i \ \text{is assigned to group $k$ such that}\ \bpi_i=\bbe_k\}|}{n}\non
		&=\frac{|\mathcal{M}^c|}{n}=1-\frac{|\mathcal{M}|}{n}\,,
	\end{align*}
	in which the first equality follows from Lemma \ref{t5}.  By Lemma \ref{t3}, we have with  probability tending to 1, $|\mathcal{M}| = O(1 / p_n^2)$. Then $p_n \gg (1 / n)^{1/2}$ implies that $|\mathcal{M}|/ n = o(1)$.

\end{proof}

\subsection{Proof of Proposition \ref{prop1}}
Actually, \eqref{sg18} follows from Lemma 4.1 of \cite{JA15} directly. Therefore we only need to prove the first statement of Proposition \ref{prop1}.  Conditions \ref{ncond1} and \ref{cond10}, together with Lemma \ref{le1} imply that
\begin{equation}\label{s0125.1}
	n\bbI\lesssim \bPi^\top\E\Theta^2\bPi\lesssim n\bbI, \ \|\bPi^\top\left(\Theta^2-\E\Theta^2\right)\bPi\|=o(n)\,,
\end{equation}
holds with high probability. By \eqref{s0125.1}, the remaining proof of Proposition \ref{prop1} is almost the same as the proof of Lemma \ref{le4} and thus we omit it.

\subsection{Proof of Theorem \ref{consist} }

Similar to the way we prove Theorem~\ref{t5},  we first need some auxilliary results regarding the theoretical properties of the clustering step. The next lemma is analogous to Lemma~\ref{le3}.

\begin{lem}\label{sle3}
	Under Conditions   \ref{ncond1} and \ref{cond10}, with high probability, there exists some positive constant $c_2$ such that
	\begin{equation}\label{sg24}
		\mathcal{Q}\mathcal{Q}^\top \ge (c_2 n)^{-1}\bbI\,,\quad \mathcal{D}\mathcal{Q}_2\mathcal{Q}_2^\top \mathcal{D}^\top\ge (c_2 n)^{-1}\bbI\,,\quad \text{and} \quad p_n \mathcal{D}\mathcal{Q}_1\mathcal{Q}_1^\top\mathcal{D}^\top \ge (c_2 n)^{-1}\bbI\,.
	\end{equation}
\end{lem}
\begin{proof}
	By checking the proof of Lemma \ref{le3} carefully, we can see that the order of the matrices in \eqref{sg24} mainly depends on the order of $\bbV^\top\bbS\bbV$. Actually, by almost the same proof as Lemma \ref{le2}, the conclusion of Lemma \ref{le2} holds by Proposition \ref{prop1}  and  Condition \ref{cond10}. Therefore our proof is finished.
\end{proof}

Let $$\widehat\bbE_{\bbS}=\min_{\bbE\in\{\text{$K\times K$ permutation matrices}\}}\|\bbS\widehat\bPi\bbE-\bbS\bPi\|_0,$$ $$\widehat\bbE_{\bbI-\bbS}=\min_{\bbE\in\{\text{$K\times K$ permutation matrices}\}} \|(\bbI-\bbS)\widehat\bPi\bbE-(\bbI-\bbS)\bPi\|_0$$  and $\mathrm{M}=\{i, a_{1i}=1, \widehat\bPi(i)\widehat\bbE_{\bbS}\neq\bPi(i)\}\cup \{i, a_{1i}=0,\widehat\bPi(i)\widehat\bbE_{\bbI-\bbS}\neq\bPi(i)\}$.
Similar to Theorem 4.2 of \cite{JA15} we have the following result.
\begin{lem}\label{gdcsbm}
	Under Conditions   \ref{ncond1} and \ref{cond10}, if $p_n\gg \frac{1}{\sqrt n f_s}$, with  probability tending to 1  we have
	$$|\mathrm{M}|\lesssim \frac{\sqrt n}{f_sp_n}\,.$$
\end{lem}
\begin{proof}
	Let $\bbQ'$ is the row normalized version of $\bbQ$ and $\bbQ'_{\mathcal{I}_1}$ be the sub matrix of $\bbQ'$ corresponding to the non zero rows of $\bbW$. Then by the inequality in front of the proof of Theorem 4.2 in \cite{JA15}, with high probability we have
	\begin{align}\label{s0130.1}
		&\|\bbW' -\bbQ'_{\mathcal{I}_1}\bbO\|_{2,1}\le 2\sum_{i=1}^n\frac{\|\bbW(i) -\bbQ_{\mathcal{I}_1}(i)\bbO\|}{\|\bbQ_{\mathcal{I}_1}(i)\|}\non
		&\le 2\sqrt{\sum_{i=1}^n\|\bbW(i) -\bbQ_{\mathcal{I}_1}(i)\bbO\|^2\sum_{i=1}^n\|\bbQ_{\mathcal{I}_1}(i)\|^{-2}}\non
		&\lesssim nf_s^{-1}\|\bbW-\bbQ\bbO\|_F\,.
	\end{align}
	Similar to the proof of Lemma \ref{t3}, the misclustered nodes can be bounded by considering $\bbS\bbW$ and $(\bbI-\bbS)\bbW$ separately. The rest of the proof is essentially the same as the proof of Theorem 4.2 in \cite{JA15}; the key points are \eqref{s0130.1}, Lemma \ref{sle3} and Lemma \ref{le5} in this setting. Therefore we omit the proof.
\end{proof}

The next result justifies our merging strategy in Algorithm \ref{alg3}.

\begin{prop}\label{t8}
	Let $\mathfrak{C}=\{\mathfrak{c}_1,\ldots,\mathfrak{c}_K\}$ and $\mathfrak{D}=\{\mathfrak{d}_1,\ldots,\mathfrak{d}_K\}$ be two ordered sets of size $K$ and let $f_0:[K]\rightarrow [K]$ be the unique permutation such that $\mathfrak{c}_{f_0(i)}=\mathfrak{d}_i$ for all $i\in[K]$.  Suppose that the $K$ elements in these ordered sets  have been endowed with pairwise connection probabilities. Let $\bbC=(c_{ij})$ be a $K\times K$ matrix such that the $(i,j)$-th entry $c_{ij}$ is the connection probability between $\mathfrak{c}_i$ and  $\mathfrak{c}_j$, and $\widetilde\bbD=(d_{ij}/e_i)$ be a $K\times K$ matrix such that the $(i,j)$-th entry $d_{ij}$ is the connection probability between  $\mathfrak{c}_i$ and  $\mathfrak{d}_j$. Let $\bbX=diag(x_1,\ldots,x_K)$ be a $K\times K$ diagonal matrix, where the entries of $\bbX$ are bounded below and above.
	If $\emph{rank}(\bbC)=K$, then for a permutation function $f:[K]\rightarrow [K]$,
	\begin{equation*}\label{g29}
		f=f_0, x_i=e_{f(i)}, i=1,\ldots,K\Longleftrightarrow \bbC_{(f,f)}=\widetilde\bbD_{(f,*)}\bbX\,,
	\end{equation*}
	where the $(i,j)$-th entries of  $\bbC_{(f,f)}$ and $\widetilde\bbD_{(f,*)}$ are  $c_{f(i)f(j)}$ and $d_{f(i)j}/e_{f(i)}$, respectively.
\end{prop}
\begin{proof}
	By similar argument as the proof of Proposition \ref{t4}, we can also imply the columns (rows) of $\bbD$ $(\widetilde \bbD)$ are distinct. Therefore, by the definition of $\bbD$, if
	$$\bbC_{(f,f)}\neq \bbD_{(f,*)}.$$
	It is not hard to see that there does not exist a diagonal matrix $\bbX$ such that
	$$\bbC_{(f,f)}=\widetilde\bbD_{(f,*)}\bbX\,.$$
	Hence, our proof is reduced to the proof of Proposition \ref{t4} and thus we omit the remaining part.
\end{proof}

Now to prove Theorem \ref{consist}, it suffices for us to prove the following result.
\begin{thm}\label{s4}
	Under Conditions \ref{ncond1}--\ref{cond10} and $p_n\gg \frac{1}{\sqrt nf_s}$, it holds with  probability tending to 1, uniformly for $i,j\in \mathrm{M}^c$, that
	\begin{equation*}
		\text{individuals } i \ \text{and} \ j\ \text{are assigned to the same community by Algorithm \ref{alg3}} \Leftrightarrow \bpi_i=\bpi_j\,.
	\end{equation*}
\end{thm}

\begin{proof}

	By Lemma \ref{gdcsbm},  we have
	\begin{equation}\label{sgtq5}
		|\mathrm{M}|=O\left(\frac{\sqrt n}{ f_sp_n}\right)\,.
	\end{equation}
	Similar to the proof of Theorem \ref{t5}, we say the $1$st group is $\{i\in[n]: a_{1i}=1\}$ and the $2$nd group is $\{i\in[n]: a_{1i}=0\}$.
	Let $\bbS_l(k)$ be the collection of individuals belonging to $k$-th  community in the $l$-th group, and $\widehat\bbS_l(k)$ be the $k$-th cluster of the $l$-th group returned by step $2$. Let $\widehat p^{(1)}_{k_1k_2}$ and $\widehat p^{(2)}_{k_1k_2}$ be the $(k_1,k_2)$-th entry of $\widehat\bbP^{\bbS,\bbS}$ and $\widehat\bbP^{\bbS,\bbI-\bbS}$ (defined by equation \eqref{eqn:prob estimate}) respectively. Let $\bbP^{\bbS,\bbS}=(p_{k_1k_2}^{(1)})$ and $\bbP^{\bbS,\bbI-\bbS}=(p_{k_1k_2}^{(2)})$ be the corresponding population versions. Note that there exists an unique permutation function $f_0$ and the corresponding diagonal matrix $g$ such that
	\begin{equation}\label{sgts22}
		\bbP^{\bbS,\bbS}_{f_0,f_0}=\bbP^{\bbS,\bbI-\bbS}_{f_0,*}g(\bbP^{\bbS,\bbS}_{(f_0,f_0)},\bbP_{(f_0,*)}^{\bbS,\bbI-\bbS})\,.
	\end{equation}
	Without loss of generality, we assume $\bbP^{\bbS,\bbI-\bbS}=\bbP$. 
	\begin{align}\label{seqn:decomposition diff}
		\widehat p^{(l)}_{k_1k_2}-p^{(l)}_{k_1k_2}=&\frac{1}{|\widehat\bbS_1(k_1)||\widehat\bbS_l(k_2)|}\Big[\sum_{i\in \bbS_1(k_1)\cap \widehat\bbS_1(k_1), j\in \bbS_l(k_2)\cap \widehat\bbS_l(k_2)}(b_{ij}-p^{(l)}_{k_1k_2})\non
		&+\sum_{i\in \widehat\bbS_1(k_1)\setminus\bbS_1(k_1),\ j\in \bbS_l(k_2)\cap \widehat\bbS_l(k_2)}(b_{ij}-p^{(l)}_{k_1k_2})\non
		&+\sum_{i\in \bbS_1(k_1)\cap \widehat\bbS_1(k_1),\ j\in \widehat\bbS_l(k_2)\setminus\bbS_l(k_2)}(b_{ij}-p^{(l)}_{k_1k_2})\non
		&+\sum_{i\in \widehat\bbS_1(k_1)\setminus\bbS_1(k_1),\ j\in \widehat\bbS_l(k_2)\setminus\bbS_l(k_2)}(b_{ij}-p^{(l)}_{k_1k_2})\Big]\,,
	\end{align}
	and
	\begin{align}\label{seqn:decomposition diffs}
		&\frac{1}{|\widehat\bbS_1(k_1)||\widehat\bbS_l(k_2)|}\sum_{i\in \bbS_1(k_1), j\in \bbS_l(k_2)}(b_{ij}-p^{(l)}_{k_1k_2})\non
&=\frac{1}{|\widehat\bbS_1(k_1)||\widehat\bbS_l(k_2)|}\Big[\sum_{i\in \bbS_1(k_1)\cap \widehat\bbS_1(k_1), j\in \bbS_l(k_2)\cap \widehat\bbS_l(k_2)}(b_{ij}-p^{(l)}_{k_1k_2})\non
		&+\sum_{i\in \bbS_1(k_1)\setminus\widehat\bbS_1(k_1),\ j\in \bbS_l(k_2)\cap \widehat\bbS_l(k_2)}(b_{ij}-p^{(l)}_{k_1k_2})\non
		&+\sum_{i\in \bbS_1(k_1)\cap \widehat\bbS_1(k_1),\ j\in \bbS_l(k_2)\setminus\widehat\bbS_l(k_2)}(b_{ij}-p^{(l)}_{k_1k_2})\non
		&+\sum_{i\in \bbS_1(k_1)\setminus\widehat\bbS_1(k_1),\ j\in \bbS_l(k_2)\setminus\widehat\bbS_l(k_2)}(b_{ij}-p^{(l)}_{k_1k_2})\Big]\,.
	\end{align}

	By Lemma \ref{le1}, with high probability, we have
	\begin{align}\label{st6e3}
		|\bbS_1(k)|=\sum_{i\in[n]: \bpi_i=\bbe_{k}}a_{1i}\sim  \sum_{i\in[n]: \bpi_i=\bbe_{k}}\E a_{1i}\sim np_n, k \in [K]\,,
	\end{align}
	and
	\begin{align}\label{st6e4}
		|\bbS_2(k)|=\sum_{i\in[n]: \bpi_i=\bbe_k}(1-a_{1i})\sim \sum_{i\in[n]: \bpi_i=\bbe_{k}}\E (1-a_{1i})\sim n, k \in [K]\,,
	\end{align}
	Moreover, by  \eqref{sgtq5}, with high probability we have
	\begin{equation}\label{st6e5}
		\left||\widehat\bbS_l(k)|-|\bbS_l(k)|\right|\le |\widehat\bbS_l(k)\setminus\bbS_l(k)|+|\bbS_l(k)\setminus\widehat\bbS_l(k)|\le |M|=O\left(\frac{\sqrt n}{p_nf_s}\right),l=1,2,\ k\in [K]\,.
	\end{equation}

	It follows from \eqref{st6e3}--\eqref{st6e5} that with high probability,
	\begin{equation}\label{sgtq6}
		\frac{1}{|\widehat\bbS_1(k_1)||\widehat\bbS_l(k_2)|}=\frac{1}{|\bbS_1(k_1)||\bbS_l(k_2)|}+O\left(\frac{\sqrt n}{f_sn^3p_n^4}\right)\,.
	\end{equation}

	By $p_n\gg 1/(\sqrt n f_s)$ and \eqref{sgtq6}  with  probability tending to 1, we have
	\begin{equation}\label{sgtq7}
		\frac{1}{|\widehat\bbS_1(k_1)||\widehat\bbS_l(k_2)|}\sum_{i\in \bbS_1(k_1), j\in \bbS_l(k_2)}(b_{ij}-p^{(l)}_{k_1k_2})=O\left(\frac{1}{np_n}\right)\ll p_n\,.
	\end{equation}

	Similar to \eqref{gtqhx1}--\eqref{gtqhx2}, By \eqref{sgtq5}, the condition that $p_n\gg 1/(\sqrt n f_s)$, we have
	\begin{align}
		\label{gtq8}
		&\frac{1}{|\widehat\bbS_1(k_1)||\widehat\bbS_l(k_2)|}\sum_{i\in \widehat\bbS_1(k_1)\setminus\bbS_1(k_1),\ j\in \bbS_l(k_2)}|b_{ij}-p^{(l)}_{k_1k_2}|\non
		&=O_p\left(\frac{1}{\sqrt{n} f_s}\right)\ll p_n\,.
	\end{align}
	Similar to \eqref{gtq8}, with  probability tending to 1 we have
	\begin{equation}\label{sgtq9}
		\frac{1}{|\widehat\bbS_1(k_1)||\widehat\bbS_l(k_2)|}\sum_{i\in \bbS_1(k_1),\ j\in \widehat\bbS_l(k_2)\setminus\bbS_l(k_2)}|b_{ij}-p_{k_1k_2}|=O\left(\frac{1}{\sqrt{n} f_s}\right)\ll p_n\,,
	\end{equation}
	
	\begin{equation}\label{sgtq10}
		\frac{1}{|\widehat\bbS_1(k_1)||\widehat\bbS_l(k_2)|}\sum_{i\in \widehat\bbS_1(k_1)\setminus\bbS_1(k_1),\ j\in \widehat\bbS_l(k_2)\setminus\bbS_l(k_2)}|b_{ij}-p_{k_1k_2}|=O\left(\frac{1}{\sqrt{n} f_s}\right)\ll p_n\,.
	\end{equation}
	and
	\begin{align}\label{sgtq10s}
		&\frac{1}{|\widehat\bbS_1(k_1)||\widehat\bbS_l(k_2)|}\Big[\sum_{i\in \bbS_1(k_1)\setminus\widehat\bbS_1(k_1),\ j\in \bbS_l(k_2)\cap \widehat\bbS_l(k_2)}|b_{ij}-p^{(l)}_{k_1k_2}|\non
		&+\sum_{i\in \bbS_1(k_1)\cap \widehat\bbS_1(k_1),\ j\in \bbS_l(k_2)\setminus\widehat\bbS_l(k_2)}|b_{ij}-p^{(l)}_{k_1k_2}|+\sum_{i\in \bbS_1(k_1)\setminus\widehat\bbS_1(k_1),\ j\in \bbS_l(k_2)\setminus\widehat\bbS_l(k_2)}|b_{ij}-p^{(l)}_{k_1k_2}|\Big]\non
&=O\left(\frac{1}{\sqrt{n} f_s}\right)\ll p_n\,.
	\end{align}
	By \eqref{seqn:decomposition diffs} and \eqref{sgtq10s}, we have
	\begin{align}\label{sgtq10ss}
		&\frac{1}{|\widehat\bbS_1(k_1)||\widehat\bbS_l(k_2)|}\left|\sum_{i\in \bbS_1(k_1)\cap \widehat\bbS_1(k_1), j\in \bbS_l(k_2)\cap \widehat\bbS_l(k_2)}(b_{ij}-p^{(l)}_{k_1k_2})\right|\ll p_n\,.
	\end{align}
	Considering the estimation of $g$, by Lemma \ref{le1}, with  probability tending to 1 we have
	\begin{align}\label{sgtq11ss}
		\|g(\widehat\bbP^{\bbS,\bbS}_{(f,f)},\widehat\bbP_{(f,*)}^{\bbS,\bbI-\bbS})-g(\bbP^{\bbS,\bbS}_{(f,f)},\bbP_{(f,*)}^{\bbS,\bbI-\bbS})\|=O(\frac{\sqrt{\log n}}{np_n})\,.
	\end{align}

	Therefore, by almost the same arguments as the proof of Theorem \ref{t5}, our conclusion holds.
\end{proof}

\subsection{Proof of Lemma \ref{le10}}

By Lemma 4.1 of \cite{JA15} and Condition \ref{cond9}, $\bbV(i)\sim \frac{\theta_i}{\sqrt n}\bbH_k$, $i$ belongs to community $k$, where $\{\bbH_k,k=1,\ldots,K\}$ are orthonormal vectors. Therefore, we imply that
$$\bbV^T\bbS\bbV=\sum_{\bbS_{ii}=1}\bbV(i)\bbV^T(i)\sim \sum_{\bbS_{ii}=1, \text{$i$ belongs to community $k$}}\frac{\theta_i^2}{n}\bbH_k\bbH^\top_k.$$
Combining this with the assumption that there are  $n_{1k}$ $\theta_i$'s generated from $F_1$ and $n_{2k}$ $\theta_i$'s from $F_2$, it is easy to see that our conclusion holds.

\subsection{Proof of of Theorem \ref{consist_fixedS}}

To prove Theorem \ref{consist_fixedS}, we first need to show results analogous to the unconditional case hold for the conditonal setting. First note that  by Lemma \ref{le1}, we have that
\begin{equation}\label{simp}
	\max_{k\in[K]}(n_{1k}+n_{2k})=O(np_n)\,,
\end{equation}
with high probability. Using this result and by almost the same proof as the first part of Theorem \ref{thm1} and {\color{black} the concentration inequality \eqref{simp} , the non-zero eigenvalues of $\bbB_{E}$ satisfy the following.
	\begin{thm}\label{nthm1}
		Under Conditions \ref{cond9} and \ref{ncond2}, for a given $\bbS$, we have $$|x_{i}|^{-1}\sim np_n\min_{k\in[K]}\sqrt{\frac{n_{2k}\mu_2^2+n_{1k}\mu_1^2}{n}}$$ for $i\in[\pm K]$.
	\end{thm}

	By Theorem \ref{nthm1}, 
	we can conclude that Lemma \ref{errorbound} holds for a given $\bbS$ under Conditions \ref{cond9} and \ref{ncond2}, in the following sense.

\begin{lem}\label{errorbound1}
	Under Conditions  \ref{cond9} and \ref{ncond2}.  For fixed $\bbS$, it holds with high probability that
	$$\|\bbB-\bbB_E\|=O(\sqrt{np_n})\,.$$
	Moreover,  with high probability   we have
	$\|\bbB-\bbB_E\|\ll \min_{i \in [2K]}\sigma_i(\bbB_E)\,.$
\end{lem}
\begin{proof}
	The first statement of Lemma \ref{errorbound1} follows the same argument as the proof of Lemma \ref{errorbound} by noticing that $\|\bbS\|\le 1$ for fixed $\bbS$. The second statement follows from Theorem \ref{nthm1} directly and thus we omit the proof.
\end{proof}

Furthermore, Theorem \ref{nthm1} and almost the same arguments as in Lemma \ref{le5} give another analogous result below (and we omit the proof).
\begin{lem}\label{s0209-1}
	Under Conditions \ref{cond9} and \ref{ncond2}, conditioned on $\bbS$, w.h.p. we have
	\begin{equation}\label{sle5}
		\|\bbW-\bbQ\bbO\|_F=O\left(\frac{1}{\sqrt{\min_{k\in[K]} (n_{2k}\mu_2^2+n_{1k}\mu_1^2) p_n}}\right)\,,
	\end{equation}
	where $\bbO$ is the same as defined in Lemma \ref{le5}.
\end{lem}

The next lemma is analogous to Lemma \ref{sle3}. We omit the proof as it is almost identical.

\begin{lem}\label{sle3_fixedS}
	Under Conditions   \ref{cond9} and \ref{ncond2}, for fixed $\bbS$, with high probability, there exists some positive constant $c_2$ such that
	\begin{equation}\label{sg25}
		\mathcal{D}\mathcal{Q}_2\mathcal{Q}_2^\top \mathcal{D}^\top\ge (c_2 n)^{-1}\bbI\,,\quad \text{and} \quad \frac{\min_k n_{2k}}{n} \mathcal{D}\mathcal{Q}_1\mathcal{Q}_1^\top\mathcal{D}^\top \ge (c_2 n)^{-1}\bbI\,.
	\end{equation}
\end{lem}

Then corresponding to Lemma \ref{gdcsbm}, we also have the following result, whose proof we again omit due to its similarity.

\begin{lem}\label{gdcsbm_fixedS}
	Under Conditions \ref{cond9} and \ref{ncond2}, if $p_n\gg \frac{1}{\min_{k\in [K]}(n_{2k}\mu_2^2+n_{1k}\mu_1^2)\mu_1^2}$,  then conditioned on $\bbS$ and $\Theta$, with high probability   we have
	$$\left|\mathrm{M}\right|\lesssim \mu_1^{-1}n\sqrt{\frac{1}{p_n\min_{k\in[K]}(n_{2k}\mu_2^2+n_{1k}\mu_1^2)}}\,.$$
\end{lem}

Now to prove Theorem \ref{consist_fixedS}, it suffices to prove the following, the proof of which is omitted due to its similarity to Theorem \ref{s4}.

\begin{thm}\label{s6}
	Under Conditions \ref{cond6}, \ref{cond9} and \ref{ncond2}, the same conclusion as Theorem \ref{s4} holds, conditioned on $\bbS$ and $\Theta$, if
	$$\mu_1^{-2}\frac{1}{\min_{k\in[K]}(n_{2k}\mu_2^2+n_{1k}\mu_1^2)}\ll p_n\,.$$
\end{thm}

\subsection{Concentration inequalities}

The following two concentration inequalities are used throughout the paper.

\begin{lem}[Bernstein inequality]\label{le1}
	Suppose that $\{y_i\}_{i=1}^n$ are independent bernoulli random variables, then for any non-random series $\{a_i\}_{i=1}^n$ such that $|a_i|\le L$ for some positive constant $L$, we have
	\begin{equation}\label{g32}
		\p\left(\left|\sum_{i=1}^na_i(y_i-\E y_i)\right|\ge t\right)\le \exp\left(-\frac{t^2/2}{\sum_{i=1}^n(a_i^2\E(y_i-\E y_i)^2)+\frac{Lt}{3}}\right),\quad  t>0\,.
	\end{equation}
\end{lem}

\begin{lem}[Matrix Bernstein inequality, Theorem 6.2 of \cite{T12}]\label{Tropp}
	Consider a finite sequence $\left\{\boldsymbol{X}_{k}\right\}$ of independent, random, self-adjoint $d\times d$ matrices. Assume that
	\[
	\E \boldsymbol{X}_{k}=\mathbf{0} \quad \text { and } \quad \E\left(\boldsymbol{X}_{k}^{p}\right) \leq \frac{p !}{2} \cdot R^{p-2} \boldsymbol{A}_{k}^{2} \quad \text { for } p=2,3,4, \ldots
	\]
	Compute the variance parameter
	\[
	\sigma^{2}:=\left\|\sum_{k} \boldsymbol{A}_{k}^{2}\right\|\,.
	\]
	Then the following chain of inequalities holds for all $t \geq 0$
	\[
	\begin{aligned}
		\p\left\{\lambda_{\max }\left(\sum_{k} \boldsymbol{X}_{k}\right) \geq t\right\} & \leq d \cdot \exp \left(\frac{-t^{2} / 2}{\sigma^{2}+R t}\right)
		\leq\left\{\begin{array}{ll}
			d \cdot \exp \left(-t^{2} / 4 \sigma^{2}\right) & \text { for } t \leq \sigma^{2} / R\,, \\
			d \cdot \exp (-t / 4 R) & \text { for } t \geq \sigma^{2} / R\,.
		\end{array}\right.
	\end{aligned}
	\]
\end{lem}

\section{Extensions of theoretical results}
\subsection{Exact Recovery}
\label{subsec:exact_recov}
We prove the exact recovery result for stochastic block model under stronger condition by applying similar approach as \cite{su2019strong}. By checking the proof of Theorem 2.3 in \cite{su2019strong}, the crucial step is to prove the almost sure convergence for the entries of spiked eigenvectors.  By \cite{wu2021} and the Borel-Cantelli Lemma, noticing that the eigenvector bound $O_p(x_K^2\sqrt{np_n})$ is calculated by second moment of the small order terms, we have the following Lemma.
\begin{cond}\label{conder}
 Condition \ref{ncond1} holds. Moreover, we assume that $p_n\ge (n/\log n)^{-2/7}$ and $|x_i/x_{j}-1|\ge c$ for some positive constant $c$.
\end{cond}
\begin{lem}\label{ler}
  Under Condition \ref{conder}, we have the following expansion.
  \begin{equation}\label{er1}
    \bbe_i^\top\bbw_k=\bbe_i^\top\bbq_k+o_{a.s.}(\frac{1}{n^{1/2}}),
  \end{equation}
  uniformly for all $1\le i\le n$.
\end{lem}
\begin{proof}
  The proof of Lemma \ref{ler} is essentially a modification of Theorem 2.2.1 of \cite{wu2021}. Noticing that the proof of Theorem 2.2.1 in \cite{wu2021} essentially rely on the upper bound of  $Var(\bbx^\top(\bbB-\bbB_E)^l\bby)$, where $\bbx$ and $\bby$ are unit vectors(maybe depending on $\bbS$). Following almost the same steps as Theorem 2.2.1 of \cite{wu2021}, we calculate the fourth moment of $\bbe_i^\top(\bbB-\bbB_E)^2-\E(\bbB-\bbB_E)^2\bbq_k$ and have the following result
  $$\bbe_i^\top\left((\bbB-\bbB_E)^2-\E(\bbB-\bbB_E)^2\right)\bbq_k=O_{L_4}(\sqrt{np_n}).$$
  Therefore, we have an improved version of Theorem 2.2.1 in \cite{wu2021} such that
    \begin{equation}\label{er2}
    \bbe_i^\top\bbw_k=\bbe_i^\top\bbq_k+x_i\bbe_i^\top(\bbB-\bbB_E)\bbq_k+O_{L_4}(|x_k|^2\sqrt{np_n})+O_{L_2}(|x_k|^3np_n)+O_{a.s.}(\frac{1}{n}),
  \end{equation}
  where $X_n=O_{L_{2k}}(s_n)$ means that $\E X_n^{2k}=O(s_n^{2k})$, $k\in \mathbb{N}$, the last term $O_{a.s}(\frac{1}{n})$ holds  uniformly for all $1\le i\le n$.
  By Theorem \ref{thm1}, we have $|x_i|^{-1}\sim np_n^{3/2}$. By the condition that $p_n\ge (n/\log n)^{-2/7}$,  we have $|x_i|^2\sqrt{np_n}=o(n^{-1/2-1/4-1/36})$ and $|x_i|^3np_n=O(n^{-1/2-1/2}\log^{-1}n)$. Combining this with the Borel-Cantelli Lemma, we imply that \eqref{er2} holds uniformly for $i$ and $k$, such that
      \begin{equation}\label{er3}
    \bbe_i^\top\bbw_k=\bbe_i^\top\bbq_k+x_k\bbe_i^\top(\bbB-\bbB_E)\bbq_k+o_{a.s.}(\frac{1}{\sqrt n}).
  \end{equation}
  Recalling that $\bbq_k=\bbS\bbV\bbq_{1k}+(\bbI-\bbS)\bbV\bbq_{2k}$, by Corollary 2.6.1 of \cite{wu2021} we have
  $$\bbq_{1k}=O_{a.s.}(1/\sqrt{p_n}),\ \bbq_{2k}=O_{a.s.}(\sqrt{p_n}).$$
  It follows from Lemma \ref{le1} that
  $$\p(\bbe_i^\top(\bbB-\bbB_E)\bbS\bbV\bbq_{1k}>4\sqrt{p_n} \log n)\le \frac{1}{n^4}.$$
  Similarly we have
  $$\p(\bbe_i^\top(\bbB-\bbB_E)(\bbI-\bbS)\bbV\bbq_{2k}>4\sqrt{p_n} \log n)\le \frac{1}{n^4}.$$
  Therefore we imply that $x_k\bbe_i^\top(\bbB-\bbB_E)\bbq_k=o_{a.s.}(1/\sqrt{n})$   uniformly for all $1\le i\le n$.
\end{proof}
According to Lemma \ref{ler}, the deviation $(\bbe_i^\top\bbw_k-\bbe_i^\top\bbq_k)$ can be bounded by controlling the order of $x_k\bbe_i^\top(\bbB-\bbB_E)\bbq_k$. In fact, given $\bbS$, $\bbq_k$ is a non-random vector, and it is not hard to see that $x_k\bbe_i^\top(\bbB-\bbB_E)\bbq_k=o_{a.s.}(\frac{1}{n^{1/2}})$ by Lemma \ref{le1}. Combining this with Lemma \ref{le6} in the main paper, we immediately obtain the following result bounding the distances between different types of rows in $\bbQ$.

\begin{lem}\label{ler2}
  Under Condition \ref{conder}, we imply the following results:
  \begin{itemize}
    \item There exists a positive constant $C$ such that $\sqrt n|\bbe_i^\top\bbq_k|\le C$.
    \item There exists  deterministic sequences $c_{2n}=o(1)$ and $c_{1n}\sim 1$ such that $\sqrt n|\bbe_i^\top(\bbw_k-\bbq_k)|\le c_{2n}$, a.s. $\sqrt n\|\bbe_i^\top\bbQ-\bbe_j\bbQ\|\ge c_{1n}$, $\pi_i\neq \pi_j$ or $a_{1i}\neq a_{1j}$.
  \end{itemize}
\end{lem}
The validity of Assumption 4 in \cite{su2019strong} is implied by Lemma \ref{ler2} above. Combining this with Condition \ref{conder}, it is easy to check that Lemma 2.2 in \cite{su2019strong} holds. Following the same argument as in the proof of Theorem 2.3 in \cite{su2019strong}, we establish the exact recovery result for clustering the two groups of nodes $\{i\in[n],a_{1i}=1\}$ and $\{i\in[n],a_{1i}=0\}$ using Algorithm \ref{alg1}. Finally, we can apply the rest of arguments in the proof of Theorem \ref{t5} together with the exact recovery result and show that
$$\|\widehat\bbP^{\bbS,\bbS}_{(f_0,f_0)}-\widehat\bbP_{(f_0,*)}^{\bbS,\bbI-\bbS}\|_F=o_{a.s.}(p_n)\,,$$
and
	$$\|\widehat\bbP^{\bbS,\bbS}_{(\widehat f_0,\widehat f_0)}-\widehat\bbP_{(\widehat f_0,*)}^{\bbS,\bbI-\bbS}\|_F\sim p_n,\ \  a.s\, .$$
Therefore Theorem \ref{t5} holds almost surely and  the proof of exact recovery is completed.

Finally, we make the following remarks about the exact recovery result above. (i) Since Condition \ref{ler} is significantly more stringent than Condition 4 in the main paper, we have chosen to extend our theoretical results to the SBM setting only for illustrative purposes. (ii) While the lower bound on $p_n$ in Condition \ref{ler} may not be optimal, we note that the arguments already involve novel theoretical results from \cite{wu2021}, which is work currently in submission by the authors. We will leave refining this bound as one possible direction for future work. (iii) The distinct eigenvalue assumption in Condition \ref{ler} is made in our proof of Lemma \ref{ler}, as we refer to \cite{wu2021} for establishing exact recovery, where this assumption is required. In other words, if the eigenvector expansion can be extended to allow eigenvalue multiplicity, exact recovery without this assumption can be proved. Moreover, considering the treatment of eigenvalue multiplicity in establishing asymptotic expansions of eigenvectors in \cite{han2019universal}, we believe that both the results in \cite{wu2021} and our Lemma \ref{ler} still hold without assuming distinct eigenvalues. We plan to adopt a similar approach as presented in the proof of \cite{han2019universal} to improve the eigenvector expansion of \cite{wu2021}. Investigating this extension would require a significant amount of work and thus we will leave it for future research.

\subsection{Extension to $K$ growing with $n$}
\label{subsec:increasing_K}

It is possible to extend our results to the case of $K\to\infty$ using the current approaches. However, as mentioned earlier, observing only a partial network has already resulted in a significant amount of signal loss, and an increasing $K$ would only make the conditions more stringent. For clarity reasons, we choose to focus on $K=O(1)$ as this paper is the first attempt at investigating the proposed partial information framework. To get a sense of how conditions on $K$ and $p_n$ change for the main theorems, we give the following outline of key points.

\begin{itemize}
    \item
   Theorem \ref{thm2}, which describes the eigenvalues and eigenvectors of $\bbB_{E}$, only requires $\bbV^\top\bbS\bbV$ and $\bbI-\bbV^\top\bbS\bbV$ to be invertible. Since Theorem \ref{thm1} further establishes the order of the eigenvalues and thus requires a stronger condition on $K$,  we will first consider Theorem \ref{thm1} for $K\to\infty$. Theorem \ref{thm1} relies on Lemma \ref{le2}; inspecting the proof of this lemma, a crucial step involves establishing the entrywise inequality
$|\bbv_i^\top(\bbS-\E\bbS)\bbv_j|\le Cp_n\sqrt{\frac{\log n}{np_n}}$ with high probability, for some positive constant $C$. It follows then $\|\bbV^\top(\bbS-\E\bbS)\bbV\|_F\le KCp_n\sqrt{\frac{\log n}{np_n}}$, and requiring $K\ll \sqrt{\frac{np_n}{\log n}}$ would lead to the key result in Eq \eqref{g3b} for proving the lemma.

\item In Condition  \ref{cond1}, given that
$
\sum_{i=1}^Kd_i^2\le n^2p_n^2$, the condition should be revised as $|d_1|\sim|d_2|\sim\ldots\sim |d_K|\sim np_n/\sqrt{K}$. Consequently, the order of $|x_i|^{-1}$ now becomes  $np_n^{3/2}/\sqrt{K}$ in Theorem \ref{thm1}.

\item For community detection, now we have $\sigma_{2K}(\bbB_E)=O({\frac{np_n^{3/2}}{\sqrt K}})$ for the signal part. The order of the noise $\|\bbB-\bbB_E\|$ remains $O_p(\sqrt{np_n})$. Thus for the signal to dominate over noise, we would further require $p_n\gg \sqrt{K}/\sqrt{n}$.

\end{itemize}

\section{Additional algorithm}
\label{sec:app_add_alg}
Both Algorithm \ref{alg2} and \ref{alg3} use combinatorial optimization to resolve the label permutation problem arising from matching the $2K$ clusters from Algorithm \ref{alg1}, which scales exponentially in $K$. we propose the following heuristic algorithm as a faster alternative to exact combinatorial optimization. With a slight abuse of notation, we use $\widehat\bPi^{\bbS}$ to denote the submatrix of $\widehat\bPi$, consisting of row $i$ with $\{i: a_{1i}=1\}$; similarly $\widehat\bPi^{\bbI-\bbS}$ consists of row $i$ with $\{i: a_{1i}=0\}$. $\bbB^{\bbI-\bbS, \bbS}$ is similarly defined as a submatrix of $\bbB$.

We note that on the population submatrix, $(\E \bbB)^{\bbS, \bbI-\bbS} = \bPi^{\bbS} \bbP (\bPi^{\bbI-\bbS})^T$. Plugging in the estimated $\widehat\bPi^{\bbS}$ (from Algorithm \ref{alg1}) and $\widehat\bbP^{\bbS,\bbS}$, and replacing $\E \bbB$ by $\bbB$, we can estimate $\bPi^{\bbI-\bbS}$ using the least squares equation,
\begin{align}
\widehat\bPi^{\bbI-\bbS}_{\text{reg}} = \bbB^{\bbI-\bbS, \bbS} \widehat\bPi^{\bbS} \left( (\widehat\bPi^{\bbS} )^T \widehat\bPi^{\bbS} \right)^{-1} (\widehat\bbP^{\bbS,\bbS})^{-1}.
\label{eq:reg_est_rev}
\end{align}
Taking $\argmax\,\widehat\bPi^{\bbI-\bbS}_{\text{reg}} (i, :)$ for each row $i$ gives one version of estimated labels for the non-neighbor nodes,  which are consistent with $\widehat\bPi^{\bbS}$ in terms of group labeling. Recalling that Algorithm \ref{alg1} gives another version of estimated labels $\widehat\bPi^{\bbI-\bbS}_{1}$, we resolve the label permutation between $\widehat\bPi^{\bbI-\bbS}_{1}$ and $\widehat\bPi^{\bbI-\bbS}_{\text{reg}} $ through a majority vote. Any remaining unresolved labels will go through the same combinatorial optimization as in the original Algorithm \ref{alg2} (in a significantly reduced space). The details of this heuristic algorithm are presented in Algorithm \ref{alg:alg4}. For DCSBM, the same algorithm can be used as a faster alternative to exact combinatorial optimization for the label matching step.

Empirically, we observe that combining this algorithm with Algorithm \ref{alg1} gives almost identical accuracy rates on all simulated and real data. Figure \ref{fig:runtime_K} compares the running time of Algorithm \ref{alg1} and Algorithm \ref{alg:alg4}, the two main components of our community detection algorithm, as $K$ increases. For each datapoint, 10 networks are generated under Model 1 with $200$ nodes in each block. The running time of Algorithm \ref{alg:alg4} (the matching step) is negligible compared to Algorithm \ref{alg1} (the main spectral clustering part).

Finally, we make two remarks about Algorithm \ref{alg:alg4}. (i) Eq \eqref{eq:reg_est_rev} itself in general is not a good estimate of $\bPi^{\bbI-\bbS}$ because, as shown in Lemma \ref{errorbound}, $\E\bbB$ is not the signal term of $\bbB$. Thus, we only use this estimate to guide the search for the optimal label permutation. (ii) We have used $\widehat\bPi^{\bbS}$ as the ``reference'' set in Eq \eqref{eq:reg_est_rev} instead of $\widehat\bPi^{\bbI-\bbS}_1$, since the neighbor nodes have more observed edges than the non-neighbor nodes and should contain more information.
}

\begin{algorithm}[h!]
	\caption{Community detection under the SBM (faster alternative for matching)}

\textbf{Input}: $\widehat\bPi^{\bbS}$, $\widehat\bPi^{\bbI-\bbS}_{1}$ and its corresponding clusters  $\{\mathfrak{d}_1,\ldots,\mathfrak{d}_K\}$ (from Algorithm \ref{alg1}); $\widehat\bbP^{\bbS,\bbS}$ and $\widehat\bbP^{\bbS,\bbI-\bbS}$ from \eqref{eqn:prob estimate}.\\
\textbf{Output}: optimal permutation $\hat{f}_0$. \\
\textbf{Initialize}: $f=\mathbf{0}_{1\times K}$; $\tau_k^*=0$, $k\in[K]$.

 \begin{algorithmic}[1]
    \State  Compute $\widehat\bPi^{\bbI-\bbS}_{\text{reg}}$ defined in \eqref{eq:reg_est_rev}. Define estimated membership matrix $\widehat\bPi^{\bbI-\bbS}_{2}$ as $\widehat\bPi^{\bbI-\bbS}_{2}(i, \argmax \, \widehat\bPi^{\bbI-\bbS}_{\text{reg}} (i, :))=1$
    \For {$k=1, \dots, K$}
    \State $\tau_k \gets \frac{\mathbf{1}^T \widehat\bPi^{\bbI-\bbS}_{2}(i\in\mathfrak{d}_k, :)}{|\mathfrak{d}_k|}$ \Comment{\textit{determine label mapping from $\widehat\bPi^{\bbI-\bbS}_{1}$ to $\widehat\bPi^{\bbI-\bbS}_{2}$ by majority vote}}
    \State $\tau'^*_k \gets \max \tau_k(:)$
    \If {$\tau'^*_k > \tau_k^*$}
    \State $\tau_k^* \gets \tau'^*_k$
    \State $f(k) \gets \argmax \,
 \tau_k(:)$
    \EndIf
    \EndFor
    \If {number of zero elements in $f$=0}
    \State return $\hat{f}_0 \gets f$;
    \Else \Comment{\textit{any unresolved labels go through combinatorial optimization}}
    \State extract $\mathcal{K} \gets \{k\in[K]:f(k)=0\}$
    \State for all permutations $\mathcal{S}_{\mathcal{K}}$ of the labels in $\mathcal{K}$,  $\hat{f}_0 \gets \argmin_{f\in\mathcal{S}_{\mathcal{K}}}\|\widehat\bbP^{\bbS,\bbS}_{(f,f)}-\widehat\bbP_{(f,*)}^{\bbS,\bbI-\bbS}\|_F$.
    \EndIf
  \end{algorithmic}
  \label{alg:alg4}
\end{algorithm}

\begin{figure}[h!]
    \centering
    \includegraphics[height=2.8in]{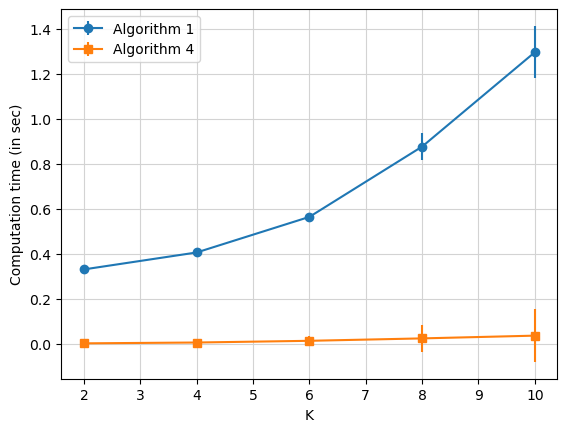}
    \caption{Running time of Algorithm \ref{alg1} and Algorithm \ref{alg:alg4} (the two main components of our community detection algorithm) under SBM as $K$ increases.}
\label{fig:runtime_K}
\end{figure}

\newpage
\section{Additional results from simulation}
\label{sec:supp_sim}

\subsection{Additional tables and figures for Model $1$}

\begin{table}
	\caption{The ratio of the  edges observed by individual $1$ out of the full networks for Model $1$, averaged over $100$ datasets for each $(n,q)$ combination.}
	\label{tab:addlabel1}
	\centering
	\begin{tabular}{c|c|c|c|c|c|c|c}
		\toprule
		$q$ $\backslash$ $n$    & 300&600& 900 & 1200 & 1500 & 1800 & 2100 \\
		\midrule
		.1     &.3590  & .3560  & .3587  & .3567  & .3560  & .3575  & .3559  \\
		$\sqrt{\log n/n}$    &.4696  & .3645  & .3147  & .2836  & .2613  & .2402  & .2237  \\
		$(\log n/n)^{1/4}/2$     & .5890  & .5280  & .4902  & .4693  & .4511  & .4368  & .4266  \\
		$1/\sqrt n$      &.2226 & .1580  & .1290  & .1117  & .1002  & .0925  & .0839  \\
		\bottomrule
	\end{tabular}
\end{table}

\begin{table}
	\caption{The fraction of the individuals within individual $1$'s knowledge depth for Model $1$, averaged over $100$ datasets for each $(n,q)$ combination. }
	\label{tab:addlabel2}
	\centering
	\begin{tabular}{c|c|c|c|c|c|c|c}
		\toprule
		$q$ $\backslash$ $n$     & 300& 600 &900 & 1200 & 1500 & 1800 & 2100 \\
		\midrule
		.1     & 1 & 1  & 1  & 1  & 1  & 1  & 1  \\
		$\sqrt{\log n/n}$    &1 & 1  & 1  & 1  & 1  & 1  & 1  \\
		$(\log n/n)^{1/4}/2$     &1 & 1  & 1  & 1  & 1  & 1  & 1  \\
		$1/\sqrt n$       & .9766  & .9732  & .9726  & .9724  & .9726  & .9726  & .9711  \\
		\bottomrule
	\end{tabular}
\end{table}

\begin{figure}[hbtp!]
	\centering
	\includegraphics[width=12cm]{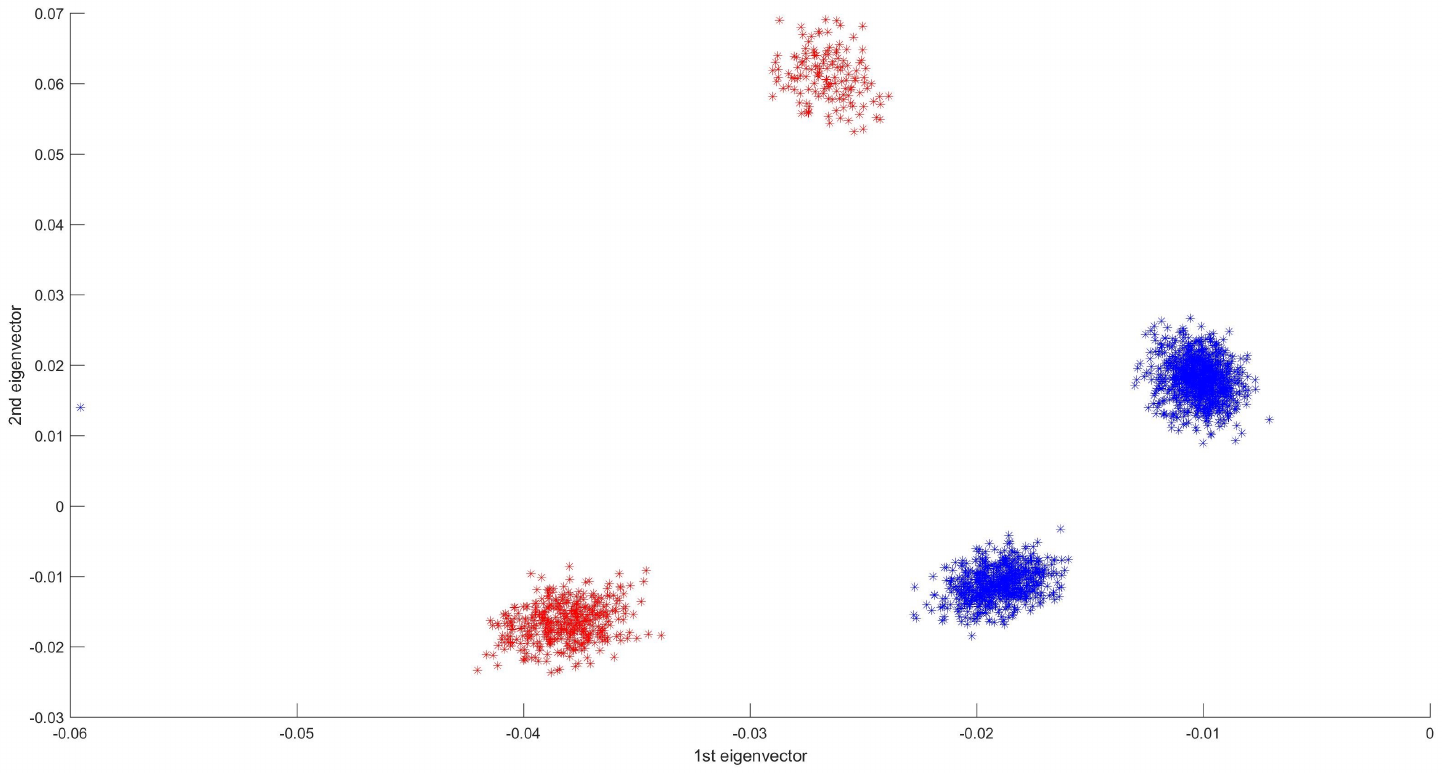}
	\caption{Scatter plot of the two eigenvectors corresponding to the positive eigenvalues of $\bbB$ for one dataset from Model $1$ when  $q=(\log n/n)^{1/4}/2$ and $n=2100$. The blue points are not adjacent to individual $1$, while the red points are adjacent to individual $1$.}
	\label{f3_supp}
\end{figure}

\begin{figure}[hbtp!]
	\centering
	\includegraphics[width=12cm]{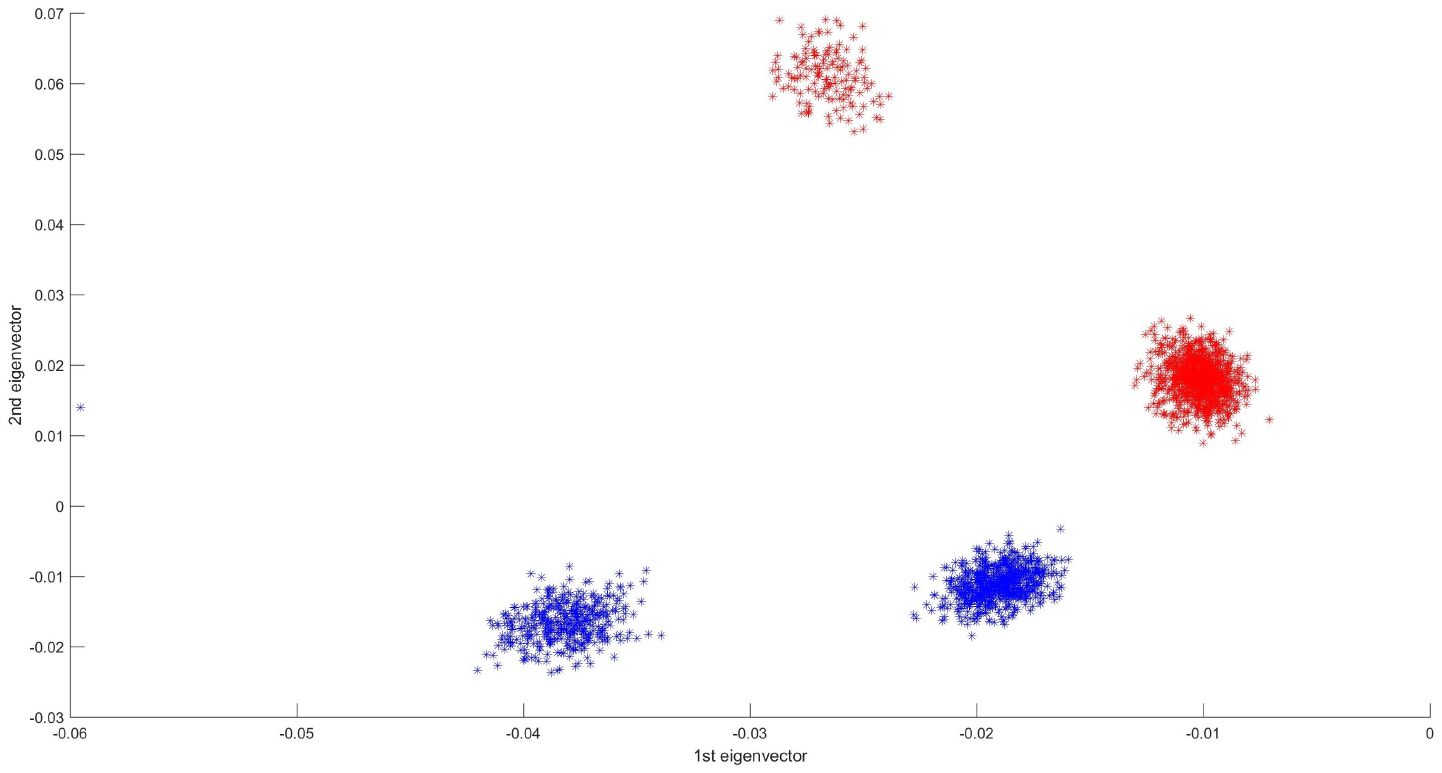}
	\caption{Scatter plot of the first two eigenvectors corresponding to positive eigenvalues of $\bbB$ for one dataset from Model $1$ when  $q=(\log n/n)^{1/4}/2$ and $n=2100$. The blue points belong to community $1$ and the red points belong to community $2$.}
	\label{f4_supp}
\end{figure}

We plot Figure \ref{f3} to provide some visualization support of Algorithm \ref{alg2}. This is a scatter plot in which the axes are the two eigenvectors corresponding to the positive eigenvalues of $\bbB$.  Recall that we have shown in previous sections that $\bbB$ for $K=2$ has two positive eigenvalues and two negative ones. The outlier point (a blue point) close to the vertical axis between $.01$ and $.02$ represents  individual $1$. The blue points are the individuals not adjacent to individual $1$, while the red points represent those who are adjacent to $1$.  Then on the same dataset,  we color the points by their true community memberships in Figure \ref{f4_supp}. Comparing Figure \ref{f3_supp} with Figure \ref{f4_supp}, one can see that it makes sense to develop a strategy to first apply $k$-means respectively  to the two groups of individuals separated by whether they are adjacent to individual $1$ and then merge the corresponding clusters across groups.

\begin{figure}[hbtp!]
	\centering
        \includegraphics[width=6cm]{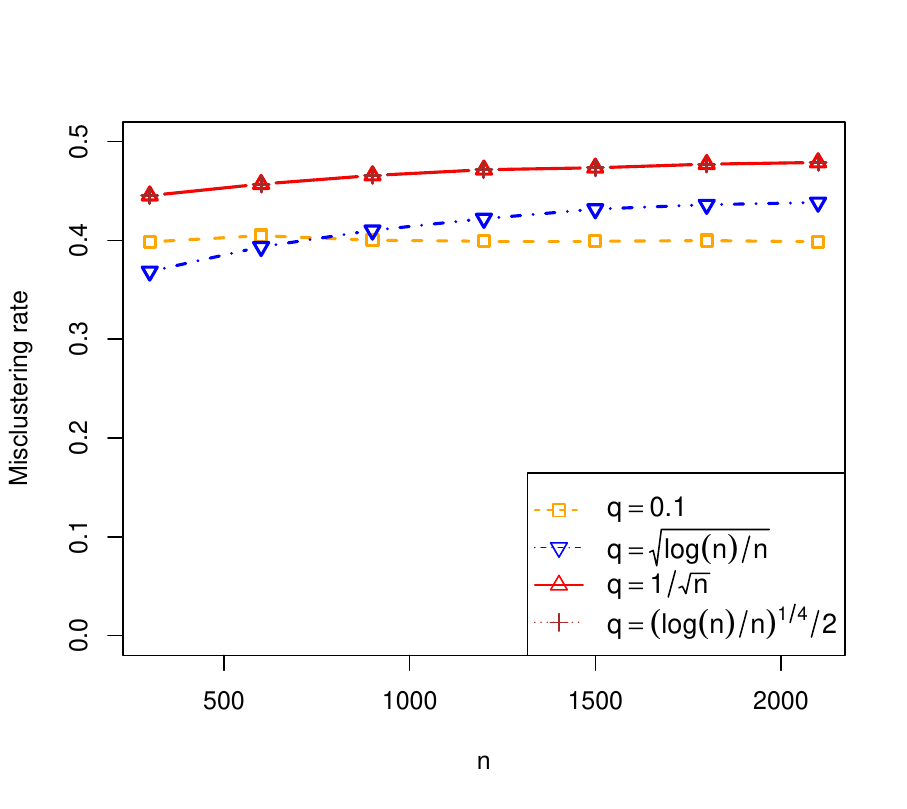}
        \caption{Misclustering rate for Model 1 ($L=2$) obtained by applying spectral clustering directly to $\bbB$, averaged over 100 datasets for each combination.}
        \label{fig:plainspec_model1}
\end{figure}

In addition, to demonstrate the need for our algorithms, we apply spectral clustering to $\bbB$ directly and plot the misclustering rates in Figure \ref{fig:plainspec_model1}. For all choices of $n$ and $q$, this approach fails to produce satisfactory clustering accuracy.

\subsection{Additional figure for Model $2$}

\begin{figure}[hbtp!]
	\centering
	\begin{subfigure}{0.5\textwidth}
		\centering
		\includegraphics[width=7cm]{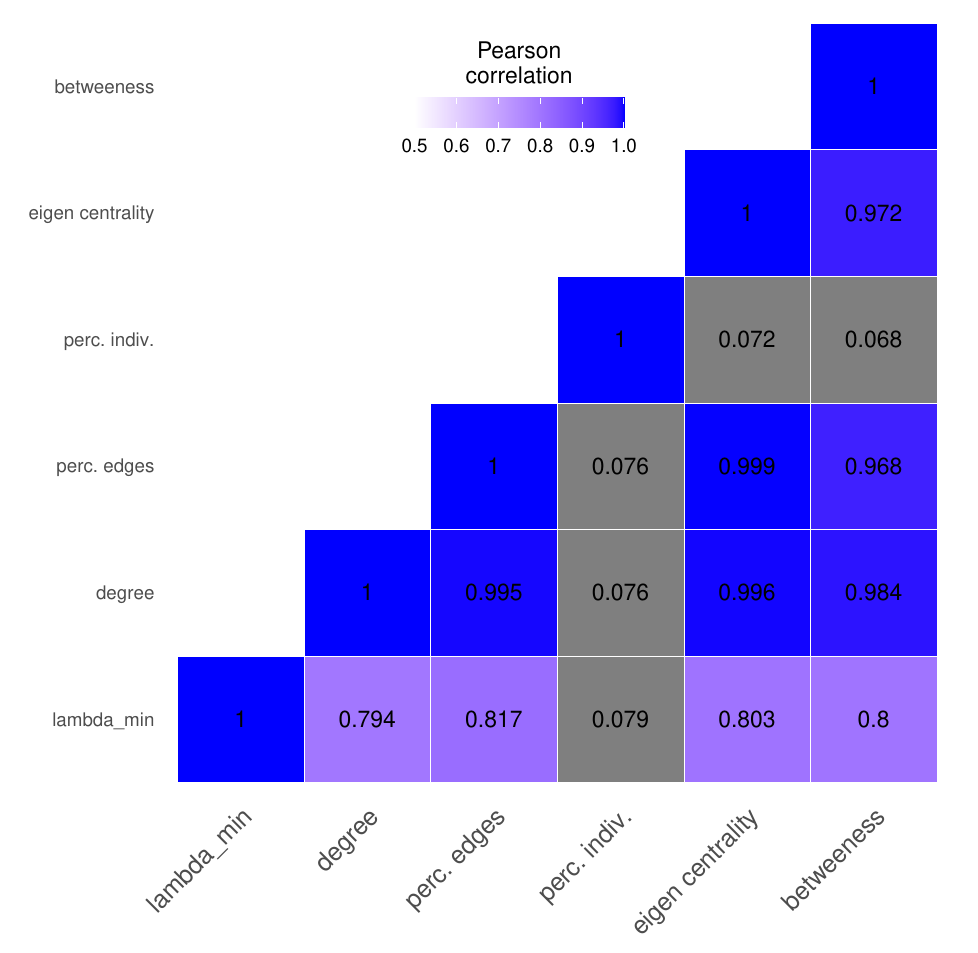}
		\caption{}
	\end{subfigure}%
	\begin{subfigure}{0.5\textwidth}
		\centering
		\includegraphics[width=7cm]{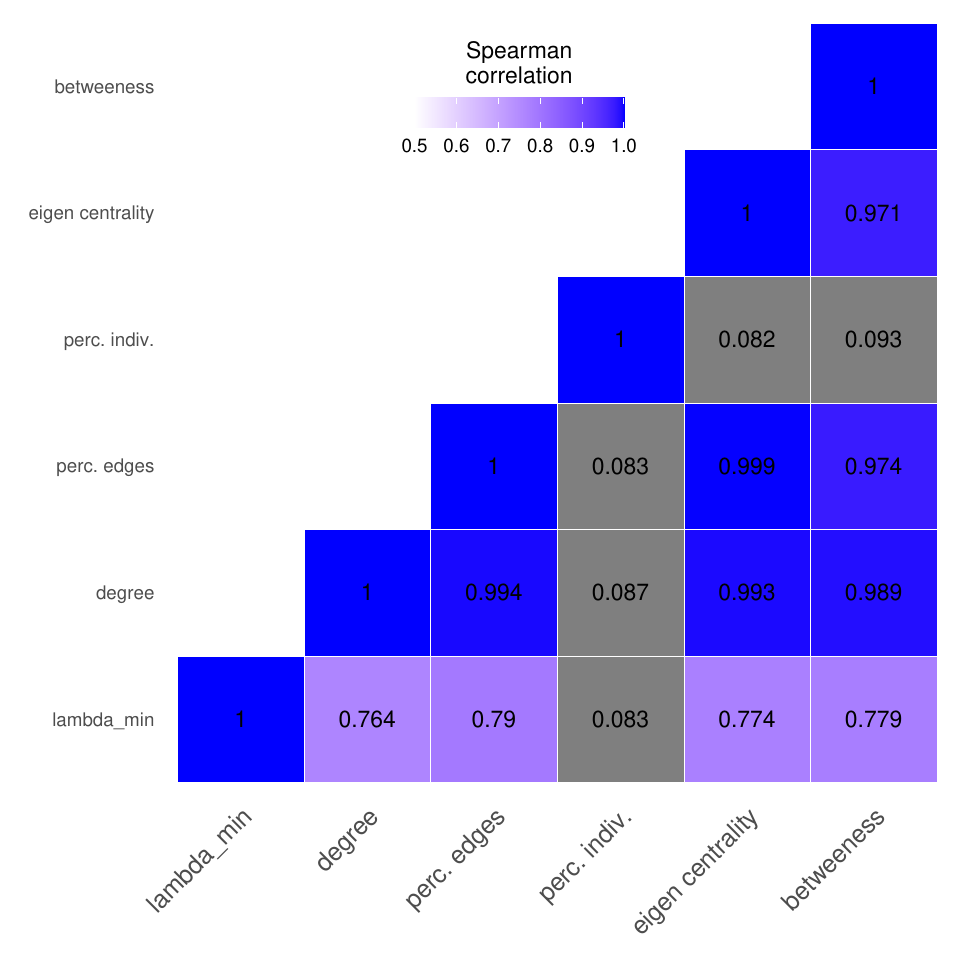}
		\caption{}
	\end{subfigure}
	\caption{Pearson (a) and Spearman (b) correlations between centrality measures under Model $2$.}
	\label{fig:cent_cor_simu}
\end{figure}

Figure \ref{fig:cent_cor_simu} computes the Pearson and Spearman correlations between five types of centrality measures using a randomly generated network from Model $2$. $\hat{\lambda}_{\min}$ is less correlated with the other four measures than those measures among themselves.

\subsection{Additional model for SBM}\label{sec:add plots model 2}

Model $3$ ($K=3$): $\bbP=\left(
\begin{array}{ccc}
	3q & 1.5q &q \\
	1.5q& 3q &1.5q\\
	q& 1.5q &3q\\
\end{array}
\right)$ and each group is of size $n/3$.

\begin{figure}[hbtp!]
	\centering
	\includegraphics[width=14cm]{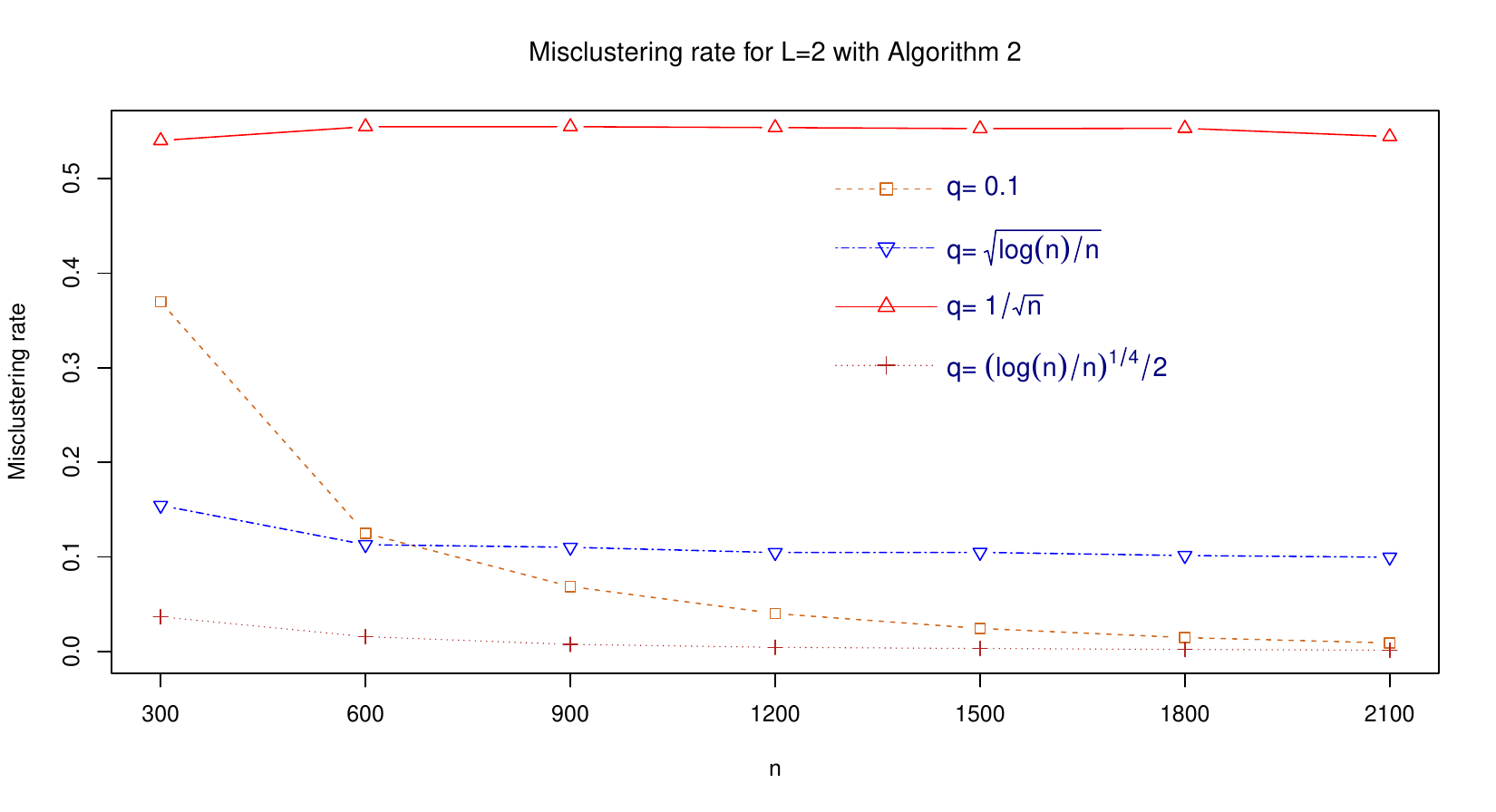}
	\caption{Misclustering rate for Model $3$ ($L=2$), averaged over $100$ datasets for each combination.}
	\label{f2s}
\end{figure}

\begin{figure}[hbtp!]
	\centering
	\includegraphics[width=14cm]{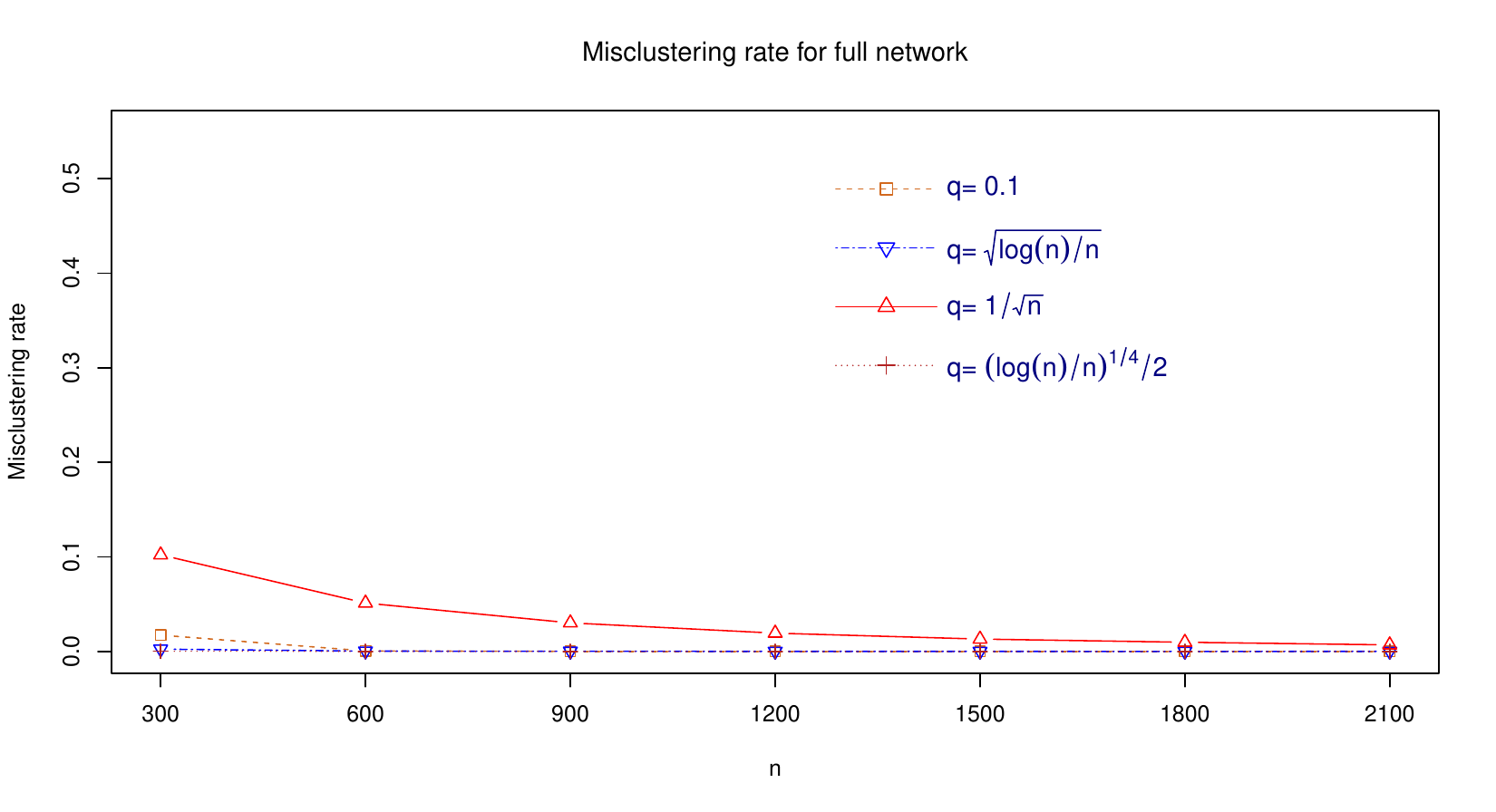}
	\caption{Misclustering rate for Model $3$ (full network), averaged over $100$ datasets for each combination.}\label{f3s}
\end{figure}

Similar to their counterparts for Model $1$, Tables \ref{tab:addlabel3} and \ref{tab:addlabel4} show that in Model $3$ ($K=3$), although individual $1$ can observe at least one edge of (almost) every other individual in the network, the proportion of total missing edges in her perspective is large. The visualization of Model $3$ data in $3$-D, similar to that of Model $1$ in $2$-D,  to support Algorithm \ref{alg2} is attached in Section \ref{sec:add plots model 2} of the Appendix.  We report the misclustering results in Figure \ref{f2s}.  This figure indicates that $K=3$  is a more challenging situation compared to $K=2$.  Algorithm \ref{alg2} for  $q= 1/\sqrt{n}$ with  Model $3$ works worse than with Model $1$. Also note that $q = \sqrt{\log n / n}$ in Model  $1$ delivers almost perfect clustering results, but the trend in Figure \ref{f2s} suggests that even as $n$ goes to infinity, the misclustering rate does not seem to go down to zero. The rate $p_n\sim q=\sqrt{\log n/n}$ is smaller than the rate in the theoretic Condition \ref{cond6}; but at this rate, Algorithm \ref{alg2} works well for Model $1$  while its performance is acceptable for Model $3$.    As a comparison, we report  in Figure \ref{f3s} the simulation results based on the adjacency matrix $\bbA$ with the usual spectral clustering algorithm. For larger $n$ in each combination, the misclustering rate is very close to $0$, which is theoretically guaranteed by a few works in the literature (c.f. \cite{abbe2017}).  Although we did not work on the boundary condition, we conjecture that the boundary condition under the new partial information framework for almost exact recovery is at least of order   $\sqrt{\log n/ n}$.

\begin{table}
	\caption{ \label{tab:addlabel3} The ratio of the  edges observed by individual $1$ out of the full networks for Model $3$,  averaged over $100$ datasets for each $(n,q)$ combination. }
	\centering
	\begin{tabular}{c|c|c|c|c|c|c|c}
		\toprule
		$q$ $\backslash$ $n$     & 300& 600 &900 & 1200 & 1500 & 1800 & 2100 \\
		\midrule
		.1     & .3298  & .3302  & .3309  & .3314  & .3286  & .3308  & .3294  \\
		$\sqrt{\log n/n}$    &.4387  & .3407  & .2932  & .2619  & .2361  & .2221  & .2069  \\
		$(\log n/n)^{1/4}/2$     & .5531  & .4941  & .4623  & .4413  & .4199  & .4064  & .3948  \\
		$1/\sqrt n$       & .2018  & .1441  & .1204  & .1047  & .0930  & .0849  & .0797  \\
		\bottomrule
	\end{tabular}%
	
\end{table}%

\begin{table}
	\caption{\label{tab:addlabel4} The fraction of the individuals within individual $1$'s knowledge depth for Model $3$, averaged over $100$ datasets for each $(n,q)$ combination.}
	\centering
	\begin{tabular}{c|c|c|c|c|c|c|c}
		\toprule
		$q$ $\backslash$ $n$     & 300& 600 &900 & 1200 & 1500 & 1800 & 2100 \\
		\midrule
		.1     & .9999 & 1  & 1  & 1  & 1  & 1  & 1  \\
		$\sqrt{\log n/n}$    &1 & 1  & 1  & 1  & 1  & 1  & 1  \\
		$(\log n/n)^{1/4}/2$     &1 & 1  & 1  & 1  & 1  & 1  & 1  \\
		$1/\sqrt n$       & .9653  & .9632  & .9619  & .9642  & .9632  & .9656  & .9648  \\
		\bottomrule
	\end{tabular}%
	
\end{table}%

\section{Additional results from real data analysis}

\subsection{Karate club data}

Table \ref{z2} shows the community detection accuracy for selected individuals in the karate club network, using Algorithm  \ref{alg2} and plain spectral clustering. The latter in general performs worse than our algorithm. 

\begin{table}
	\caption{\label{z2} The network information and detection results for chosen individuals using Algorithm \ref{alg2} and plain spectral clustering. }
	\centering
	
	\begin{tabular}{l| c c c c c c}
		\hline
		individual of interest    & H & 2 &3 & A & 20 & 32\\
		\hline
		accuracy (Algorithm \ref{alg2})  & .676 & .559 & .676   & .706  & .971  & 1  \\
accuracy (spectral clustering) &.706 & .529 & .529 &
.529 &
.559 &
.500 \\
		\hline
	\end{tabular}%
	
\end{table}%

\subsection{Indian villages}
\label{sec:indian_supp}

\begin{figure}[hbtp!]
	\centering
	\begin{subfigure}[b]{0.5\textwidth}
		\centering
		\includegraphics[width=8cm]{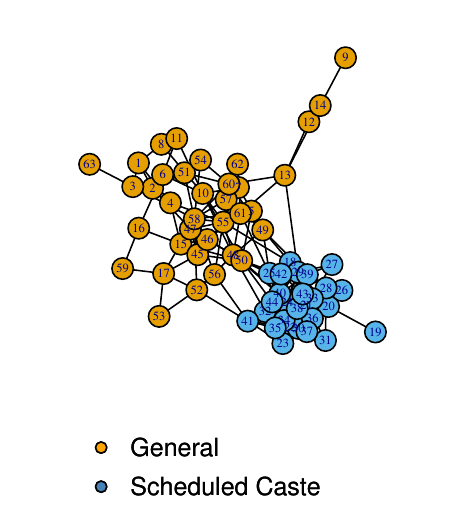}
		\caption{}
	\end{subfigure}%
	\begin{subfigure}[b]{0.5\textwidth}
		\centering
		\includegraphics[width=6cm]{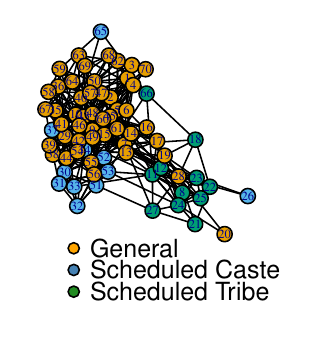}
		\caption{}
	\end{subfigure}
	\caption{Social networks in villages 2 (a) and 49 (b), colored by caste information.}
	\label{fig:vil_networks}
\end{figure}

Figure \ref{fig:vil_networks} shows two examples of social networks from the dataset, with nodes (households) colored by their caste.

\noindent \textbf{Data preprocessing}.
For each of the 43 villages, the dataset contains households sampled for individual surveys, which contain  their meta information such as caste, religion, language and occupation. Within each village, we first remove households whose caste information is ``NA'' or inconsistent among the surveyed household members (both constituting only a small number of the total). To avoid imbalanced cluster sizes, in each village we also remove castes with fewer than 10 households. After filtering out these households, we include villages with $K\geq 2$. The process results in 39 villages to be included in the analysis.

Both the eigenvector and betweenness centrality are normalized to make them comparable across different villages.

\begin{figure}[hbtp!]
	\centering
	\begin{subfigure}[b]{0.5\textwidth}
		\centering
		\includegraphics[width=7cm]{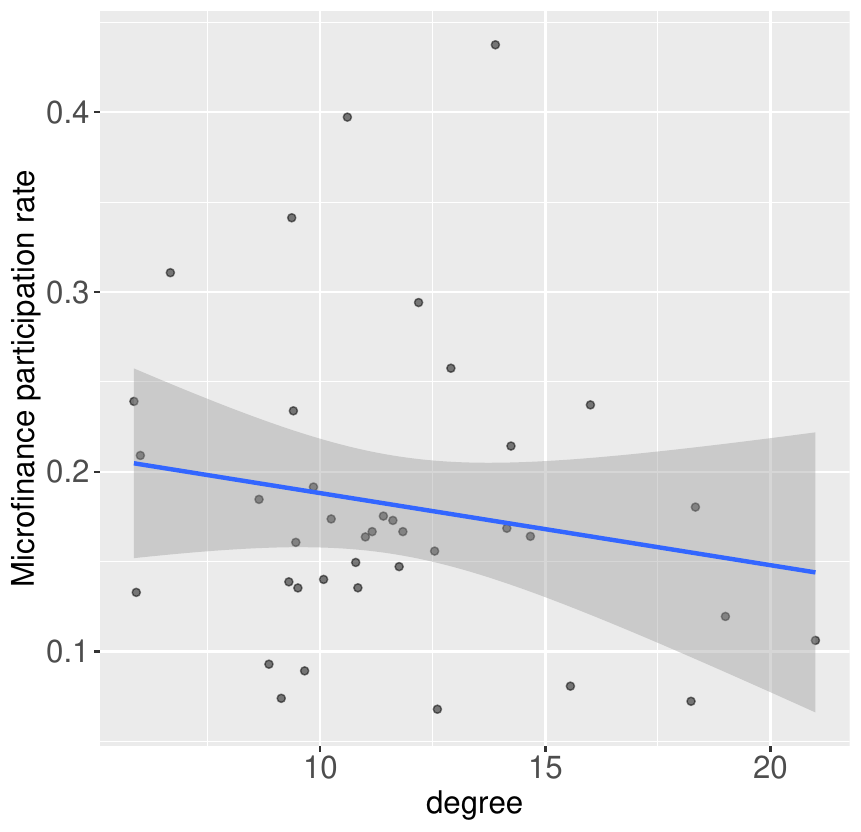}
		\caption{}
	\end{subfigure}%
	\begin{subfigure}[b]{0.5\textwidth}
		\centering
		\includegraphics[width=7cm]{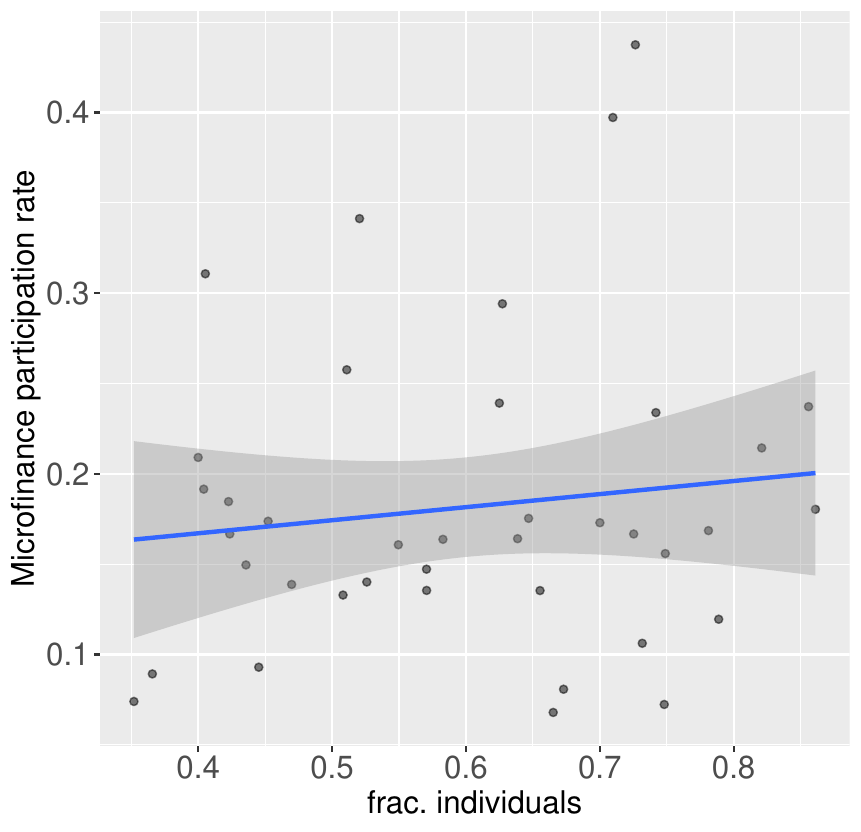}
		\caption{}
	\end{subfigure} \\

 \begin{subfigure}[b]{0.5\textwidth}
		\centering
		\includegraphics[width=7cm]{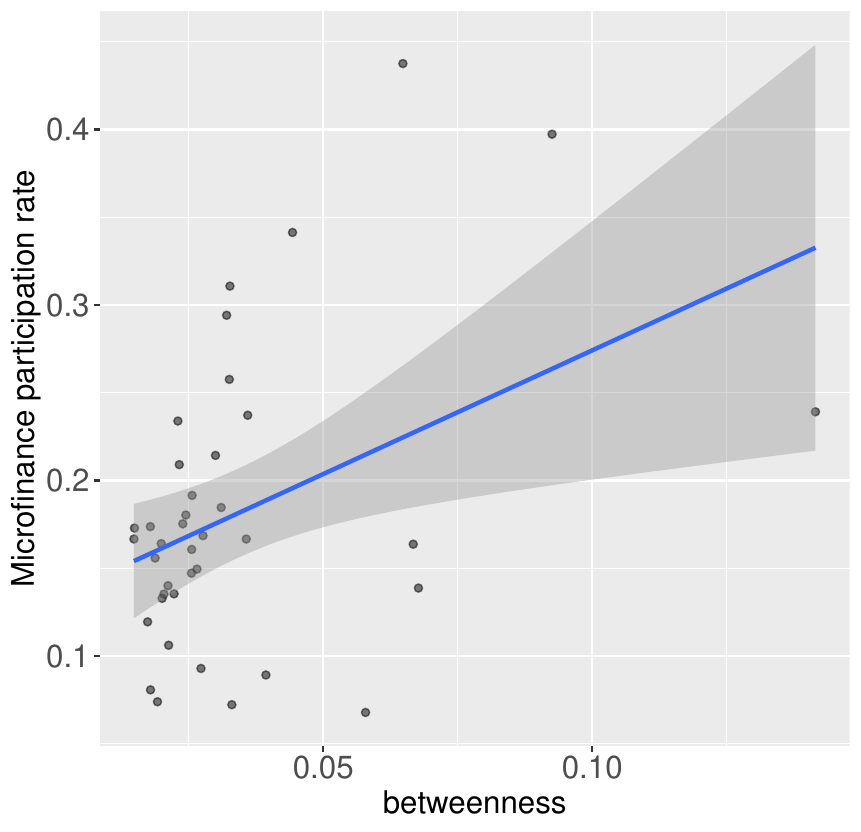}
		\caption{}
	\end{subfigure}%
	\begin{subfigure}[b]{0.5\textwidth}
		\centering
		\includegraphics[width=7cm]{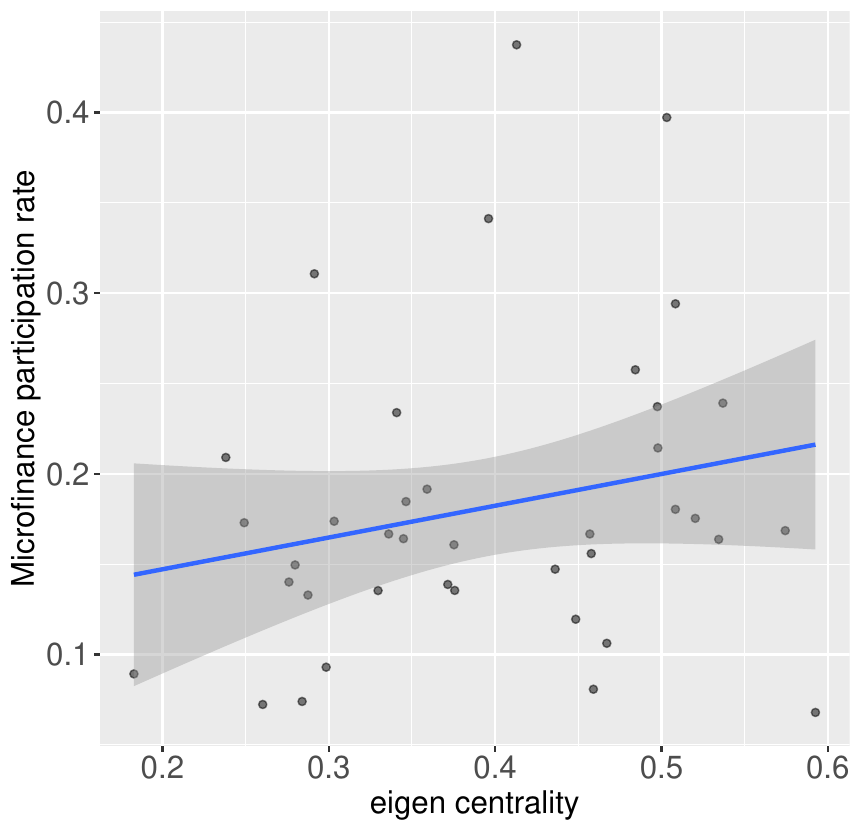}
		\caption{}
	\end{subfigure}
	\caption{Program participation rate as a function of (a) node degree; (b) fraction of individuals reached by the partial network; (c) betweenness centrality; (d) eigenvector centrality. The centrality measures are calculated for leaders in each village and an average is taken across each village.}\label{fig:mfrate_reg_app}
\end{figure}

\subsection{Political blogs}

\begin{table}[h!]
	\caption{\label{z5} The network information and detection results for individuals (blogs) in the political blog data using Algorithm \ref{alg2}.}
	\centering
	
	\begin{tabular}{l| c c c c c c}
		\hline
		node of interest   & 1073 & 1074 &1075 & 1076 & 1077 &1078\\
		
		\hline
		the ratio of the edges observed  & .1145 & .1295 & .3362  & .1553  &.2761 & .4116  \\
		$\#$ of the nodes observed & 476 & 485 & 880 & 715 & 808  & 793  \\
		clustering accuracy  & .5819 & .5505 & .9090 & .9035  & .9331& .8601  \\
		\hline
	\end{tabular}%
	
\end{table}%

\vskip 0.2in

\bibliographystyle{plainnat}
\bibliography{references}

\end{document}